\documentclass[aps,amsmath,amssymb,twocolumn,groupedaddress,longbibliography,floats,final,postprint,superscriptaddress]{revtex4-2}
\usepackage{titlesec}
\usepackage{anyfontsize}
\usepackage{enumerate}
\usepackage{graphicx}
\usepackage{amsmath,amssymb,amsthm,enumitem}

\usepackage{color}
\definecolor{LinkColor}{RGB}{199,21,133}
\usepackage[colorlinks=true,linkcolor=blue,anchorcolor=blue,citecolor=blue,urlcolor=blue]{hyperref}
\usepackage[polutonikogreek,english]{babel}

\newcommand{\headline}[1]{\textit{\textcolor{blue}{#1.}}}

\usepackage{hyperref}

\usepackage{epstopdf}
\usepackage{type1cm}
\usepackage{multirow}
\usepackage{times}

\begin{document}
\title{Long-Range Order in a Strictly Short-Range Quasi-2D XY Model: When Critical Fluctuations Matter}

\author{Minghui Hu}
\thanks{M.H. and C.Z. contributed equally to this work.}
\affiliation{Department of Physics and Anhui Province Key Laboratory for Control and Applications of Optoelectronic Information Materials, Anhui Normal University, Wuhu, Anhui 241000, China}
\author{Chao Zhang}
\thanks{M.H. and C.Z. contributed equally to this work.}
\affiliation{Department of Physics and Anhui Province Key Laboratory for Control and Applications of Optoelectronic Information Materials, Anhui Normal University, Wuhu, Anhui 241000, China}
\author{Dajun Zhang}
\affiliation{Department of Physics and Anhui Province Key Laboratory for Control and Applications of Optoelectronic Information Materials, Anhui Normal University, Wuhu, Anhui 241000, China}
\author{Yanan Sun}
\affiliation{Department of Physics and Anhui Province Key Laboratory for Control and Applications of Optoelectronic Information Materials, Anhui Normal University, Wuhu, Anhui 241000, China}
\author{Youjin Deng}
\email{yjdeng@ustc.edu.cn}
\affiliation{Department of Modern Physics, University of Science and Technology of China, Hefei, Anhui 230026, China}
\affiliation{Hefei National Laboratory, University of Science and Technology of China, Hefei 230088, China}
\author{Jian-Ping Lv}
\email{jplv2014@ahnu.edu.cn}
\affiliation{Department of Physics and Anhui Province Key Laboratory for Control and Applications of Optoelectronic Information Materials, Anhui Normal University, Wuhu, Anhui 241000, China}
\affiliation{Key Laboratory of Functional Molecular Solids, Ministry of Education, Wuhu, Anhui 241000, China}

\begin{abstract}
The phase of spins in the quasi-two-dimensional (q2D) XY model has emerged as a topic of significant interest across multiple subfields of physics. Conventional wisdom, rooted in the Mermin-Wagner theorem and supported by existing paradigms, asserts that true long-range (LR) order is prohibited in q2D systems with continuous symmetries and short-range (SR) interactions. In this Letter, we propose a strictly SR q2D XY model defined on a plane perpendicularly intersected by a group of parallel planes, where each plane consists of XY spins coupled via nearest-neighbor interactions. Through large-scale Monte Carlo simulations complemented by finite-size scaling analysis, we establish the complete phase diagram of the setup. A LR ordered phase emerges in the q2D model when the spins on the parallel planes develop a Berezinskii-Kosterlitz-Thouless critical phase. The LR ordered phase is anisotropic: true LR correlations develop exclusively along the direction of the intersection lines, while the perpendicular direction exhibits quasi-long-range order. Furthermore, the LR order exhibits Goldstone-mode physics. Our findings reveal a mechanism for stabilizing LR order in low-dimensional systems with continuous symmetries, thereby establishing a new platform for studying exotic superfluidity.
\end{abstract}

\maketitle

\headline{Introduction}
The two-dimensional (2D) XY model---a prototype model of superfluidity---plays a vital role in the study of critical phenomena. While the nearest-neighbor spin-spin coupling $J$ of the 2D XY model varies continuously, the Berezinskii-Kosterlitz-Thouless (BKT) transition occurs at $J=J_{\rm BKT}$~\cite{Kosterlitz17}. Although the Mermin-Wagner theorem forbids long-range (LR) order at any finite $J$, quasi-long-range (QLR) order emerges for $J>J_{\rm BKT}$. When LR couplings are turned on, LR order survives at finite couplings~\cite{Kunz76}, though debates persist regarding phase diagrams~\cite{Giachetti2021,Giachetti2022,Xiao2024,yao2024nonclassical}.

\begin{figure}[h!]
	\centering
	\includegraphics[height=6.9cm,width=8.2cm]{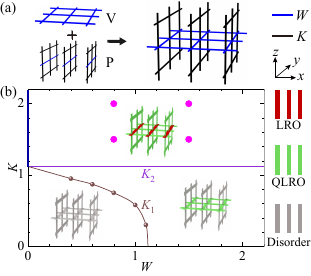}
	\caption{Setup and phase diagram. (a) Construction of the setup, which consists of a vertical (V) plane and $L$ parallel (P) planes, with each plane containing $L^2$ sites. The V plane is now a q2D XY model of size $L=3$. The blue and black edges represent the couplings in and out of V plane, with the coupling strengths $W$ and $K$, respectively. (b) Phase diagram of the setup. For $0<W<J_{\rm BKT}$, the V plane experiences a transition at $K_1$ ($K_1<J_{\rm BKT}$) and enters a LR ordered phase at $K_2$ ($K_2=J_{\rm BKT}$), with $J_{\rm BKT} \approx 1.11996$ the critical coupling of standard square-lattice XY model. The brown spheres represent the transitions at $K_1$, which are located at $(W,K) \approx (0.4,0.95)$, $(0.6,0.87)$, $(0.8,0.75)$, $(1,0.58)$ and $(1.10,0.3)$. For a finite $W>J_{\rm BKT}$, the V plane experiences a transition from the QLR to LR ordered phase at $K_2$. Each P plane experiences a BKT transition at $K_2$ from the disordered to QLR ordered phase. The columnar patterns mark the phases of V and P planes, which can be the LR and QLR ordered phases as well as the disordered phase. Nevertheless, for the limiting case $W=0$ and finite $K \geq K_2$ (dark blue line), the $x$ lines are disordered, while the $y$ lines are critical. The phases of V plane at the magenta solid dots are analyzed with Figs.~\ref{Fig3} and~\ref{Fig4}.}~\label{Fig1}
\end{figure}

Recently, encouraging progress has been made in the study of surface and interface critical phenomena, giving rise to a new concept of critical phase and motivating the search for LR order---a long-standing pivotal issue that continues to attract attention~\cite{bianchi2023conformal,trepanier2023surface,Raviv2023,Giombi2023,Bolla2023,Cuomo2024,sun25,sun25B}. The systems of interest can be viewed as quasi-two-dimensional (q2D) XY models coupled to a critical three-dimensional (3D) bulk. LR order is forbidden by a recently proposed no-go theorem in the context of boundary conformal theory for continuous symmetry breaking~\cite{Cuomo2024}. Indeed, the extraordinary-log (E-Log) phase---characterized by the non-trivial power-law decay with the logarithm of distance for the two-point correlation function---has been found in the surfaces and interfaces of XY models~\cite{Metlitski,Hu2021,ToldinMetlitski22,Krishnan2023,Sun2023} and other O($n$) systems~\cite{ParisenToldin21,Zhang22,Zou22,Sun22A,Sun22B,Zhang23}. The E-Log phase was also theoretically realized with quantum measurements~\cite{PRXQ.4.030317,baweja2024postmeasurement} and quantum Hall bilayers~\cite{ZhangXY23}. In contrast to the no-go theorem, for the O($n$) model with $n \ge 2$, it was only recently predicted~\cite{sun25} that the surface of tricritical 3D bulk can host LR order. 

While the prevalence of E-Log phase in the q2D systems underscores the challenge of stabilizing LR order, the existence of LR order in the quasi-one-dimensional (q1D) quantum systems with continuous symmetries is less controversial~\cite{Werner2005,Cazalilla2006,Sperstad2012,Weber2022,Ribeiro2024,Radzihovsky23,Kuklov24a,Kuklov24b,ZhangPhysRevB109,kuklov2024transverse,KuklovPhysRevResearch6,Radzihovsky24}. In quantum dissipative systems, the coupling to a dissipative bosonic bath leads to retarded interactions stabilizing a LR ordered ground state~\cite{Werner2005,Cazalilla2006,Sperstad2012,Weber2022,Ribeiro2024}. Very recently, the concepts ``transverse quantum fluid (TQF)" and ``incoherent TQF (iTQF)" were proposed~\cite{Radzihovsky23,Kuklov24a,Kuklov24b,ZhangPhysRevB109,kuklov2024transverse,KuklovPhysRevResearch6,Radzihovsky24}. TQF is a one-dimensional (1D) quantum phase of matter originally associated with edge dislocations of superfluid $^4$He~\cite{Radzihovsky23,Kuklov24a,Kuklov24b}. Intriguingly, TQF exhibits infinite compressibility as well as off-diagonal LR order and demonstrates the irrelevance of Landau criterion~\cite{kuklov2024transverse}. iTQF shares some defining features of TQF but is distinguished by a unique characteristic---diffusive dynamics---stemming from gapless transverse particle reservoirs. The signatures of iTQF have been found in 1D quantum lattice models of interacting bosons coupled to a bath~\cite{Kuklov24a,Kuklov24b,ZhangPhysRevB109,Radzihovsky24}.

Hence, the study of LR superfluid order in low-dimensional systems represents a topic of broad interdisciplinary interest, spanning not only surface and interface critical phenomena, but also dissipative dynamics, TQF and iTQF. A central question bridging these fields is whether LR order can persist in a q2D classical superfluid at finite coupling. To resolve this issue, a rigorously defined q2D system---constructed from a strictly short-range (SR) XY model---is essential.

\begin{figure}
	\includegraphics[height=7cm,width=8cm]{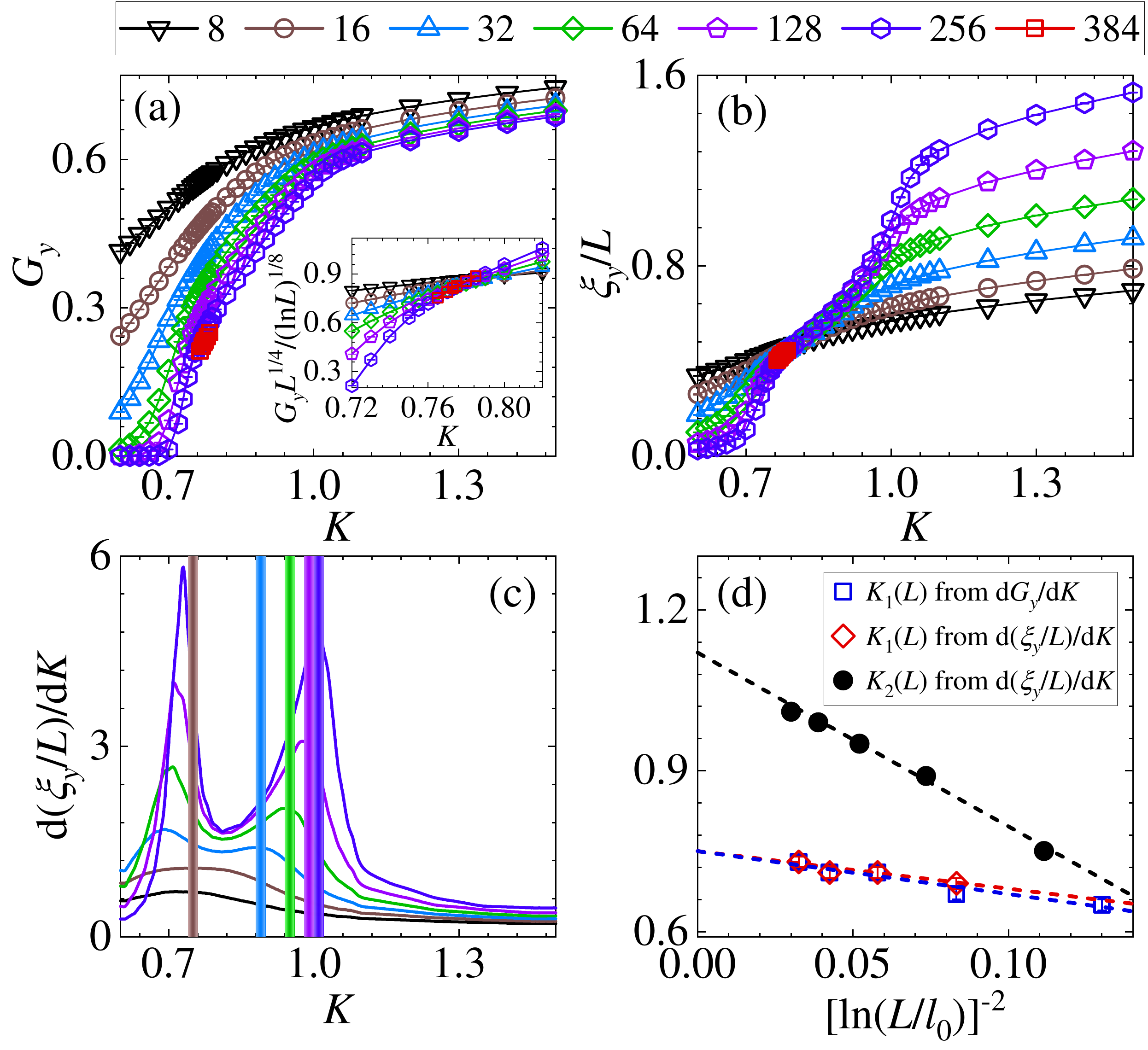}
	\caption{Phase transitions of a $y$ line in the V plane with $W=0.8$. (a) The large-distance spin-spin correlation $G_y$ versus $K$. Inset: $G_yL^{1/4}/({\rm ln}L)^{1/8}$ versus $K$. (b) The reduced correlation length $\xi_y/L$ versus $K$. (c) The derivative ${\rm d}(\xi_y/L)/{\rm d}K$ versus $K$. The shaded areas, whose widths represent two error bars, denote $K_2(L)$ (the coordinate of the peak at larger $K$). (d) $K_1(L)$ from ${\rm d}G_y/{\rm d}K$ and ${\rm d}(\xi_y/L)/{\rm d}K$ as well as $K_2(L)$ from ${\rm d}(\xi_y/L)/{\rm d}K$ versus $[{\rm ln}(L/l_0)]^{-2}$. Conforming to fits, $l_0=1$ and $0.8$ are set for $K_1(L)$ and $K_2(L)$, respectively. The intercepts $K_1=0.75$ and $K_2=1.11996$ represent the thermodynamic limits of $K_1(L)$ and $K_2(L)$, respectively.}~\label{Fig2}
\end{figure}

\headline{Main Results}
The construction of our setup is illustrated by Fig.~\ref{Fig1}(a), where a square-lattice q2D XY model (V plane) is perpendicularly intersected by a group of square-lattice XY models (P planes). The XY spins in the setup are coupled via ferromagnetic nearest-neighbor interactions. The Hamiltonian of the setup reads
\begin{equation}~\label{Ham0}	
\mathcal{H}/(k_{\rm B}T)=-\sum_{\langle {\bf rr'} \rangle}J_{{\bf rr'}}\vec{S}_{\bf r} \cdot \vec{S}_{\bf r'},	
\end{equation}
where $\vec{S}_{\bf r}$ represents an XY spin at lattice site ${\bf r}$, $J_{\bf rr'}$ denotes coupling strength, and the summation runs over pairs of nearest-neighbor sites. For the nearest-neighbor sites ${\bf r}$ and ${\bf r'}$, $J_{{\bf rr'}}$ is defined as
\begin{equation}\label{tp}
J_{{\bf rr'}} = \begin{cases} W, & {\bf r} \in {\rm V} ~~{\rm and}~~ {\bf r'} \in {\rm V},  \\
K,  & {\bf r} \notin {\rm V} ~~~~{\rm or}~~~ {\bf r'} \notin {\rm V}. \end{cases}
\end{equation}
Notably, the V plane is intersected by the P planes, with a coupling strength of $W$ along intersections. For each of the ($L+1$) square lattices, a V plane and $L$ P planes, periodic boundary conditions are imposed along ($x$, $y$)- or ($y$, $z$)-directions. The setup is not inherently 3D: the coupling between the P planes and the V plane is confined to subextensive lines, while the vast majority of sites---$O(L^3)$ in number---have a coordination number of four. As a result, every P plane remains macroscopically equivalent to a standard 2D XY model, precluding the development of LR order along the $z$-direction at any finite $K$. For $K=0$, the V plane reduces to a standard square-lattice XY model, which features a BKT transition at $W=J_{\rm BKT}$ ($J_{\rm BKT} \approx 1.11996$~\cite{Komura2012Large}).

Using extensive Monte Carlo simulations, we map out the full phase diagram of the setup, which is shown in Fig.~\ref{Fig1}(b). For $0<W<J_{\rm BKT}$, the q2D XY model---V plane---experiences successive transitions as $K$ varies continuously, whereas for a finite $W>J_{\rm BKT}$, it exhibits a single transition. For finite $K \geq J_{\rm BKT}$ and $W>0$, the P planes are critical, while the q2D XY model enters a LR ordered phase. Furthermore, our observations constitute strong evidence for anisotropic LR order: true LR coherence develops along the $y$-direction, whereas the QLR order emerges along the $x$-direction. The LR order of $y$ lines further exhibits Goldstone-mode physics. Hence, the present scenario is drastically different from iTQF. Now, the ``transverse'' superfluid (flowing along $x$-direction) is critical rather than exhibiting LR order.

\headline{Method}
We perform equilibrium Monte Carlo simulations of model~(\ref{Ham0}) for \textit{finite} $W$ \textit{and} $K$ \textit{couplings} using Wolff's single-cluster algorithm~\cite{Wolff89}, with linear lattice sizes up to $L=384$ (maximum number of spins: $L^3=56623104$). In the process of cluster growth, for the edge between a pair of nearest-neighbor sites (say ${\bf r}$ and ${\bf r'}$), a bond is randomly placed on with the probability $p=\max[0,1-e^{-2 J_{{\bf rr'}}S^{(\alpha)}_{\bf r}S^{(\alpha)}_{\bf r'}}]$, where $\alpha$ is a randomly chosen direction. Upon completion of cluster growth, the spins in the cluster are flipped along $\alpha$-direction. For $L=384$, the maximum number of created clusters reaches $1.2 \times 10^7$; details are given in Supplemental Material (SM).

\headline{Successive Phase Transitions}
We start by analyzing $y$ lines. In the case $W=0.8$ and $K=0$, the V plane is in a disordered phase. We turn on the coupling $K>0$ within the P planes. For a line along $y$-direction of the V plane, we sample the large-distance spin-spin correlation $G_y = \langle \vec{S}_{{\bf r}} \cdot \vec{S}_{{\bf r}+(0,L/2)} \rangle$. Figure~\ref{Fig2}(a) shows that $G_y$ increases with $K$. Using the finite-size scaling (FSS) form $G_y \sim L^{-1/4}({\rm ln}L)^{1/8}$~\cite{Janke1997,Chen2022} of a BKT transition, where the term $({\rm ln}L)^{1/8}$ represents a multiplicative logarithmic correction, we infer a transition point at $K_1 \approx 0.77$ [inset of Fig.~\ref{Fig2}(a)]. We then compute the second-moment correlation length $\xi_y=(2{\rm sin}\frac{\pi}{L})^{-1}\sqrt{\frac{\langle M^2_y \rangle}{\langle M^2_{yk} \rangle}-1}$ with $M_y=L^{-1} |\sum_{{\bf r} \in {\rm a} \, y \, {\rm line}} \vec{S}_{\bf r}| $ and $M_{yk}=L^{-1} |\sum_{{\bf r} \in {\rm a} \, y \, {\rm line}}\vec{S}_{\bf r}e^{i\frac{2\pi}{L}y} | $, where the summation runs over sites in a $y$ line. Figure~\ref{Fig2}(b) implies that, as $L$ increases, the reduced correlation length $\xi_y/L$ changes drastically at $K_1$ and $K_2 \approx 1.1$. In the intermediate regime $K_1< K< K_2$, $\xi_y/L$ tends to be scale-invariant, indicating the existence of QLR order. We infer that the BKT transition in the P planes at $K=J_{\rm BKT}$ is the cause of the change of criticality in the V plane; in other words, the criticality at $K_2$ in the V plane is inherited from the simultaneous BKT transition occurring in the P planes.

Figure~\ref{Fig2}(c) demonstrates that the derivative ${\rm d}(\xi_y/L)/{\rm d}K$ has a double-peaked structure. For each $L$, we define a pair of pseudo-critical points $K_1(L)$ and $K_2(L)$ with $K_1(L)<K_2(L)$, which are the $K$ coordinates of the two peaks. For an interpolation-based procedure used to obtain the derivative and pseudo-critical points (including estimated non-statistical error bars), see SM. We observe the slow convergence of $K_1(L)$ and $K_2(L)$ in the $L \rightarrow \infty$ limit, conforming to the characteristic BKT-type finite-size drift. We perform least-squares fits according to the BKT scaling formula~\cite{Tomita2002Probability}
\begin{equation}~\label{Equ_K1}
	K_n(L)=K_n+a[{\rm ln}(L/l_0)]^{-2}
\end{equation}
with $n=1$ and $2$ respectively, where $a$ denotes a non-universal constant and $l_0$ represents a reference length. Throughout this Letter, we examine the quality and stability of fits by monitoring $\chi^2/{\rm DOF}$ upon gradually increasing $L_{\rm min}$, with $\chi^2$ standard chi-squared, DOF the degree of freedom, and $L_{\rm min}$ the lower bound of $L$. We prefer the fits with $\chi^2/{\rm DOF}=O(1)$. Assuming the transition at $K_1$ is of BKT type, by fixing $l_0=1$---neglecting effective finite-size corrections and reducing uncertainties---we obtain $K_1=0.75(1)$ and $\chi^2/{\rm DOF} \approx 0.5$ with $L_{\rm min}=32$. For $K_2(L)$, the fit with free $l_0$ yields $K_2=1.07(2)$ and $\chi^2/{\rm DOF} \approx 0.1$ with $L_{\rm min}=16$. The estimated $K_2$ is close to $J_{\rm BKT} \approx 1.11996$~\cite{Komura2012Large}. If $K_2=1.11996$ is fixed, a preferred fit is obtained, producing $l_0=0.8(2)$ and $\chi^2/{\rm DOF} \approx 0.9$ with $L_{\rm min}=16$. Indeed, the FSS ansatz for the conventional BKT transition proves effective in locating $K_2$. Additionally, Fig.~\ref{Fig2}(a) indicates that ${\rm d}G_y/{\rm d}K$ is peaked at $K_1(L)$, for which a preferred fit produces $K_1=0.750(10)$ and $\chi^2/{\rm DOF} \approx 1.0$ with $L_{\rm min}=16$. Hence, the methodology is verified via the consistency of estimated $K_1$ from $G_y$, ${\rm d}(\xi_y/L)/{\rm d}K$ and ${\rm d}G_y/{\rm d}K$. For $n=1$ and $2$, Fig.~\ref{Fig2}(d) illustrates the linear dependence of $K_n(L)$ on $[{\rm ln}(L/l_0)]^{-2}$ as well as the $L \rightarrow \infty$ limit of $K_n(L)$.

Summarizing, we find successive transitions at $K_1$ and $K_2$ for $0<W<J_{\rm BKT}$. In SM, we verify that the transition at $K_1$ obeys the BKT scaling even when $W$ is much smaller, providing strong evidence that the $K_1$ critical line is BKT-type. For $W>J_{\rm BKT}$, the V plane is already in a QLR ordered phase at $K=0$. In SM, by increasing $K$, we confirm the existence of transition at $K_2$. The transition lines for $K_1$ and $K_2$ can be found in Fig.~\ref{Fig1}. 

\begin{figure}
\includegraphics[height=10cm,width=8cm]{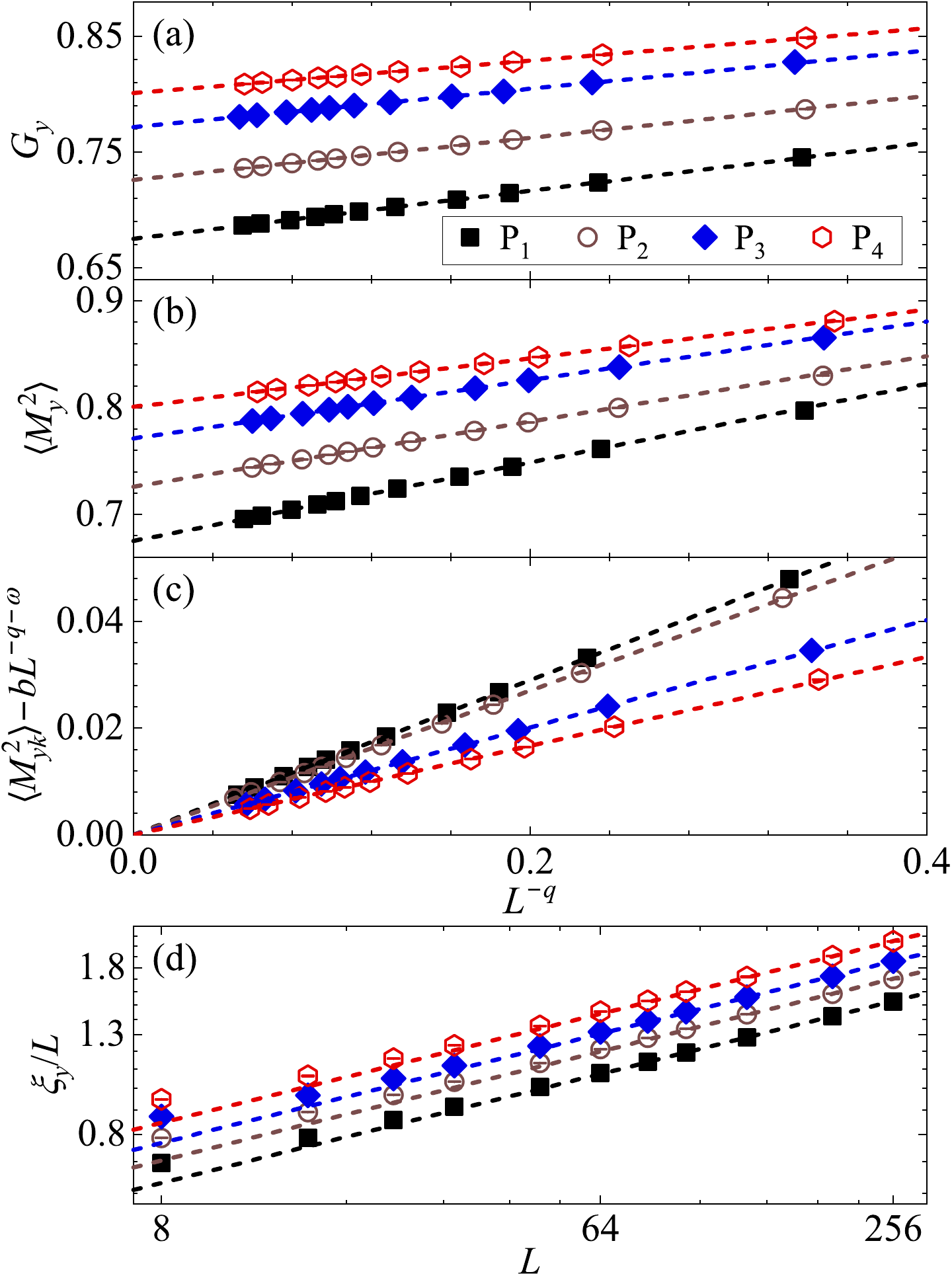}
\caption{Goldstone-mode effects of the LR order of a $y$ line in the V plane. (a)-(b) The large-distance spin-spin correlation $G_y$ and the zero-mode magnetic fluctuations $\langle M^2_y \rangle$ versus $L^{-q}$. The values of $q$ are obtained from preferred fits (Table~\ref{tab1}). The intercepts stand for the values of $G_y$ or $\langle M^2_y \rangle$ in the thermodynamic limit. (c) The magnetic fluctuations at smallest non-zero mode (with finite-size corrections being considered), $\langle M^2_{yk} \rangle-b L^{-q-\omega}$, versus $L^{-q}$. For ${\rm P_1}$, ${\rm P_2}$, ${\rm P_3}$ and ${\rm P_4}$, the values $\omega=0.62$, $0.39$, $1.1$ and $1.16$ are from preferred fits, respectively. (d) Log-log plot of the reduced correlation length $\xi_y/L$ versus $L$. The slope $0.255$ of dashed lines stands for $q/2$.}~\label{Fig3}
\end{figure}

\headline{Long-range Ordered Phase}
For $K \ge K_2$, Fig.~\ref{Fig2}(b) implies that $\xi_y/L$ grows rapidly with $L$, indicating the possibility of LR order along $y$-direction. In what follows, for $y$ and $x$ lines, we explore ordering by using four parameter sets ${\rm P_1}: (W=0.8, K=1.5)$, ${\rm P_2}: (W=0.8, K=2)$, ${\rm P_3}: (W=1.5, K=1.5)$ and ${\rm P_4}: (W=1.5, K=2)$.

If the scenario is the LR order associated with the breaking of O(2) symmetry, Goldstone mode probably plays a role and spin-spin correlation function scales as~\cite{kardar2007statistical}
\begin{equation}~\label{gy}
g(r) = a+ b r^{-q}
\end{equation}
with $r$ the spatial distance, where $q$ is an exponent controlling leading decay and $b$ denotes a non-universal constant. Further, the value of $q$, arising from the Gaussian expansion of the transverse (Goldstone) mode around the LR spontaneous order $m$, is expected to be independent of couplings $W$ and $K$, which only affect the magnitude of $m$. It follows that $G_y$ and the magnetic fluctuations $\langle M^2_y \rangle$ respectively obey the FSS formulae~\cite{yao2024nonclassical}
\begin{equation}~\label{Equ_Gy}
	G_y=a+ b L^{-q} \,\,\,\, \mbox{and} \,\,\,\,  \langle M^2_y \rangle=a+b L^{-q}.
\end{equation}
The non-universal constant $a$ now represents the value of $G_y$ or $\langle M^2_y \rangle$ in the thermodynamic limit. By performing fits of $G_y$ according to Eq.~(\ref{Equ_Gy}), we obtain $q=0.523(1)$, $0.5204(8)$, $0.528(2)$ and $0.52(2)$ as well as $\chi^2/{\rm DOF} \approx 1.0$, $1.2$, $1.1$ and $1.0$, with $L_{\rm min}=24$, $16$, $24$ and $96$, for ${\rm P_1}$, ${\rm P_2}$, ${\rm P_3}$ and ${\rm P_4}$, respectively. From $\langle M^2_y \rangle$, we obtain $q=0.521(4)$, $0.508(4)$, $0.5074(8)$ and $0.500(2)$ as well as $\chi^2/{\rm DOF} \approx 1.3$, $0.8$, $0.8$ and $1.0$, with $L_{\rm min}=80$, $80$, $32$ and $64$, for ${\rm P_1}$, ${\rm P_2}$, ${\rm P_3}$ and ${\rm P_4}$, respectively. The FSS behaviors of $G_y$ and $\langle M^2_y \rangle$ [Figs.~\ref{Fig3}(a) and \ref{Fig3}(b)], particularly their non-zero thermodynamic limits, provide strong evidence of LR order. According to Eq.~(\ref{gy}), we further predict that constant contribution is vanishing in the magnetic fluctuations $\langle M^2_{yk} \rangle$ at smallest non-zero mode, which scales as~\cite{yao2024nonclassical}
\begin{equation}~\label{Equ_f1}
	\langle M^2_{yk} \rangle=L^{-q}(a+b L^{-\omega}),
\end{equation}
where $\omega$ is a correction exponent. By performing fits, we obtain $q=0.5321(10)$, $0.537(7)$, $0.516(2)$ and $0.511(1)$ as well as $\chi^2/{\rm DOF} \approx 1.1$, $2.3$, $0.3$ and $1.4$, with $L_{\rm min}=8$, $24$, $24$ and $24$, for ${\rm P_1}$, ${\rm P_2}$, ${\rm P_3}$ and ${\rm P_4}$, respectively. With finite-size corrections being handled, the linear dependence of $\langle M^2_{yk} \rangle$ on $L^{-q}$ is shown in Fig.~\ref{Fig3}(c).

\begin{table}
	\caption{The power-law exponent $q$ of Goldstone mode for the 1D LR order in the V plane. The estimates are extracted from preferred least-squares fits of $G_y$, $\langle M^2_y \rangle$, $\langle M^2_{yk} \rangle$ and $\xi_y$ according to Eqs.~(\ref{Equ_Gy})-(\ref{Equ_xi}).}~\label{tab1}
	\centering
	\scalebox{1.0}{
		\tabcolsep 0.04in
		\begin{tabular}{c|c|c|c|c}
			\hline
			{\rm Parameters}  & $q$ from $G_y$ & $q$ from $\langle M^2_y \rangle$ & $q$ from $\langle M^2_{yk} \rangle$ & $q$ from $\xi_y$\\
			\hline
			${\rm P_1}$  & 0.523(1)\phantom{0}      & 0.521(4)\phantom{0}      & 0.5321(10) 				& 0.533(2)\\
			${\rm P_2}$  & 0.5204(8)    			& 0.508(4)\phantom{0}      & 0.537(7)\phantom{00}   	& 0.52(1)\phantom{0}\\
			${\rm P_3}$  & 0.528(2)\phantom{0}      & 0.5074(8)     		   & 0.516(2)\phantom{00}   	& 0.52(1)\phantom{0}\\
			${\rm P_4}$  & 0.52(2)\phantom{00}      & 0.500(2)\phantom{0}      & 0.511(1)\phantom{00}   	& 0.505(9)\\
			\hline
		\end{tabular}
	}
\end{table}

We explain the FSS behaviors of $\xi_y/L$. According to the definition of $\xi_y$, an asymptotic relation reads $(\xi_y/L)^2 \sim \langle M^2_y \rangle/\langle M^2_{yk} \rangle$. Using Eqs.~(\ref{Equ_Gy}) and (\ref{Equ_f1}), the scaling formula of $(\xi_y/L)^2$ is written as~\cite{yao2024nonclassical}
\begin{equation}~\label{Equ_xi}
	(\xi_y/L)^2=L^{q}(a+b L^{-\omega})+c
\end{equation}
with $c$ a background term. By fits, we obtain $q=0.533(2)$, $0.52(1)$, $0.52(1)$ and $0.505(9)$ as well as $\chi^2/{\rm DOF} \approx 0.3$, $1.7$, $0.3$ and $1.5$, with $L_{\rm min}=8$, $16$, $24$ and $16$, for ${\rm P_1}$, ${\rm P_2}$, ${\rm P_3}$ and ${\rm P_4}$, respectively. The power-law divergence of $\xi_y/L$ upon increasing $L$, shown in Fig.~\ref{Fig3}(d), is therefore related to the LR order.

The values of $q$ extracted from different parameter sets are close to each other (Table~\ref{tab1}). Such a universality provides further evidence of Goldstone-mode effects. A final estimate of $q$ is $q=0.51(2)$, which might be exactly identical to $1/2$ and remains to be theoretically derived.

\begin{figure}[t]
\includegraphics[height=7cm,width=8cm]{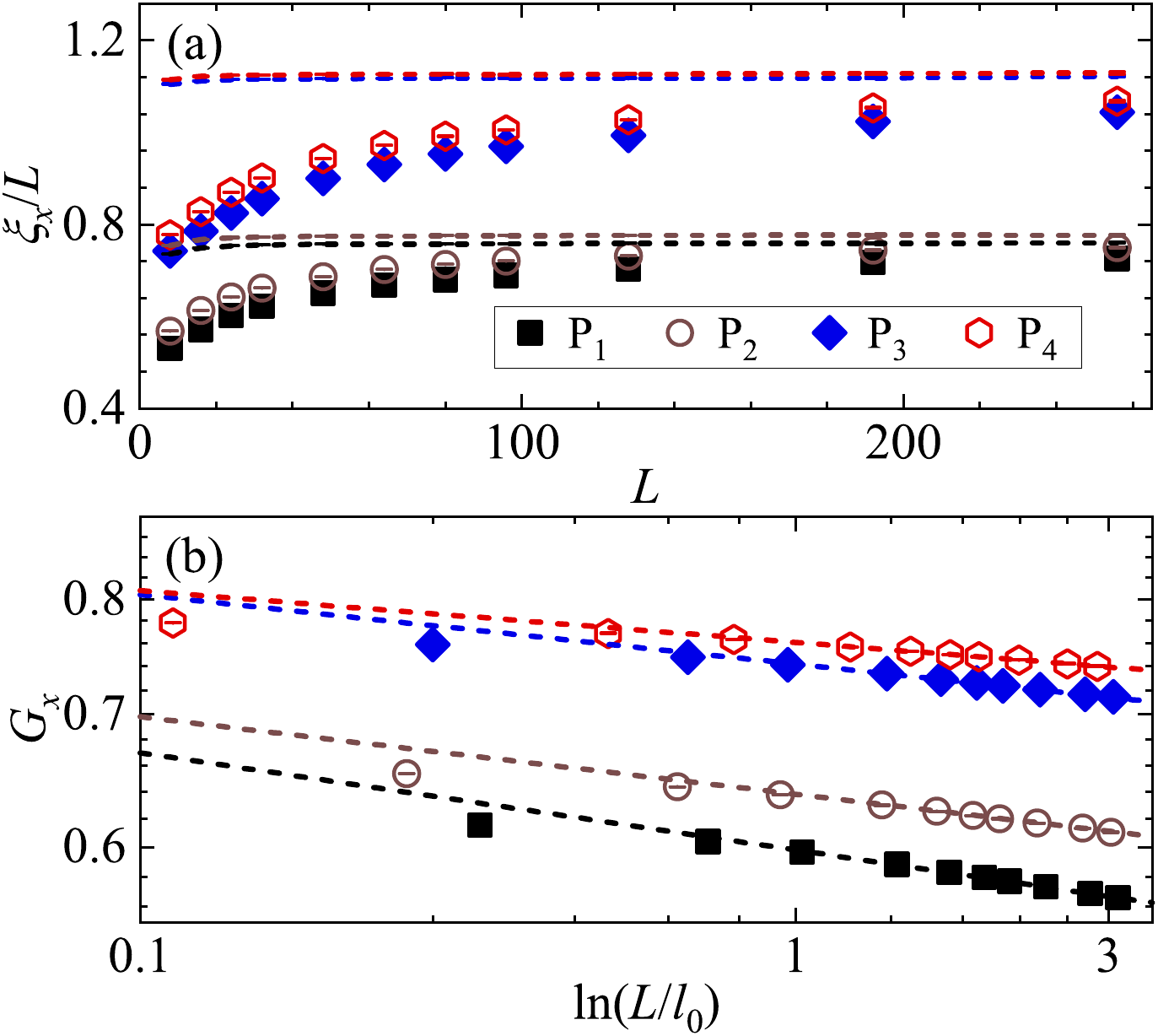}
\caption{Critical phase of an $x$ line in the V plane. (a) Reduced correlation length $\xi_x/L$ versus $L$. The dashed lines represent $\xi'_x/L$. (b) Log-log plot of the large-distance spin-spin correlation $G_x$ versus ${\rm ln}(L/l_0)$. The dashed lines represent preferred fits according to Eq.~(\ref{Equ_Gx}). For ${\rm P_1}$, ${\rm P_2}$, ${\rm P_3}$ and ${\rm P_4}$, the values $l_0=11.5$, $12.4$, $12.1$ and $14.3$ are from preferred fits, respectively.}~\label{Fig4}
\end{figure}

A distinct scenario is observed for the $x$ lines. We compute the second-moment correlation length $\xi_x=(2{\rm sin}\frac{\pi}{L})^{-1}\sqrt{\frac{\langle M^2_x \rangle}{\langle M^2_{xk} \rangle}-1}$ with $M_x=L^{-1} |\sum_{{\bf r} \in {\rm an} \, x \, {\rm line}} \vec{S}_{\bf r}| $ and $M_{xk}=L^{-1} |\sum_{{\bf r} \in {\rm an} \, x \, {\rm line}}\vec{S}_{\bf r}e^{i\frac{2\pi}{L}x}|$. Figure~\ref{Fig4}(a) shows that, $\xi_x/L$ converges in the thermodynamic limit. The size-dependent behavior of $\xi_x/L$ can be quantified by extrapolation to constant supplemented by logarithmic or power-law finite-size corrections (SM). Moreover, we find that the finite-size corrections are largely suppressed if a slightly different definition of correlation length is utilized. This definition reads $\xi'_x=(2{\rm sin}\frac{\pi}{L})^{-1}\sqrt{\frac{\Gamma_{\rm V}(0,0)}{\Gamma_{\rm V}(\frac{2\pi}{L},0)}-1}$, with $\Gamma_{\rm V}({\bf k})=L^{-4}\langle |\sum_{{\bf r} \in {\rm V}}\vec{S}_{\bf r}e^{i{\bf k}\cdot{\bf r}} |^2 \rangle$, where the summation runs over sites in the V plane. Figure~\ref{Fig4}(a) demonstrates that $\xi'_x/L$ converges fast to constant as $L \rightarrow \infty$. Therefore, the possibility of LR order along the $x$-direction, reminiscent of the LR order along the $y$-direction described by Eqs.~(\ref{gy})-(\ref{Equ_xi}), is precluded. Meanwhile, the existence of a critical phase is indicated for the $x$ lines. Furthermore, we find that the large-distance spin-spin correlation $G_x = \langle \vec{S}_{{\bf r}} \cdot \vec{S}_{{\bf r}+(L/2,0)} \rangle$ can be described by
\begin{equation}~\label{Equ_Gx}
	G_x=a[{\rm ln}(L/l_0)]^{-\hat{q}}
\end{equation}
with $\hat{q}$ a critical exponent. By fits, we obtain $\hat{q}=0.049(2)$, $0.039(8)$, $0.035(3)$ and $0.026(2)$ as well as $\chi^2/{\rm DOF} \approx 0.8$, $0.7$, $1.9$ and $1.3$, with $L_{\rm min}=48$, $80$, $64$ and $64$, for ${\rm P_1}$, ${\rm P_2}$, ${\rm P_3}$ and ${\rm P_4}$, respectively. The dependence of $G_x$ on ${\rm ln}(L/l_0)$ is illustrated by Fig.~\ref{Fig4}(b). It is of practical difficulty to completely preclude the power-law scaling $G_x \sim L^{-\tilde{q}}$ with $\tilde{q}$ a critical exponent, though the possibility is small. In SM, we present the spin-spin correlation function along the $x$-direction, which decreases much more quickly with distance than that along the $y$-direction. Finally, we point out that, although the logarithmic scenario may exist, it differs from the E-Log scenario of surface and interface criticality, where the correlation length diverges logarithmically with $L$~\cite{Metlitski} owing to the two-length scaling of correlation function~\cite{Hu2021}.

In short, when $K \ge K_2$, the $y$ lines enter a LR ordered phase, whereas a critical phase emerges in the $x$ lines. Hence, the ordering in the V plane is anisotropic. An intuitive picture for such anisotropy is that the ordering along $y$ (resp. $x$) lines is directly (resp. indirectly) enhanced by the effective interactions mediated by P planes.

\headline{Discussions}
The search for LR order in a SR q2D XY model is a topic of mutual interest, motivated by the recent advances in the surface and interface critical phenomena, the rapid development of defect conformal field theory, the ordering by effective retarded interactions from dissipations, and the novel concepts TQF and iTQF. By proposing a minimalistic setup with only two parameters, we obtain the first conclusive evidence of emergent LR order in a q2D XY model with strictly SR finite couplings. A comprehensive scaling analysis reveals Goldstone-mode effects in the LR ordered phase. Our study reveals a mechanism that circumvents the conditions of the Mermin-Wagner theorem: the coupling to critical P planes with large fluctuations mediates effective interactions along $y$-direction, thereby taking the system beyond the scope where the theorem applies and enabling true LR order.

Finally, we emphasize that the LR order here is not a conventional 2D one; rather, it is a directional phenomenon induced by a critical environment: emergent 1D LR order develops exclusively along the $y$-direction, while the $x$-direction exhibits QLR order. Although partly motivated by the iTQF in the comb-lattice Bose-Hubbard model~\cite{Radzihovsky24}, our LR ordered phase is conceptually distinct. It arises from a finite-temperature classical mechanism, offering a new paradigm for low-dimensional LR order based on classical critical fluctuations rather than quantum effects. In addition to numerical
studies, a field-theoretical investigation is called for to characterize the mediated effective interactions.

\headline{Acknowledgments}
We acknowledge support from the National Natural Science Foundation of China under Grants No.~12275002, 12275263, 12505035 and 12204173, the Quantum Science and Technology-National Science and Technology Major Project under Grant No.~2021ZD0301900, and the Natural Science Foundation of Anhui Province under Grant No.~2508085QA032. Y.D. is also supported by the Natural Science Foundation of Fujian Province under Grant No.~2023J02032.

\headline{Data Availability}
The data that support the findings of this article are openly available at~\cite{hu_2026_19901475}.

\bibliography{papers}

\begin{thebibliography}{53}%
\makeatletter
\providecommand \@ifxundefined [1]{%
 \@ifx{#1\undefined}
}%
\providecommand \@ifnum [1]{%
 \ifnum #1\expandafter \@firstoftwo
 \else \expandafter \@secondoftwo
 \fi
}%
\providecommand \@ifx [1]{%
 \ifx #1\expandafter \@firstoftwo
 \else \expandafter \@secondoftwo
 \fi
}%
\providecommand \natexlab [1]{#1}%
\providecommand \enquote  [1]{``#1''}%
\providecommand \bibnamefont  [1]{#1}%
\providecommand \bibfnamefont [1]{#1}%
\providecommand \citenamefont [1]{#1}%
\providecommand \href@noop [0]{\@secondoftwo}%
\providecommand \href [0]{\begingroup \@sanitize@url \@href}%
\providecommand \@href[1]{\@@startlink{#1}\@@href}%
\providecommand \@@href[1]{\endgroup#1\@@endlink}%
\providecommand \@sanitize@url [0]{\catcode `\\12\catcode `\$12\catcode
  `\&12\catcode `\#12\catcode `\^12\catcode `\_12\catcode `\%12\relax}%
\providecommand \@@startlink[1]{}%
\providecommand \@@endlink[0]{}%
\providecommand \url  [0]{\begingroup\@sanitize@url \@url }%
\providecommand \@url [1]{\endgroup\@href {#1}{\urlprefix }}%
\providecommand \urlprefix  [0]{URL }%
\providecommand \Eprint [0]{\href }%
\providecommand \doibase [0]{https://doi.org/}%
\providecommand \selectlanguage [0]{\@gobble}%
\providecommand \bibinfo  [0]{\@secondoftwo}%
\providecommand \bibfield  [0]{\@secondoftwo}%
\providecommand \translation [1]{[#1]}%
\providecommand \BibitemOpen [0]{}%
\providecommand \bibitemStop [0]{}%
\providecommand \bibitemNoStop [0]{.\EOS\space}%
\providecommand \EOS [0]{\spacefactor3000\relax}%
\providecommand \BibitemShut  [1]{\csname bibitem#1\endcsname}%
\let\auto@bib@innerbib\@empty
\bibitem [{\citenamefont {Kosterlitz}(2017)}]{Kosterlitz17}%
  \BibitemOpen
  \bibfield  {author} {\bibinfo {author} {\bibfnamefont {J.~M.}\ \bibnamefont
  {Kosterlitz}},\ }\bibfield  {title} {\bibinfo {title} {Nobel lecture:
  Topological defects and phase transitions},\ }\href
  {https://doi.org/10.1103/RevModPhys.89.040501} {\bibfield  {journal}
  {\bibinfo  {journal} {Rev. Mod. Phys.}\ }\textbf {\bibinfo {volume} {89}},\
  \bibinfo {pages} {040501} (\bibinfo {year} {2017})}\BibitemShut {NoStop}%
\bibitem [{\citenamefont {Kunz}\ and\ \citenamefont {Pfister}(1976)}]{Kunz76}%
  \BibitemOpen
  \bibfield  {author} {\bibinfo {author} {\bibfnamefont {H.}~\bibnamefont
  {Kunz}}\ and\ \bibinfo {author} {\bibfnamefont {C.~E.}\ \bibnamefont
  {Pfister}},\ }\bibfield  {title} {\bibinfo {title} {First order phase
  transition in the plane rotator ferromagnetic model in two dimensions},\
  }\href {https://link.springer.com/article/10.1007/BF01609121} {\bibfield
  {journal} {\bibinfo  {journal} {Comm. Math. Phys.}\ }\textbf {\bibinfo
  {volume} {46}},\ \bibinfo {pages} {245} (\bibinfo {year} {1976})}\BibitemShut
  {NoStop}%
\bibitem [{\citenamefont {Giachetti}\ \emph {et~al.}(2021)\citenamefont
  {Giachetti}, \citenamefont {Defenu}, \citenamefont {Ruffo},\ and\
  \citenamefont {Trombettoni}}]{Giachetti2021}%
  \BibitemOpen
  \bibfield  {author} {\bibinfo {author} {\bibfnamefont {G.}~\bibnamefont
  {Giachetti}}, \bibinfo {author} {\bibfnamefont {N.}~\bibnamefont {Defenu}},
  \bibinfo {author} {\bibfnamefont {S.}~\bibnamefont {Ruffo}},\ and\ \bibinfo
  {author} {\bibfnamefont {A.}~\bibnamefont {Trombettoni}},\ }\bibfield
  {title} {\bibinfo {title} {Berezinskii-kosterlitz-thouless phase transitions
  with long-range couplings},\ }\href
  {https://doi.org/10.1103/PhysRevLett.127.156801} {\bibfield  {journal}
  {\bibinfo  {journal} {Phys. Rev. Lett.}\ }\textbf {\bibinfo {volume} {127}},\
  \bibinfo {pages} {156801} (\bibinfo {year} {2021})},\ \Eprint
  {https://arxiv.org/abs/2104.13217} {arXiv:2104.13217 [cond-mat]} \BibitemShut
  {NoStop}%
\bibitem [{\citenamefont {Giachetti}\ \emph {et~al.}(2022)\citenamefont
  {Giachetti}, \citenamefont {Trombettoni}, \citenamefont {Ruffo},\ and\
  \citenamefont {Defenu}}]{Giachetti2022}%
  \BibitemOpen
  \bibfield  {author} {\bibinfo {author} {\bibfnamefont {G.}~\bibnamefont
  {Giachetti}}, \bibinfo {author} {\bibfnamefont {A.}~\bibnamefont
  {Trombettoni}}, \bibinfo {author} {\bibfnamefont {S.}~\bibnamefont {Ruffo}},\
  and\ \bibinfo {author} {\bibfnamefont {N.}~\bibnamefont {Defenu}},\
  }\bibfield  {title} {\bibinfo {title} {Berezinskii-kosterlitz-thouless
  transitions in classical and quantum long-range systems},\ }\href
  {https://doi.org/10.1103/PhysRevB.106.014106} {\bibfield  {journal} {\bibinfo
   {journal} {Phys. Rev. B}\ }\textbf {\bibinfo {volume} {106}},\ \bibinfo
  {pages} {014106} (\bibinfo {year} {2022})},\ \Eprint
  {https://arxiv.org/abs/2201.03650} {arXiv:2201.03650 [cond-mat]} \BibitemShut
  {NoStop}%
\bibitem [{\citenamefont {Xiao}\ \emph {et~al.}(2025)\citenamefont {Xiao},
  \citenamefont {Yao}, \citenamefont {Zhang}, \citenamefont {Fan},\ and\
  \citenamefont {Deng}}]{Xiao2024}%
  \BibitemOpen
  \bibfield  {author} {\bibinfo {author} {\bibfnamefont {T.}~\bibnamefont
  {Xiao}}, \bibinfo {author} {\bibfnamefont {D.}~\bibnamefont {Yao}}, \bibinfo
  {author} {\bibfnamefont {C.}~\bibnamefont {Zhang}}, \bibinfo {author}
  {\bibfnamefont {Z.}~\bibnamefont {Fan}},\ and\ \bibinfo {author}
  {\bibfnamefont {Y.}~\bibnamefont {Deng}},\ }\bibfield  {title} {\bibinfo
  {title} {Two-dimensional xy ferromagnet induced by long-range interaction},\
  }\href {https://doi.org/10.1088/0256-307X/42/7/070002} {\bibfield  {journal}
  {\bibinfo  {journal} {Chin. Phys. Lett.}\ }\textbf {\bibinfo {volume} {42}},\
  \bibinfo {pages} {070002} (\bibinfo {year} {2025})},\ \Eprint
  {https://arxiv.org/abs/2404.08498} {arXiv:2404.08498 [cond-mat]} \BibitemShut
  {NoStop}%
\bibitem [{\citenamefont {Yao}\ \emph {et~al.}(2025)\citenamefont {Yao},
  \citenamefont {Xiao}, \citenamefont {Zhang}, \citenamefont {Deng},\ and\
  \citenamefont {Fan}}]{yao2024nonclassical}%
  \BibitemOpen
  \bibfield  {author} {\bibinfo {author} {\bibfnamefont {D.}~\bibnamefont
  {Yao}}, \bibinfo {author} {\bibfnamefont {T.}~\bibnamefont {Xiao}}, \bibinfo
  {author} {\bibfnamefont {C.}~\bibnamefont {Zhang}}, \bibinfo {author}
  {\bibfnamefont {Y.}~\bibnamefont {Deng}},\ and\ \bibinfo {author}
  {\bibfnamefont {Z.}~\bibnamefont {Fan}},\ }\bibfield  {title} {\bibinfo
  {title} {Nonclassical regime of the two-dimensional long-range xy model: A
  comprehensive monte carlo study},\ }\href {https://doi.org/10.1103/yghs-p5mg}
  {\bibfield  {journal} {\bibinfo  {journal} {Phys. Rev. B}\ }\textbf {\bibinfo
  {volume} {112}},\ \bibinfo {pages} {144429} (\bibinfo {year} {2025})},\
  \Eprint {https://arxiv.org/abs/2411.01811} {arXiv:2411.01811 [cond-mat]}
  \BibitemShut {NoStop}%
\bibitem [{\citenamefont {Bianchi}\ and\ \citenamefont
  {Bonomi}(2023)}]{bianchi2023conformal}%
  \BibitemOpen
  \bibfield  {author} {\bibinfo {author} {\bibfnamefont {L.}~\bibnamefont
  {Bianchi}}\ and\ \bibinfo {author} {\bibfnamefont {D.}~\bibnamefont
  {Bonomi}},\ }\bibfield  {title} {\bibinfo {title} {Conformal dispersion
  relations for defects and boundaries},\ }\href
  {https://www.scipost.org/SciPostPhys.15.2.055?acad_field_slug=physics}
  {\bibfield  {journal} {\bibinfo  {journal} {SciPost Phys.}\ }\textbf
  {\bibinfo {volume} {15}},\ \bibinfo {pages} {055} (\bibinfo {year} {2023})},\
  \Eprint {https://arxiv.org/abs/2205.09775} {arXiv:2205.09775 [hep-th]}
  \BibitemShut {NoStop}%
\bibitem [{\citenamefont {Tr{\'e}panier}(2023)}]{trepanier2023surface}%
  \BibitemOpen
  \bibfield  {author} {\bibinfo {author} {\bibfnamefont {M.}~\bibnamefont
  {Tr{\'e}panier}},\ }\bibfield  {title} {\bibinfo {title} {Surface defects in
  the o(n) model},\ }\href
  {https://link.springer.com/article/10.1007/JHEP09(2023)074} {\bibfield
  {journal} {\bibinfo  {journal} {J. High Energ. Phys.}\ }\textbf {\bibinfo
  {volume} {2023}},\ \bibinfo {pages} {74}},\ \Eprint
  {https://arxiv.org/abs/2305.10486} {arXiv:2305.10486 [hep-th]} \BibitemShut
  {NoStop}%
\bibitem [{\citenamefont {Raviv-Moshe}\ and\ \citenamefont
  {Zhong}(2023)}]{Raviv2023}%
  \BibitemOpen
  \bibfield  {author} {\bibinfo {author} {\bibfnamefont {A.}~\bibnamefont
  {Raviv-Moshe}}\ and\ \bibinfo {author} {\bibfnamefont {S.}~\bibnamefont
  {Zhong}},\ }\bibfield  {title} {\bibinfo {title} {Phases of surface defects
  in scalar field theories},\ }\href {https://doi.org/10.1007/JHEP08(2023)143}
  {\bibfield  {journal} {\bibinfo  {journal} {J. High Energ. Phys.}\ }\textbf
  {\bibinfo {volume} {2023}},\ \bibinfo {pages} {143}},\ \Eprint
  {https://arxiv.org/abs/2305.11370} {arXiv:2305.11370 [hep-th]} \BibitemShut
  {NoStop}%
\bibitem [{\citenamefont {Giombi}\ and\ \citenamefont
  {Liu}(2023)}]{Giombi2023}%
  \BibitemOpen
  \bibfield  {author} {\bibinfo {author} {\bibfnamefont {S.}~\bibnamefont
  {Giombi}}\ and\ \bibinfo {author} {\bibfnamefont {B.}~\bibnamefont {Liu}},\
  }\bibfield  {title} {\bibinfo {title} {Notes on a surface defect in the o(n)
  model},\ }\href {https://doi.org/10.1007/JHEP12(2023)004} {\bibfield
  {journal} {\bibinfo  {journal} {J. High Energ. Phys.}\ }\textbf {\bibinfo
  {volume} {2023}},\ \bibinfo {pages} {4}},\ \Eprint
  {https://arxiv.org/abs/2305.11402} {arXiv:2305.11402 [hep-th]} \BibitemShut
  {NoStop}%
\bibitem [{\citenamefont {Bolla}\ \emph {et~al.}(2023)\citenamefont {Bolla},
  \citenamefont {Rodriguez-Gomez},\ and\ \citenamefont {Russo}}]{Bolla2023}%
  \BibitemOpen
  \bibfield  {author} {\bibinfo {author} {\bibfnamefont {I.~C.}\ \bibnamefont
  {Bolla}}, \bibinfo {author} {\bibfnamefont {D.}~\bibnamefont
  {Rodriguez-Gomez}},\ and\ \bibinfo {author} {\bibfnamefont {J.~G.}\
  \bibnamefont {Russo}},\ }\bibfield  {title} {\bibinfo {title} {Defects, rigid
  holography, and $c$-theorems},\ }\href
  {https://doi.org/10.1103/PhysRevD.108.L041701} {\bibfield  {journal}
  {\bibinfo  {journal} {Phys. Rev. D}\ }\textbf {\bibinfo {volume} {108}},\
  \bibinfo {pages} {L041701} (\bibinfo {year} {2023})},\ \Eprint
  {https://arxiv.org/abs/2306.11796} {arXiv:2306.11796 [hep-th]} \BibitemShut
  {NoStop}%
\bibitem [{\citenamefont {Cuomo}\ and\ \citenamefont
  {Zhang}(2024)}]{Cuomo2024}%
  \BibitemOpen
  \bibfield  {author} {\bibinfo {author} {\bibfnamefont {G.}~\bibnamefont
  {Cuomo}}\ and\ \bibinfo {author} {\bibfnamefont {S.}~\bibnamefont {Zhang}},\
  }\bibfield  {title} {\bibinfo {title} {Spontaneous symmetry breaking on
  surface defects},\ }\href {https://doi.org/10.1007/JHEP03(2024)022}
  {\bibfield  {journal} {\bibinfo  {journal} {J. High Energ. Phys.}\ }\textbf
  {\bibinfo {volume} {2024}},\ \bibinfo {pages} {22}},\ \Eprint
  {https://arxiv.org/abs/2306.00085} {arXiv:2306.00085 [hep-th]} \BibitemShut
  {NoStop}%
\bibitem [{\citenamefont {Sun}\ and\ \citenamefont
  {Jian}(2025{\natexlab{a}})}]{sun25}%
  \BibitemOpen
  \bibfield  {author} {\bibinfo {author} {\bibfnamefont {X.}~\bibnamefont
  {Sun}}\ and\ \bibinfo {author} {\bibfnamefont {S.-K.}\ \bibnamefont {Jian}},\
  }\bibfield  {title} {\bibinfo {title} {Boundary operator expansion and
  extraordinary phase transition in the tricritical o(n) model},\ }\href
  {https://doi.org/10.21468/SciPostPhys.18.6.210} {\bibfield  {journal}
  {\bibinfo  {journal} {SciPost Phys.}\ }\textbf {\bibinfo {volume} {18}},\
  \bibinfo {pages} {210} (\bibinfo {year} {2025}{\natexlab{a}})},\ \Eprint
  {https://arxiv.org/abs/2501.06287} {arXiv:2501.06287 [cond-mat]} \BibitemShut
  {NoStop}%
\bibitem [{\citenamefont {Sun}\ and\ \citenamefont
  {Jian}(2025{\natexlab{b}})}]{sun25B}%
  \BibitemOpen
  \bibfield  {author} {\bibinfo {author} {\bibfnamefont {X.}~\bibnamefont
  {Sun}}\ and\ \bibinfo {author} {\bibfnamefont {S.-K.}\ \bibnamefont {Jian}},\
  }\bibfield  {title} {\bibinfo {title} {Boundary operator expansion and
  extraordinary-log phase in the tricritical o(n) model},\ }\href
  {https://summit.aps.org/events/MAR-W25/14} {\bibfield  {journal} {\bibinfo
  {journal} {APS Global Physics Summit}\ } (\bibinfo {year}
  {2025}{\natexlab{b}})}\BibitemShut {NoStop}%
\bibitem [{\citenamefont {Metlitski}(2022)}]{Metlitski}%
  \BibitemOpen
  \bibfield  {author} {\bibinfo {author} {\bibfnamefont {M.~A.}\ \bibnamefont
  {Metlitski}},\ }\bibfield  {title} {\bibinfo {title} {{Boundary criticality
  of the O(N) model in d = 3 critically revisited}},\ }\href
  {https://doi.org/10.21468/SciPostPhys.12.4.131} {\bibfield  {journal}
  {\bibinfo  {journal} {SciPost Phys.}\ }\textbf {\bibinfo {volume} {12}},\
  \bibinfo {pages} {131} (\bibinfo {year} {2022})},\ \Eprint
  {https://arxiv.org/abs/2009.05119} {arXiv:2009.05119 [cond-mat]} \BibitemShut
  {NoStop}%
\bibitem [{\citenamefont {Hu}\ \emph {et~al.}(2021)\citenamefont {Hu},
  \citenamefont {Deng},\ and\ \citenamefont {Lv}}]{Hu2021}%
  \BibitemOpen
  \bibfield  {author} {\bibinfo {author} {\bibfnamefont {M.}~\bibnamefont
  {Hu}}, \bibinfo {author} {\bibfnamefont {Y.}~\bibnamefont {Deng}},\ and\
  \bibinfo {author} {\bibfnamefont {J.-P.}\ \bibnamefont {Lv}},\ }\bibfield
  {title} {\bibinfo {title} {Extraordinary-log surface phase transition in the
  three-dimensional $xy$ model},\ }\href
  {https://doi.org/10.1103/PhysRevLett.127.120603} {\bibfield  {journal}
  {\bibinfo  {journal} {Phys. Rev. Lett.}\ }\textbf {\bibinfo {volume} {127}},\
  \bibinfo {pages} {120603} (\bibinfo {year} {2021})},\ \Eprint
  {https://arxiv.org/abs/2104.05152} {arXiv:2104.05152 [cond-mat]} \BibitemShut
  {NoStop}%
\bibitem [{\citenamefont {Parisen~Toldin}\ and\ \citenamefont
  {Metlitski}(2022)}]{ToldinMetlitski22}%
  \BibitemOpen
  \bibfield  {author} {\bibinfo {author} {\bibfnamefont {F.}~\bibnamefont
  {Parisen~Toldin}}\ and\ \bibinfo {author} {\bibfnamefont {M.~A.}\
  \bibnamefont {Metlitski}},\ }\bibfield  {title} {\bibinfo {title} {Boundary
  criticality of the 3d o(n) model: from normal to extraordinary},\ }\href
  {https://doi.org/10.1103/PhysRevLett.128.215701} {\bibfield  {journal}
  {\bibinfo  {journal} {Phys. Rev. Lett.}\ }\textbf {\bibinfo {volume} {128}},\
  \bibinfo {pages} {215701} (\bibinfo {year} {2022})},\ \Eprint
  {https://arxiv.org/abs/2111.03613} {arXiv:2111.03613 [cond-mat]} \BibitemShut
  {NoStop}%
\bibitem [{\citenamefont {Krishnan}\ and\ \citenamefont
  {Metlitski}(2023)}]{Krishnan2023}%
  \BibitemOpen
  \bibfield  {author} {\bibinfo {author} {\bibfnamefont {A.}~\bibnamefont
  {Krishnan}}\ and\ \bibinfo {author} {\bibfnamefont {M.~A.}\ \bibnamefont
  {Metlitski}},\ }\bibfield  {title} {\bibinfo {title} {{A plane defect in the
  3d O(N) model}},\ }\href {https://doi.org/10.21468/SciPostPhys.15.3.090}
  {\bibfield  {journal} {\bibinfo  {journal} {SciPost Phys.}\ }\textbf
  {\bibinfo {volume} {15}},\ \bibinfo {pages} {090} (\bibinfo {year} {2023})},\
  \Eprint {https://arxiv.org/abs/2301.05728} {arXiv:2301.05728 [cond-mat]}
  \BibitemShut {NoStop}%
\bibitem [{\citenamefont {Sun}\ \emph {et~al.}(2023)\citenamefont {Sun},
  \citenamefont {Hu}, \citenamefont {Deng},\ and\ \citenamefont
  {Lv}}]{Sun2023}%
  \BibitemOpen
  \bibfield  {author} {\bibinfo {author} {\bibfnamefont {Y.}~\bibnamefont
  {Sun}}, \bibinfo {author} {\bibfnamefont {M.}~\bibnamefont {Hu}}, \bibinfo
  {author} {\bibfnamefont {Y.}~\bibnamefont {Deng}},\ and\ \bibinfo {author}
  {\bibfnamefont {J.-P.}\ \bibnamefont {Lv}},\ }\bibfield  {title} {\bibinfo
  {title} {Extraordinary-log universality of critical phenomena in plane
  defects},\ }\href {https://doi.org/10.1103/PhysRevLett.131.207101} {\bibfield
   {journal} {\bibinfo  {journal} {Phys. Rev. Lett.}\ }\textbf {\bibinfo
  {volume} {131}},\ \bibinfo {pages} {207101} (\bibinfo {year} {2023})},\
  \Eprint {https://arxiv.org/abs/2301.11720} {arXiv:2301.11720 [cond-mat]}
  \BibitemShut {NoStop}%
\bibitem [{\citenamefont {Parisen~Toldin}(2021)}]{ParisenToldin21}%
  \BibitemOpen
  \bibfield  {author} {\bibinfo {author} {\bibfnamefont {F.}~\bibnamefont
  {Parisen~Toldin}},\ }\bibfield  {title} {\bibinfo {title} {Boundary critical
  behavior of the three-dimensional heisenberg universality class},\ }\href
  {https://doi.org/10.1103/PhysRevLett.126.135701} {\bibfield  {journal}
  {\bibinfo  {journal} {Phys. Rev. Lett.}\ }\textbf {\bibinfo {volume} {126}},\
  \bibinfo {pages} {135701} (\bibinfo {year} {2021})},\ \Eprint
  {https://arxiv.org/abs/2012.00039} {arXiv:2012.00039 [cond-mat]} \BibitemShut
  {NoStop}%
\bibitem [{\citenamefont {Zhang}\ \emph {et~al.}(2022)\citenamefont {Zhang},
  \citenamefont {Ding}, \citenamefont {Deng},\ and\ \citenamefont
  {Zhang}}]{Zhang22}%
  \BibitemOpen
  \bibfield  {author} {\bibinfo {author} {\bibfnamefont {L.-R.}\ \bibnamefont
  {Zhang}}, \bibinfo {author} {\bibfnamefont {C.}~\bibnamefont {Ding}},
  \bibinfo {author} {\bibfnamefont {Y.}~\bibnamefont {Deng}},\ and\ \bibinfo
  {author} {\bibfnamefont {L.}~\bibnamefont {Zhang}},\ }\bibfield  {title}
  {\bibinfo {title} {Surface criticality of the antiferromagnetic potts
  model},\ }\href {https://doi.org/10.1103/PhysRevB.105.224415} {\bibfield
  {journal} {\bibinfo  {journal} {Phys. Rev. B}\ }\textbf {\bibinfo {volume}
  {105}},\ \bibinfo {pages} {224415} (\bibinfo {year} {2022})},\ \Eprint
  {https://arxiv.org/abs/2204.11692} {arXiv:2204.11692 [cond-mat]} \BibitemShut
  {NoStop}%
\bibitem [{\citenamefont {Zou}\ \emph {et~al.}(2022)\citenamefont {Zou},
  \citenamefont {Liu},\ and\ \citenamefont {Guo}}]{Zou22}%
  \BibitemOpen
  \bibfield  {author} {\bibinfo {author} {\bibfnamefont {X.}~\bibnamefont
  {Zou}}, \bibinfo {author} {\bibfnamefont {S.}~\bibnamefont {Liu}},\ and\
  \bibinfo {author} {\bibfnamefont {W.}~\bibnamefont {Guo}},\ }\bibfield
  {title} {\bibinfo {title} {Surface critical properties of the
  three-dimensional clock model},\ }\href
  {https://doi.org/10.1103/PhysRevB.106.064420} {\bibfield  {journal} {\bibinfo
   {journal} {Phys. Rev. B}\ }\textbf {\bibinfo {volume} {106}},\ \bibinfo
  {pages} {064420} (\bibinfo {year} {2022})},\ \Eprint
  {https://arxiv.org/abs/2204.13612} {arXiv:2204.13612 [cond-mat]} \BibitemShut
  {NoStop}%
\bibitem [{\citenamefont {Sun}\ and\ \citenamefont {Lv}(2022)}]{Sun22A}%
  \BibitemOpen
  \bibfield  {author} {\bibinfo {author} {\bibfnamefont {Y.}~\bibnamefont
  {Sun}}\ and\ \bibinfo {author} {\bibfnamefont {J.-P.}\ \bibnamefont {Lv}},\
  }\bibfield  {title} {\bibinfo {title} {Quantum extraordinary-log universality
  of boundary critical behavior},\ }\href
  {https://doi.org/10.1103/PhysRevB.106.224502} {\bibfield  {journal} {\bibinfo
   {journal} {Phys. Rev. B}\ }\textbf {\bibinfo {volume} {106}},\ \bibinfo
  {pages} {224502} (\bibinfo {year} {2022})},\ \Eprint
  {https://arxiv.org/abs/2205.00878} {arXiv:2205.00878 [cond-mat]} \BibitemShut
  {NoStop}%
\bibitem [{\citenamefont {Sun}\ \emph {et~al.}(2022)\citenamefont {Sun},
  \citenamefont {Lyu},\ and\ \citenamefont {Lv}}]{Sun22B}%
  \BibitemOpen
  \bibfield  {author} {\bibinfo {author} {\bibfnamefont {Y.}~\bibnamefont
  {Sun}}, \bibinfo {author} {\bibfnamefont {J.}~\bibnamefont {Lyu}},\ and\
  \bibinfo {author} {\bibfnamefont {J.-P.}\ \bibnamefont {Lv}},\ }\bibfield
  {title} {\bibinfo {title} {Classical-quantum correspondence of special and
  extraordinary-log criticality: Villain's bridge},\ }\href
  {https://doi.org/10.1103/PhysRevB.106.174516} {\bibfield  {journal} {\bibinfo
   {journal} {Phys. Rev. B}\ }\textbf {\bibinfo {volume} {106}},\ \bibinfo
  {pages} {174516} (\bibinfo {year} {2022})},\ \Eprint
  {https://arxiv.org/abs/2211.11376} {arXiv:2211.11376 [cond-mat]} \BibitemShut
  {NoStop}%
\bibitem [{\citenamefont {Zhang}\ \emph
  {et~al.}(2023{\natexlab{a}})\citenamefont {Zhang}, \citenamefont {Ding},
  \citenamefont {Zhang},\ and\ \citenamefont {Zhang}}]{Zhang23}%
  \BibitemOpen
  \bibfield  {author} {\bibinfo {author} {\bibfnamefont {L.~R.}\ \bibnamefont
  {Zhang}}, \bibinfo {author} {\bibfnamefont {C.}~\bibnamefont {Ding}},
  \bibinfo {author} {\bibfnamefont {W.}~\bibnamefont {Zhang}},\ and\ \bibinfo
  {author} {\bibfnamefont {L.}~\bibnamefont {Zhang}},\ }\bibfield  {title}
  {\bibinfo {title} {Sublattice extraordinary-log phase and new special point
  of the antiferromagnetic potts model},\ }\href
  {https://doi.org/10.1103/PhysRevB.108.024402} {\bibfield  {journal} {\bibinfo
   {journal} {Phys. Rev. B}\ }\textbf {\bibinfo {volume} {108}},\ \bibinfo
  {pages} {024402} (\bibinfo {year} {2023}{\natexlab{a}})},\ \Eprint
  {https://arxiv.org/abs/2301.08926} {arXiv:2301.08926 [cond-mat]} \BibitemShut
  {NoStop}%
\bibitem [{\citenamefont {Lee}\ \emph {et~al.}(2023)\citenamefont {Lee},
  \citenamefont {Jian},\ and\ \citenamefont {Xu}}]{PRXQ.4.030317}%
  \BibitemOpen
  \bibfield  {author} {\bibinfo {author} {\bibfnamefont {J.~Y.}\ \bibnamefont
  {Lee}}, \bibinfo {author} {\bibfnamefont {C.-M.}\ \bibnamefont {Jian}},\ and\
  \bibinfo {author} {\bibfnamefont {C.}~\bibnamefont {Xu}},\ }\bibfield
  {title} {\bibinfo {title} {Quantum criticality under decoherence or weak
  measurement},\ }\href {https://doi.org/10.1103/PRXQuantum.4.030317}
  {\bibfield  {journal} {\bibinfo  {journal} {PRX Quantum}\ }\textbf {\bibinfo
  {volume} {4}},\ \bibinfo {pages} {030317} (\bibinfo {year} {2023})},\ \Eprint
  {https://arxiv.org/abs/2301.05238} {arXiv:2301.05238 [cond-mat]} \BibitemShut
  {NoStop}%
\bibitem [{\citenamefont {Baweja}\ \emph {et~al.}()\citenamefont {Baweja},
  \citenamefont {Luitz},\ and\ \citenamefont
  {Garratt}}]{baweja2024postmeasurement}%
  \BibitemOpen
  \bibfield  {author} {\bibinfo {author} {\bibfnamefont {K.}~\bibnamefont
  {Baweja}}, \bibinfo {author} {\bibfnamefont {D.~J.}\ \bibnamefont {Luitz}},\
  and\ \bibinfo {author} {\bibfnamefont {S.~J.}\ \bibnamefont {Garratt}},\
  }\bibfield  {title} {\bibinfo {title} {Post-measurement quantum monte
  carlo},\ }\href {https://arxiv.org/abs/2410.13844} {\ }\Eprint
  {https://arxiv.org/abs/2410.13844} {arXiv:2410.13844 [cond-mat]} \BibitemShut
  {NoStop}%
\bibitem [{\citenamefont {Zhang}\ \emph
  {et~al.}(2023{\natexlab{b}})\citenamefont {Zhang}, \citenamefont {Zhu},\ and\
  \citenamefont {Vishwanath}}]{ZhangXY23}%
  \BibitemOpen
  \bibfield  {author} {\bibinfo {author} {\bibfnamefont {Y.-H.}\ \bibnamefont
  {Zhang}}, \bibinfo {author} {\bibfnamefont {Z.}~\bibnamefont {Zhu}},\ and\
  \bibinfo {author} {\bibfnamefont {A.}~\bibnamefont {Vishwanath}},\ }\bibfield
   {title} {\bibinfo {title} {Xy* transition and extraordinary boundary
  criticality from fractional exciton condensation in quantum hall bilayer},\
  }\href {https://doi.org/10.1103/PhysRevX.13.031023} {\bibfield  {journal}
  {\bibinfo  {journal} {Phys. Rev. X}\ }\textbf {\bibinfo {volume} {13}},\
  \bibinfo {pages} {031023} (\bibinfo {year} {2023}{\natexlab{b}})},\ \Eprint
  {https://arxiv.org/abs/2302.03703} {arXiv:2302.03703 [cond-mat]} \BibitemShut
  {NoStop}%
\bibitem [{\citenamefont {Werner}\ \emph {et~al.}(2005)\citenamefont {Werner},
  \citenamefont {Troyer},\ and\ \citenamefont {Sachdev}}]{Werner2005}%
  \BibitemOpen
  \bibfield  {author} {\bibinfo {author} {\bibfnamefont {P.}~\bibnamefont
  {Werner}}, \bibinfo {author} {\bibfnamefont {M.}~\bibnamefont {Troyer}},\
  and\ \bibinfo {author} {\bibfnamefont {S.}~\bibnamefont {Sachdev}},\
  }\bibfield  {title} {\bibinfo {title} {Quantum spin chains with site
  dissipation},\ }\href {https://journals.jps.jp/doi/abs/10.1143/JPSJS.74S.67}
  {\bibfield  {journal} {\bibinfo  {journal} {J. Phys. Soc. Jpn.}\ }\textbf
  {\bibinfo {volume} {74}},\ \bibinfo {pages} {67} (\bibinfo {year} {2005})},\
  \Eprint {https://arxiv.org/abs/cond-mat/0412529} {arXiv:cond-mat/0412529
  [cond-mat]} \BibitemShut {NoStop}%
\bibitem [{\citenamefont {Cazalilla}\ \emph {et~al.}(2006)\citenamefont
  {Cazalilla}, \citenamefont {Sols},\ and\ \citenamefont
  {Guinea}}]{Cazalilla2006}%
  \BibitemOpen
  \bibfield  {author} {\bibinfo {author} {\bibfnamefont {M.~A.}\ \bibnamefont
  {Cazalilla}}, \bibinfo {author} {\bibfnamefont {F.}~\bibnamefont {Sols}},\
  and\ \bibinfo {author} {\bibfnamefont {F.}~\bibnamefont {Guinea}},\
  }\bibfield  {title} {\bibinfo {title} {Dissipation-driven quantum phase
  transitions in a tomonaga-luttinger liquid electrostatically coupled to a
  metallic gate},\ }\href {https://doi.org/10.1103/PhysRevLett.97.076401}
  {\bibfield  {journal} {\bibinfo  {journal} {Phys. Rev. Lett.}\ }\textbf
  {\bibinfo {volume} {97}},\ \bibinfo {pages} {076401} (\bibinfo {year}
  {2006})},\ \Eprint {https://arxiv.org/abs/cond-mat/0603176}
  {arXiv:cond-mat/0603176 [cond-mat]} \BibitemShut {NoStop}%
\bibitem [{\citenamefont {Sperstad}\ \emph {et~al.}(2012)\citenamefont
  {Sperstad}, \citenamefont {Stiansen},\ and\ \citenamefont
  {Sudb{\o}}}]{Sperstad2012}%
  \BibitemOpen
  \bibfield  {author} {\bibinfo {author} {\bibfnamefont {I.~B.}\ \bibnamefont
  {Sperstad}}, \bibinfo {author} {\bibfnamefont {E.~B.}\ \bibnamefont
  {Stiansen}},\ and\ \bibinfo {author} {\bibfnamefont {A.}~\bibnamefont
  {Sudb{\o}}},\ }\bibfield  {title} {\bibinfo {title} {Quantum criticality in
  spin chains with non-ohmic dissipation},\ }\href
  {https://doi.org/10.1103/PhysRevB.85.214302} {\bibfield  {journal} {\bibinfo
  {journal} {Phys. Rev. B}\ }\textbf {\bibinfo {volume} {85}},\ \bibinfo
  {pages} {214302} (\bibinfo {year} {2012})},\ \Eprint
  {https://arxiv.org/abs/1204.2538} {arXiv:1204.2538 [cond-mat]} \BibitemShut
  {NoStop}%
\bibitem [{\citenamefont {Weber}\ \emph {et~al.}(2022)\citenamefont {Weber},
  \citenamefont {Luitz},\ and\ \citenamefont {Assaad}}]{Weber2022}%
  \BibitemOpen
  \bibfield  {author} {\bibinfo {author} {\bibfnamefont {M.}~\bibnamefont
  {Weber}}, \bibinfo {author} {\bibfnamefont {D.~J.}\ \bibnamefont {Luitz}},\
  and\ \bibinfo {author} {\bibfnamefont {F.~F.}\ \bibnamefont {Assaad}},\
  }\bibfield  {title} {\bibinfo {title} {Dissipation-induced order: the s= 1/2
  quantum spin chain coupled to an ohmic bath},\ }\href
  {https://doi.org/10.1103/PhysRevLett.129.056402} {\bibfield  {journal}
  {\bibinfo  {journal} {Phys. Rev. Lett.}\ }\textbf {\bibinfo {volume} {129}},\
  \bibinfo {pages} {056402} (\bibinfo {year} {2022})},\ \Eprint
  {https://arxiv.org/abs/2112.02124} {arXiv:2112.02124 [cond-mat]} \BibitemShut
  {NoStop}%
\bibitem [{\citenamefont {Ribeiro}\ \emph {et~al.}(2024)\citenamefont
  {Ribeiro}, \citenamefont {McClarty}, \citenamefont {Ribeiro},\ and\
  \citenamefont {Weber}}]{Ribeiro2024}%
  \BibitemOpen
  \bibfield  {author} {\bibinfo {author} {\bibfnamefont {A.~L.~S.}\
  \bibnamefont {Ribeiro}}, \bibinfo {author} {\bibfnamefont {P.}~\bibnamefont
  {McClarty}}, \bibinfo {author} {\bibfnamefont {P.}~\bibnamefont {Ribeiro}},\
  and\ \bibinfo {author} {\bibfnamefont {M.}~\bibnamefont {Weber}},\ }\bibfield
   {title} {\bibinfo {title} {Dissipation-induced long-range order in the
  one-dimensional bose-hubbard model},\ }\href
  {https://doi.org/10.1103/PhysRevB.110.115145} {\bibfield  {journal} {\bibinfo
   {journal} {Phys. Rev. B}\ }\textbf {\bibinfo {volume} {110}},\ \bibinfo
  {pages} {115145} (\bibinfo {year} {2024})},\ \Eprint
  {https://arxiv.org/abs/2311.07683} {arXiv:2311.07683 [cond-mat]} \BibitemShut
  {NoStop}%
\bibitem [{\citenamefont {Radzihovsky}\ \emph {et~al.}(2023)\citenamefont
  {Radzihovsky}, \citenamefont {Kuklov}, \citenamefont {Prokof'ev},\ and\
  \citenamefont {Svistunov}}]{Radzihovsky23}%
  \BibitemOpen
  \bibfield  {author} {\bibinfo {author} {\bibfnamefont {L.}~\bibnamefont
  {Radzihovsky}}, \bibinfo {author} {\bibfnamefont {A.}~\bibnamefont {Kuklov}},
  \bibinfo {author} {\bibfnamefont {N.}~\bibnamefont {Prokof'ev}},\ and\
  \bibinfo {author} {\bibfnamefont {B.}~\bibnamefont {Svistunov}},\ }\bibfield
  {title} {\bibinfo {title} {Superfluid edge dislocation: Transverse quantum
  fluid},\ }\href {https://doi.org/10.1103/PhysRevLett.131.196001} {\bibfield
  {journal} {\bibinfo  {journal} {Phys. Rev. Lett.}\ }\textbf {\bibinfo
  {volume} {131}},\ \bibinfo {pages} {196001} (\bibinfo {year} {2023})},\
  \Eprint {https://arxiv.org/abs/2304.03309} {arXiv:2304.03309 [cond-mat]}
  \BibitemShut {NoStop}%
\bibitem [{\citenamefont {Kuklov}\ \emph
  {et~al.}(2024{\natexlab{a}})\citenamefont {Kuklov}, \citenamefont
  {Prokof'ev}, \citenamefont {Radzihovsky},\ and\ \citenamefont
  {Svistunov}}]{Kuklov24a}%
  \BibitemOpen
  \bibfield  {author} {\bibinfo {author} {\bibfnamefont {A.}~\bibnamefont
  {Kuklov}}, \bibinfo {author} {\bibfnamefont {N.}~\bibnamefont {Prokof'ev}},
  \bibinfo {author} {\bibfnamefont {L.}~\bibnamefont {Radzihovsky}},\ and\
  \bibinfo {author} {\bibfnamefont {B.}~\bibnamefont {Svistunov}},\ }\bibfield
  {title} {\bibinfo {title} {Transverse quantum fluids},\ }\href
  {https://doi.org/10.1103/PhysRevB.109.L100502} {\bibfield  {journal}
  {\bibinfo  {journal} {Phys. Rev. B}\ }\textbf {\bibinfo {volume} {109}},\
  \bibinfo {pages} {L100502} (\bibinfo {year} {2024}{\natexlab{a}})},\ \Eprint
  {https://arxiv.org/abs/2309.02501} {arXiv:2309.02501 [cond-mat]} \BibitemShut
  {NoStop}%
\bibitem [{\citenamefont {Kuklov}\ \emph
  {et~al.}(2024{\natexlab{b}})\citenamefont {Kuklov}, \citenamefont {Pollet},
  \citenamefont {Prokof'ev}, \citenamefont {Radzihovsky},\ and\ \citenamefont
  {Svistunov}}]{Kuklov24b}%
  \BibitemOpen
  \bibfield  {author} {\bibinfo {author} {\bibfnamefont {A.}~\bibnamefont
  {Kuklov}}, \bibinfo {author} {\bibfnamefont {L.}~\bibnamefont {Pollet}},
  \bibinfo {author} {\bibfnamefont {N.}~\bibnamefont {Prokof'ev}}, \bibinfo
  {author} {\bibfnamefont {L.}~\bibnamefont {Radzihovsky}},\ and\ \bibinfo
  {author} {\bibfnamefont {B.}~\bibnamefont {Svistunov}},\ }\bibfield  {title}
  {\bibinfo {title} {Universal correlations as fingerprints of transverse
  quantum fluids},\ }\href {https://doi.org/10.1103/PhysRevA.109.L011302}
  {\bibfield  {journal} {\bibinfo  {journal} {Phys. Rev. A}\ }\textbf {\bibinfo
  {volume} {109}},\ \bibinfo {pages} {L011302} (\bibinfo {year}
  {2024}{\natexlab{b}})},\ \Eprint {https://arxiv.org/abs/2310.19875}
  {arXiv:2310.19875 [cond-mat]} \BibitemShut {NoStop}%
\bibitem [{\citenamefont {Zhang}\ \emph {et~al.}(2024)\citenamefont {Zhang},
  \citenamefont {Boninsegni}, \citenamefont {Kuklov}, \citenamefont
  {Prokof'ev},\ and\ \citenamefont {Svistunov}}]{ZhangPhysRevB109}%
  \BibitemOpen
  \bibfield  {author} {\bibinfo {author} {\bibfnamefont {C.}~\bibnamefont
  {Zhang}}, \bibinfo {author} {\bibfnamefont {M.}~\bibnamefont {Boninsegni}},
  \bibinfo {author} {\bibfnamefont {A.}~\bibnamefont {Kuklov}}, \bibinfo
  {author} {\bibfnamefont {N.}~\bibnamefont {Prokof'ev}},\ and\ \bibinfo
  {author} {\bibfnamefont {B.}~\bibnamefont {Svistunov}},\ }\bibfield  {title}
  {\bibinfo {title} {Superclimbing modes in transverse quantum fluids:
  Signature statistical and dynamical features},\ }\href
  {https://doi.org/10.1103/PhysRevB.109.214519} {\bibfield  {journal} {\bibinfo
   {journal} {Phys. Rev. B}\ }\textbf {\bibinfo {volume} {109}},\ \bibinfo
  {pages} {214519} (\bibinfo {year} {2024})},\ \Eprint
  {https://arxiv.org/abs/2404.03465} {arXiv:2404.03465 [cond-mat]} \BibitemShut
  {NoStop}%
\bibitem [{\citenamefont {Kuklov}\ \emph
  {et~al.}(2024{\natexlab{c}})\citenamefont {Kuklov}, \citenamefont {Pollet},
  \citenamefont {Prokof'ev},\ and\ \citenamefont
  {Svistunov}}]{kuklov2024transverse}%
  \BibitemOpen
  \bibfield  {author} {\bibinfo {author} {\bibfnamefont {A.}~\bibnamefont
  {Kuklov}}, \bibinfo {author} {\bibfnamefont {L.}~\bibnamefont {Pollet}},
  \bibinfo {author} {\bibfnamefont {N.}~\bibnamefont {Prokof'ev}},\ and\
  \bibinfo {author} {\bibfnamefont {B.}~\bibnamefont {Svistunov}},\ }\bibfield
  {title} {\bibinfo {title} {Transverse quantum superfluids},\ }\href
  {https://www.annualreviews.org/content/journals/10.1146/annurev-conmatphys-042924-103908}
  {\bibfield  {journal} {\bibinfo  {journal} {Annu. Rev. Condens. Matter Phys}\
  }\textbf {\bibinfo {volume} {16}} (\bibinfo {year} {2024}{\natexlab{c}})},\
  \Eprint {https://arxiv.org/abs/2404.15480} {arXiv:2404.15480 [cond-mat]}
  \BibitemShut {NoStop}%
\bibitem [{\citenamefont {Kuklov}\ \emph
  {et~al.}(2024{\natexlab{d}})\citenamefont {Kuklov}, \citenamefont
  {Prokof'ev},\ and\ \citenamefont {Svistunov}}]{KuklovPhysRevResearch6}%
  \BibitemOpen
  \bibfield  {author} {\bibinfo {author} {\bibfnamefont {A.}~\bibnamefont
  {Kuklov}}, \bibinfo {author} {\bibfnamefont {N.}~\bibnamefont {Prokof'ev}},\
  and\ \bibinfo {author} {\bibfnamefont {B.}~\bibnamefont {Svistunov}},\
  }\bibfield  {title} {\bibinfo {title} {Autonomous dynamics of two-dimensional
  insulating domain with superclimbing edges},\ }\href
  {https://doi.org/10.1103/PhysRevResearch.6.033008} {\bibfield  {journal}
  {\bibinfo  {journal} {Phys. Rev. Res.}\ }\textbf {\bibinfo {volume} {6}},\
  \bibinfo {pages} {033008} (\bibinfo {year} {2024}{\natexlab{d}})},\ \Eprint
  {https://arxiv.org/abs/2404.18290} {arXiv:2404.18290 [cond-mat]} \BibitemShut
  {NoStop}%
\bibitem [{\citenamefont {Radzihovsky}\ and\ \citenamefont
  {Pellett}(2025)}]{Radzihovsky24}%
  \BibitemOpen
  \bibfield  {author} {\bibinfo {author} {\bibfnamefont {L.}~\bibnamefont
  {Radzihovsky}}\ and\ \bibinfo {author} {\bibfnamefont {E.}~\bibnamefont
  {Pellett}},\ }\bibfield  {title} {\bibinfo {title} {Quantum phases and
  transitions of bosons on a comb lattice},\ }\href
  {https://doi.org/10.1103/dljc-j3z7} {\bibfield  {journal} {\bibinfo
  {journal} {Phys. Rev. Lett.}\ }\textbf {\bibinfo {volume} {135}},\ \bibinfo
  {pages} {016001} (\bibinfo {year} {2025})},\ \Eprint
  {https://arxiv.org/abs/2412.06915} {arXiv:2412.06915 [cond-mat]} \BibitemShut
  {NoStop}%
\bibitem [{\citenamefont {Komura}\ and\ \citenamefont
  {Okabe}(2012)}]{Komura2012Large}%
  \BibitemOpen
  \bibfield  {author} {\bibinfo {author} {\bibfnamefont {Y.}~\bibnamefont
  {Komura}}\ and\ \bibinfo {author} {\bibfnamefont {Y.}~\bibnamefont {Okabe}},\
  }\bibfield  {title} {\bibinfo {title} {Large-scale monte carlo simulation of
  two-dimensional classical xy model using multiple gpus},\ }\href
  {https://doi.org/10.1143/JPSJ.81.113001} {\bibfield  {journal} {\bibinfo
  {journal} {J. Phys. Soc. Jpn.}\ }\textbf {\bibinfo {volume} {81}},\ \bibinfo
  {pages} {113001} (\bibinfo {year} {2012})},\ \Eprint
  {https://arxiv.org/abs/1210.6116} {arXiv:1210.6116 [cond-mat]} \BibitemShut
  {NoStop}%
\bibitem [{\citenamefont {Wolff}(1989)}]{Wolff89}%
  \BibitemOpen
  \bibfield  {author} {\bibinfo {author} {\bibfnamefont {U.}~\bibnamefont
  {Wolff}},\ }\bibfield  {title} {\bibinfo {title} {Collective monte carlo
  updating for spin systems},\ }\href
  {https://journals.aps.org/prl/abstract/10.1103/PhysRevLett.62.361} {\bibfield
   {journal} {\bibinfo  {journal} {Phys. Rev. Lett.}\ }\textbf {\bibinfo
  {volume} {62}},\ \bibinfo {pages} {361} (\bibinfo {year} {1989})}\BibitemShut
  {NoStop}%
\bibitem [{\citenamefont {Janke}(1997)}]{Janke1997}%
  \BibitemOpen
  \bibfield  {author} {\bibinfo {author} {\bibfnamefont {W.}~\bibnamefont
  {Janke}},\ }\bibfield  {title} {\bibinfo {title} {Logarithmic corrections in
  the two-dimensional xy model},\ }\href
  {https://doi.org/10.1103/PhysRevB.55.3580} {\bibfield  {journal} {\bibinfo
  {journal} {Phys. Rev. B}\ }\textbf {\bibinfo {volume} {55}},\ \bibinfo
  {pages} {3580} (\bibinfo {year} {1997})}\BibitemShut {NoStop}%
\bibitem [{\citenamefont {Chen}\ \emph {et~al.}(2022)\citenamefont {Chen},
  \citenamefont {Hou}, \citenamefont {Fang},\ and\ \citenamefont
  {Deng}}]{Chen2022}%
  \BibitemOpen
  \bibfield  {author} {\bibinfo {author} {\bibfnamefont {H.}~\bibnamefont
  {Chen}}, \bibinfo {author} {\bibfnamefont {P.}~\bibnamefont {Hou}}, \bibinfo
  {author} {\bibfnamefont {S.}~\bibnamefont {Fang}},\ and\ \bibinfo {author}
  {\bibfnamefont {Y.}~\bibnamefont {Deng}},\ }\bibfield  {title} {\bibinfo
  {title} {Monte carlo study of duality and the berezinskii-kosterlitz-thouless
  phase transitions of the two-dimensional $q$-state clock model in flow
  representations},\ }\href {https://doi.org/10.1103/PhysRevE.106.024106}
  {\bibfield  {journal} {\bibinfo  {journal} {Phys. Rev. E}\ }\textbf {\bibinfo
  {volume} {106}},\ \bibinfo {pages} {024106} (\bibinfo {year} {2022})},\
  \Eprint {https://arxiv.org/abs/2205.02642} {arXiv:2205.02642 [cond-mat]}
  \BibitemShut {NoStop}%
\bibitem [{\citenamefont {Tomita}\ and\ \citenamefont
  {Okabe}(2002)}]{Tomita2002Probability}%
  \BibitemOpen
  \bibfield  {author} {\bibinfo {author} {\bibfnamefont {Y.}~\bibnamefont
  {Tomita}}\ and\ \bibinfo {author} {\bibfnamefont {Y.}~\bibnamefont {Okabe}},\
  }\bibfield  {title} {\bibinfo {title} {Probability-changing cluster algorithm
  for two-dimensional $\mathrm{XY}$ and clock models},\ }\href
  {https://doi.org/10.1103/PhysRevB.65.184405} {\bibfield  {journal} {\bibinfo
  {journal} {Phys. Rev. B}\ }\textbf {\bibinfo {volume} {65}},\ \bibinfo
  {pages} {184405} (\bibinfo {year} {2002})},\ \Eprint
  {https://arxiv.org/abs/cond-mat/0202161} {arXiv:cond-mat/0202161 [cond-mat]}
  \BibitemShut {NoStop}%
\bibitem [{\citenamefont {Kardar}(2007)}]{kardar2007statistical}%
  \BibitemOpen
  \bibfield  {author} {\bibinfo {author} {\bibfnamefont {M.}~\bibnamefont
  {Kardar}},\ }\href@noop {} {\emph {\bibinfo {title} {Statistical physics of
  fields}}}\ (\bibinfo  {publisher} {Cambridge University Press},\ \bibinfo
  {year} {2007})\BibitemShut {NoStop}%
\bibitem [{\citenamefont {Hu}(2026)}]{hu_2026_19901475}%
  \BibitemOpen
  \bibfield  {author} {\bibinfo {author} {\bibfnamefont {M.}~\bibnamefont
  {Hu}},\ }\bibfield  {title} {\bibinfo {title} {Dataset for ``long-range order
  in a strictly short-range quasi-2d xy model: when critical fluctuations
  matter"},\ }\href {https://doi.org/10.5281/zenodo.19901475}
  {10.5281/zenodo.19901475} (\bibinfo {year} {2026})\BibitemShut {NoStop}%
\bibitem [{\citenamefont {Berche}(2003)}]{Berche2003}%
  \BibitemOpen
  \bibfield  {author} {\bibinfo {author} {\bibfnamefont {B.}~\bibnamefont
  {Berche}},\ }\bibfield  {title} {\bibinfo {title} {Bulk and surface
  properties in the critical phase of the two-dimensional xy model},\ }\href
  {https://doi.org/10.1088/0305-4470/36/3/301} {\bibfield  {journal} {\bibinfo
  {journal} {J. Phys. A: Math. Gen.}\ }\textbf {\bibinfo {volume} {36}},\
  \bibinfo {pages} {585} (\bibinfo {year} {2003})},\ \Eprint
  {https://arxiv.org/abs/cond-mat/0211584} {arXiv:cond-mat/0211584 [cond-mat]}
  \BibitemShut {NoStop}%
\bibitem [{\citenamefont {Xu}\ \emph {et~al.}(2019)\citenamefont {Xu},
  \citenamefont {Sun}, \citenamefont {Lv},\ and\ \citenamefont
  {Deng}}]{Xu2019}%
  \BibitemOpen
  \bibfield  {author} {\bibinfo {author} {\bibfnamefont {W.}~\bibnamefont
  {Xu}}, \bibinfo {author} {\bibfnamefont {Y.}~\bibnamefont {Sun}}, \bibinfo
  {author} {\bibfnamefont {J.-P.}\ \bibnamefont {Lv}},\ and\ \bibinfo {author}
  {\bibfnamefont {Y.}~\bibnamefont {Deng}},\ }\bibfield  {title} {\bibinfo
  {title} {High-precision monte carlo study of several models in the
  three-dimensional u(1) universality class},\ }\href
  {https://doi.org/10.1103/PhysRevB.100.064525} {\bibfield  {journal} {\bibinfo
   {journal} {Phys. Rev. B}\ }\textbf {\bibinfo {volume} {100}},\ \bibinfo
  {pages} {064525} (\bibinfo {year} {2019})},\ \Eprint
  {https://arxiv.org/abs/1908.10990} {arXiv:1908.10990 [cond-mat]} \BibitemShut
  {NoStop}%
\bibitem [{\citenamefont {Dantchev}\ and\ \citenamefont
  {Dietrich}(2023)}]{Dantchev2023}%
  \BibitemOpen
  \bibfield  {author} {\bibinfo {author} {\bibfnamefont {D.~M.}\ \bibnamefont
  {Dantchev}}\ and\ \bibinfo {author} {\bibfnamefont {S.}~\bibnamefont
  {Dietrich}},\ }\bibfield  {title} {\bibinfo {title} {Critical casimir effect:
  exact results},\ }\href {https://doi.org/10.1016/j.physrep.2022.12.004}
  {\bibfield  {journal} {\bibinfo  {journal} {Phys. Rep.}\ }\textbf {\bibinfo
  {volume} {1005}},\ \bibinfo {pages} {1} (\bibinfo {year} {2023})},\ \Eprint
  {https://arxiv.org/abs/2203.15050} {arXiv:2203.15050 [cond-mat]} \BibitemShut
  {NoStop}%
\bibitem [{\citenamefont {Andrews}\ \emph {et~al.}(1997)\citenamefont
  {Andrews}, \citenamefont {Townsend}, \citenamefont {Miesner}, \citenamefont
  {Durfee}, \citenamefont {Kurn},\ and\ \citenamefont
  {Ketterle}}]{andrews1997observation}%
  \BibitemOpen
  \bibfield  {author} {\bibinfo {author} {\bibfnamefont {M.~R.}\ \bibnamefont
  {Andrews}}, \bibinfo {author} {\bibfnamefont {C.~G.}\ \bibnamefont
  {Townsend}}, \bibinfo {author} {\bibfnamefont {H.-J.}\ \bibnamefont
  {Miesner}}, \bibinfo {author} {\bibfnamefont {D.~S.}\ \bibnamefont {Durfee}},
  \bibinfo {author} {\bibfnamefont {D.~M.}\ \bibnamefont {Kurn}},\ and\
  \bibinfo {author} {\bibfnamefont {W.}~\bibnamefont {Ketterle}},\ }\bibfield
  {title} {\bibinfo {title} {Observation of interference between two bose
  condensates},\ }\href
  {https://www.science.org/doi/abs/10.1126/science.275.5300.637} {\bibfield
  {journal} {\bibinfo  {journal} {Science}\ }\textbf {\bibinfo {volume}
  {275}},\ \bibinfo {pages} {637} (\bibinfo {year} {1997})}\BibitemShut
  {NoStop}%
\bibitem [{\citenamefont {Bakr}\ \emph {et~al.}(2009)\citenamefont {Bakr},
  \citenamefont {Gillen}, \citenamefont {Peng}, \citenamefont {F{\"o}lling},\
  and\ \citenamefont {Greiner}}]{bakr2009quantum}%
  \BibitemOpen
  \bibfield  {author} {\bibinfo {author} {\bibfnamefont {W.~S.}\ \bibnamefont
  {Bakr}}, \bibinfo {author} {\bibfnamefont {J.~I.}\ \bibnamefont {Gillen}},
  \bibinfo {author} {\bibfnamefont {A.}~\bibnamefont {Peng}}, \bibinfo {author}
  {\bibfnamefont {S.}~\bibnamefont {F{\"o}lling}},\ and\ \bibinfo {author}
  {\bibfnamefont {M.}~\bibnamefont {Greiner}},\ }\bibfield  {title} {\bibinfo
  {title} {A quantum gas microscope for detecting single atoms in a
  hubbard-regime optical lattice},\ }\href
  {https://www.nature.com/articles/nature08482} {\bibfield  {journal} {\bibinfo
   {journal} {Nature}\ }\textbf {\bibinfo {volume} {462}},\ \bibinfo {pages}
  {74} (\bibinfo {year} {2009})},\ \Eprint {https://arxiv.org/abs/0908.0174}
  {arXiv:0908.0174 [cond-mat]} \BibitemShut {NoStop}%
\bibitem [{\citenamefont {Gauthier}\ \emph {et~al.}(2016)\citenamefont
  {Gauthier}, \citenamefont {Lenton}, \citenamefont {McKay~Parry},
  \citenamefont {Baker}, \citenamefont {Davis}, \citenamefont
  {Rubinsztein-Dunlop},\ and\ \citenamefont {Neely}}]{gauthier2016direct}%
  \BibitemOpen
  \bibfield  {author} {\bibinfo {author} {\bibfnamefont {G.}~\bibnamefont
  {Gauthier}}, \bibinfo {author} {\bibfnamefont {I.}~\bibnamefont {Lenton}},
  \bibinfo {author} {\bibfnamefont {N.}~\bibnamefont {McKay~Parry}}, \bibinfo
  {author} {\bibfnamefont {M.}~\bibnamefont {Baker}}, \bibinfo {author}
  {\bibfnamefont {M.~J.}\ \bibnamefont {Davis}}, \bibinfo {author}
  {\bibfnamefont {H.}~\bibnamefont {Rubinsztein-Dunlop}},\ and\ \bibinfo
  {author} {\bibfnamefont {T.~W.}\ \bibnamefont {Neely}},\ }\bibfield  {title}
  {\bibinfo {title} {Direct imaging of a digital-micromirror device for
  configurable microscopic optical potentials},\ }\href
  {https://doi.org/10.1364/OPTICA.3.001136} {\bibfield  {journal} {\bibinfo
  {journal} {Optica}\ }\textbf {\bibinfo {volume} {3}},\ \bibinfo {pages}
  {1136} (\bibinfo {year} {2016})},\ \Eprint {https://arxiv.org/abs/1605.04928}
  {arXiv:1605.04928 [cond-mat]} \BibitemShut {NoStop}%
\end{thebibliography}%


\begin{thebibliography}{3}%
\makeatletter
\providecommand \@ifxundefined [1]{%
 \@ifx{#1\undefined}
}%
\providecommand \@ifnum [1]{%
 \ifnum #1\expandafter \@firstoftwo
 \else \expandafter \@secondoftwo
 \fi
}%
\providecommand \@ifx [1]{%
 \ifx #1\expandafter \@firstoftwo
 \else \expandafter \@secondoftwo
 \fi
}%
\providecommand \natexlab [1]{#1}%
\providecommand \enquote  [1]{``#1''}%
\providecommand \bibnamefont  [1]{#1}%
\providecommand \bibfnamefont [1]{#1}%
\providecommand \citenamefont [1]{#1}%
\providecommand \href@noop [0]{\@secondoftwo}%
\providecommand \href [0]{\begingroup \@sanitize@url \@href}%
\providecommand \@href[1]{\@@startlink{#1}\@@href}%
\providecommand \@@href[1]{\endgroup#1\@@endlink}%
\providecommand \@sanitize@url [0]{\catcode `\\12\catcode `\$12\catcode
  `\&12\catcode `\#12\catcode `\^12\catcode `\_12\catcode `\%12\relax}%
\providecommand \@@startlink[1]{}%
\providecommand \@@endlink[0]{}%
\providecommand \url  [0]{\begingroup\@sanitize@url \@url }%
\providecommand \@url [1]{\endgroup\@href {#1}{\urlprefix }}%
\providecommand \urlprefix  [0]{URL }%
\providecommand \Eprint [0]{\href }%
\providecommand \doibase [0]{https://doi.org/}%
\providecommand \selectlanguage [0]{\@gobble}%
\providecommand \bibinfo  [0]{\@secondoftwo}%
\providecommand \bibfield  [0]{\@secondoftwo}%
\providecommand \translation [1]{[#1]}%
\providecommand \BibitemOpen [0]{}%
\providecommand \bibitemStop [0]{}%
\providecommand \bibitemNoStop [0]{.\EOS\space}%
\providecommand \EOS [0]{\spacefactor3000\relax}%
\providecommand \BibitemShut  [1]{\csname bibitem#1\endcsname}%
\let\auto@bib@innerbib\@empty
\bibitem [{\citenamefont {Gauthier}\ \emph {et~al.}(2016)\citenamefont
  {Gauthier}, \citenamefont {Lenton}, \citenamefont {McKay~Parry},
  \citenamefont {Baker}, \citenamefont {Davis}, \citenamefont
  {Rubinsztein-Dunlop},\ and\ \citenamefont {Neely}}]{gauthier2016direct}%
  \BibitemOpen
  \bibfield  {author} {\bibinfo {author} {\bibfnamefont {G.}~\bibnamefont
  {Gauthier}}, \bibinfo {author} {\bibfnamefont {I.}~\bibnamefont {Lenton}},
  \bibinfo {author} {\bibfnamefont {N.}~\bibnamefont {McKay~Parry}}, \bibinfo
  {author} {\bibfnamefont {M.}~\bibnamefont {Baker}}, \bibinfo {author}
  {\bibfnamefont {M.~J.}\ \bibnamefont {Davis}}, \bibinfo {author}
  {\bibfnamefont {H.}~\bibnamefont {Rubinsztein-Dunlop}},\ and\ \bibinfo
  {author} {\bibfnamefont {T.~W.}\ \bibnamefont {Neely}},\ }\bibfield  {title}
  {\bibinfo {title} {Direct imaging of a digital-micromirror device for
  configurable microscopic optical potentials},\ }\href
  {https://doi.org/10.1364/OPTICA.3.001136} {\bibfield  {journal} {\bibinfo
  {journal} {Optica}\ }\textbf {\bibinfo {volume} {3}},\ \bibinfo {pages}
  {1136} (\bibinfo {year} {2016})},\ \Eprint {https://arxiv.org/abs/1605.04928}
  {arXiv:1605.04928 [cond-mat]} \BibitemShut {NoStop}%
\bibitem [{\citenamefont {Bakr}\ \emph {et~al.}(2009)\citenamefont {Bakr},
  \citenamefont {Gillen}, \citenamefont {Peng}, \citenamefont {F{\"o}lling},\
  and\ \citenamefont {Greiner}}]{bakr2009quantum}%
  \BibitemOpen
  \bibfield  {author} {\bibinfo {author} {\bibfnamefont {W.~S.}\ \bibnamefont
  {Bakr}}, \bibinfo {author} {\bibfnamefont {J.~I.}\ \bibnamefont {Gillen}},
  \bibinfo {author} {\bibfnamefont {A.}~\bibnamefont {Peng}}, \bibinfo {author}
  {\bibfnamefont {S.}~\bibnamefont {F{\"o}lling}},\ and\ \bibinfo {author}
  {\bibfnamefont {M.}~\bibnamefont {Greiner}},\ }\bibfield  {title} {\bibinfo
  {title} {A quantum gas microscope for detecting single atoms in a
  hubbard-regime optical lattice},\ }\href
  {https://www.nature.com/articles/nature08482} {\bibfield  {journal} {\bibinfo
   {journal} {Nature}\ }\textbf {\bibinfo {volume} {462}},\ \bibinfo {pages}
  {74} (\bibinfo {year} {2009})},\ \Eprint {https://arxiv.org/abs/0908.0174}
  {arXiv:0908.0174 [cond-mat]} \BibitemShut {NoStop}%
\bibitem [{\citenamefont {Andrews}\ \emph {et~al.}(1997)\citenamefont
  {Andrews}, \citenamefont {Townsend}, \citenamefont {Miesner}, \citenamefont
  {Durfee}, \citenamefont {Kurn},\ and\ \citenamefont
  {Ketterle}}]{andrews1997observation}%
  \BibitemOpen
  \bibfield  {author} {\bibinfo {author} {\bibfnamefont {M.~R.}\ \bibnamefont
  {Andrews}}, \bibinfo {author} {\bibfnamefont {C.~G.}\ \bibnamefont
  {Townsend}}, \bibinfo {author} {\bibfnamefont {H.-J.}\ \bibnamefont
  {Miesner}}, \bibinfo {author} {\bibfnamefont {D.~S.}\ \bibnamefont {Durfee}},
  \bibinfo {author} {\bibfnamefont {D.~M.}\ \bibnamefont {Kurn}},\ and\
  \bibinfo {author} {\bibfnamefont {W.}~\bibnamefont {Ketterle}},\ }\bibfield
  {title} {\bibinfo {title} {Observation of interference between two bose
  condensates},\ }\href
  {https://www.science.org/doi/abs/10.1126/science.275.5300.637} {\bibfield
  {journal} {\bibinfo  {journal} {Science}\ }\textbf {\bibinfo {volume}
  {275}},\ \bibinfo {pages} {637} (\bibinfo {year} {1997})}\BibitemShut
  {NoStop}%
\end{thebibliography}%

\clearpage
\appendix

\section*{End Matter}

\begin{figure}[t!]
\includegraphics[height=7.6cm,width=8cm]{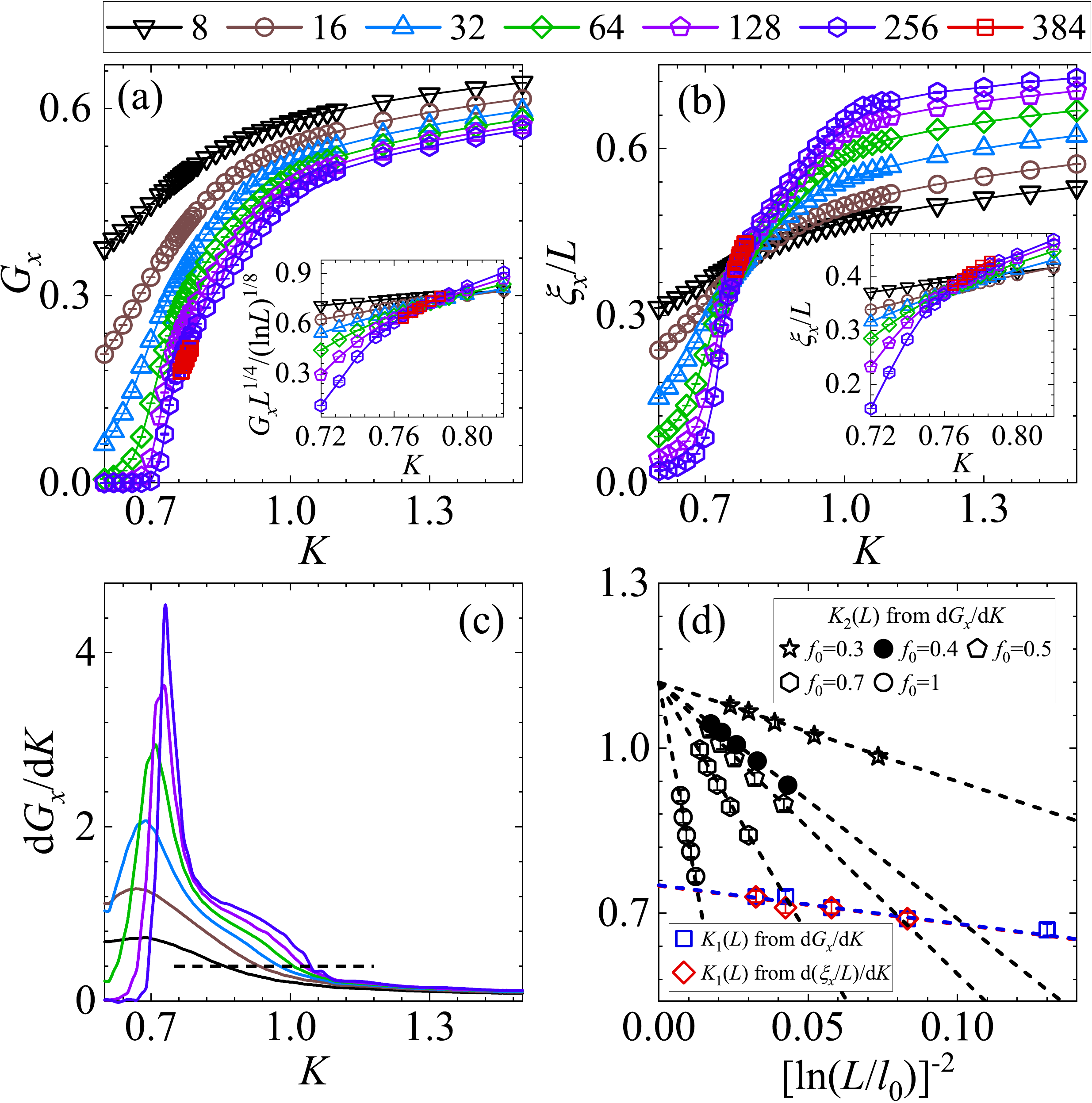}
\caption{Phase transitions of an $x$ line in the V plane with $W=0.8$. (a) The large-distance spin-spin correlation $G_x$ versus $K$. Inset: $G_xL^{1/4}/({\rm ln}L)^{1/8}$ versus $K$. (b) The reduced correlation length $\xi_x/L$ versus $K$. Inset: a zoom-in plot. (c) The derivative ${\rm d}G_x/{\rm d}K$ versus $K$. The dashed line represents $f_0=0.4$. (d) $K_1(L)$ from ${\rm d}G_x/{\rm d}K$ and ${\rm d}(\xi_x/L)/{\rm d}K$ as well as $K_2(L)$ from ${\rm d}G_x/{\rm d}K$ versus $[{\rm ln}(L/l_0)]^{-2}$. Conforming to fits, $l_0=1$ is set for $K_1(L)$. For $K_2(L)$, the values of $l_0$ are $l_0=0.4$ $(f_0=0.3)$, $0.13$ $(f_0=0.4)$, $0.12$ $(f_0=0.5)$, $0.05$ $(f_0=0.7)$ and $0.002$ $(f_0=1)$. The intercepts $K_1 \approx 0.750$ and $K_2=1.11996$ represent the thermodynamic limits of $K_1(L)$ and $K_2(L)$, respectively.}~\label{Fig5}
\end{figure}

\headline{Phase Transitions of an $x$ Line}
We analyze the transitions of an $x$ line with $W=0.8$. Figure~\ref{Fig5}(a) shows that $G_x$ varies drastically at $K_1$. For $K<K_1$, $G_x$ decays fast to zero in the $L \rightarrow \infty$ limit, indicating that the line is in a disordered phase. The FSS of $G_x$ confirms a BKT transition at $K_1$ (SM). Figure~\ref{Fig5}(b) shows that $\xi_x/L$ starts to be scale-invariant at $K_1$. Figure~\ref{Fig5}(c) demonstrates that ${\rm d}G_x/{\rm d}K$ is peaked at $K_1$ and decreases drastically at $K_2$. For each $L$, we define the pseudo-critical points $K_1(L)$ and $K_2(L)$: $K_1(L)$ is the location of the peak of ${\rm d}G_x/{\rm d}K$, while $K_2(L)$ is the numerical result of $K$ satisfying ${\rm d}G_x/{\rm d}K=f_0$ (SM), with $f_0$ a constant. For $K_1(L)$ and $K_2(L)$, we perform fits according to Eq.~(\ref{Equ_K1}) separately. Using $l_0=1$, we obtain $K_1=0.751(10)$ and $\chi^2/{\rm DOF} \approx 0.4$ with $L_{\rm min}=16$. Using $f_0=0.4$ and free $l_0$, we obtain $K_2=1.12(8)$ with $L_{\rm min}=16$. A consistency check over different values of $f_0$ is performed in SM. It is therefore confirmed that $K_2=J_{\rm BKT}$. An additional verification is that, the $L \rightarrow \infty$ limit of $K_1(L)$ extracted from ${\rm d}(\xi_x/L)/{\rm d}K$ goes to $K_1=0.75(3)$ with $\chi^2/{\rm DOF} \approx 1.0$ and $L_{\rm min}=64$, which is consistent with aforementioned estimates of $K_1$. The variations of $K_n(L)$ with $[{\rm ln}(L/l_0)]^{-2}$ ($n=1,2$) are shown in Fig.~\ref{Fig5}(d). A good agreement is found between estimated $K_n$ from $y$ and $x$ lines.

It is instructive to contrast the behavior at $K_2$ in the $y$- and $x$-directions. In the $y$-direction, $K_2$ signals the onset of LR order from a QLR ordered phase (Fig.~\ref{Fig2}). Correspondingly, the FSS form of $\xi_y/L$ changes drastically across $K_2$, and its derivative $\mathrm{d}(\xi_y/L)/\mathrm{d}K$ develops a pronounced, size-enhanced peak near $K_2$ (with the characteristic BKT-type finite-size drift). By contrast, the $x$-direction remains critical for all finite $K \ge K_1$. Consequently, while $\xi_x/L$ remains of order one on both sides of $K_2$, making it difficult to justify the divergence of its derivative, the behavior of $G_x$ offers clearer insight. In the thermodynamic limit, $G_x$ vanishes on both sides of $K_2$, and $\mathrm{d}G_x/\mathrm{d}K$ is not expected to diverge. However, the asymptotic decay form of $G_x$ changes across $K_2$: for $K_1 \le K < K_2$, $G_x \sim L^{-\eta(K)}$; for $K \ge K_2$, our data are consistent with either a logarithmic decay $G_x \sim ({\rm ln}L)^{-\hat{q}(K)}$ or a distinct power law. This change in the underlying critical behavior manifests in finite-size systems as a sharp feature---a kink---in $\mathrm{d}G_x/\mathrm{d}K$ at $K_2$ [Fig.~\ref{Fig5}(c)]. The height of this kink remains finite and ultimately tends to zero in the thermodynamic limit. We use the location of this finite-size feature to define pseudo-critical points for our scaling analysis.

\headline{Isotropic Critical Phase}
In the region between the $K_1$ and $K_2$ lines, the V plane is critical. However, it is unknown whether the anomalous dimensions are identical for critical $y$ and $x$ lines. To address this issue, we perform fits of $G_y$ and $G_x$ according to the form $G \sim L^{-\eta}$, where $\eta$ denotes the anomalous dimension. We find that the fitting results of $\eta$ are close for the $y$ and $x$ lines, with $\eta \approx 0.130$ $(W=1.5, K=0.4)$, $0.117$ $(W=1.5, K=0.6)$, $0.085$ $(W=1.5, K=0.8)$ and $0.065$ $(W=2, K=0.8)$. The details of fits are presented in SM, where the isotropy of $\eta$ is furthermore examined, as $G_x/G_y$ tends to be invariant upon increasing $L$. These results further indicate that the anisotropy of ordering in the V plane is largely suppressed with $K<K_2$.

\begin{figure}
	\centering
	\includegraphics[height=3.6cm,width=8cm]{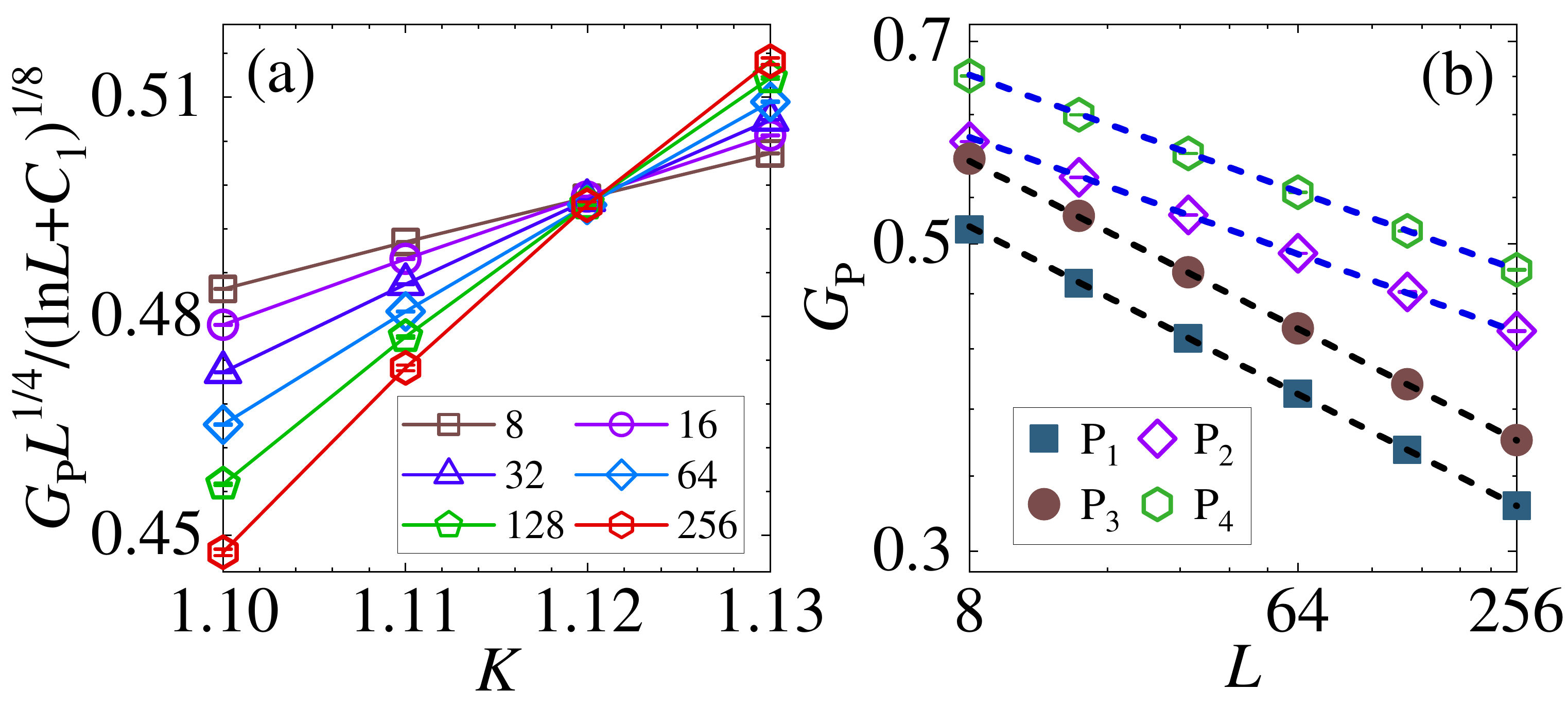}	
	\caption{Correlations between P planes. (a) Scaled large-distance spin-spin correlation $G_{\rm P}L^{1/4}/({\rm ln}L+C_1)^{1/8}$ versus $K$ for $W=0.8$. The value $C_1=7.4$ is from a preferred fit. (b) Log-log plot of $G_{\rm P}$ versus $L$ for ${\rm P_1}$, ${\rm P_2}$, ${\rm P_3}$ and ${\rm P_4}$. The slopes of black and blue lines represent $-\eta=-0.1339$ and $-0.0934$, respectively.}\label{Fig6}
\end{figure}

\begin{figure*}
\centering
\includegraphics[height=4.6cm,width=13cm]{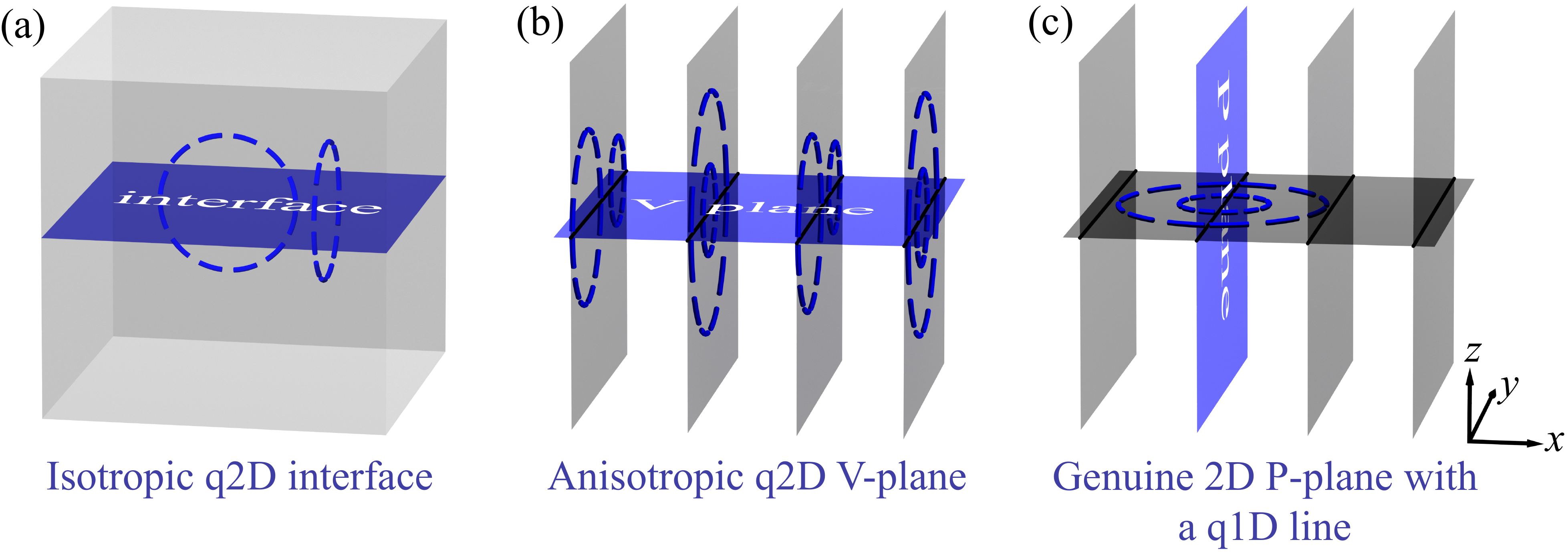}
\caption{Schematic illustration comparing different q2D XY systems. (a) Isotropic q2D interface. The $x$- and $y$-directions exhibit isotropic effective interactions (denoted by dashed lines), which are pronounced only when the bulk is tuned to the 3D XY critical point. The same applies to isotropic q2D surface. (b) Anisotropic q2D V-plane. The $y$-direction exhibits more pronounced effective interactions, which can be tuned by changing the coupling strengths within the P planes. (c) Genuine 2D P-plane with a q1D line. Pronounced effective interactions can be mediated in the q1D intersection but cannot alter the macroscopic behavior of the P plane.}~\label{Fig7}
\end{figure*}

\headline{Critical Correlations Between P Planes}
To analyze the correlations between different P planes, we sample the large-distance correlation $G_{\rm P} = \langle \vec{S}_{{\bf r}} \cdot \vec{S}_{{\bf r}+(L/2,0,0)} \rangle$ for $W=0.8$, where ${\bf r}$ is a site in a P plane that is displaced from the V plane by a distance of $L/2$ along the $z$-direction. Figure~\ref{Fig6}(a) shows that $G_{\rm P}$ obeys the BKT scaling form at $K_2$. For $K \ge K_2$, $G_{\rm P}$ scales as $G_{\rm P} \sim L^{-\eta}$. The exponent $\eta$ is found to be $\eta=0.1339(3)$ and $0.0934(5)$ for $K=1.5$ and $2$ [Fig.~\ref{Fig6}(b)], respectively, which are independent of $W$. Remarkably, our estimates of $\eta$ for $K=1.5$ and $2$ roughly agree with the results for the standard 2D XY model~\cite{Berche2003}, where $K$ is the nearest-neighbour coupling. 

We explain the scaling behavior of $G_{\rm P}$. Consider two XY spins $\vec{S}(x_1,y,z)$ and $\vec{S}(x_2,y,z)$ with the same $y$ and $z$ coordinates but different $x$. The correlation between them factorizes approximately as the product of their individual couplings to the V plane at $z=0$ and an in-plane correlation within the V plane: $\langle \vec{S}(x_1,y,z) \cdot \vec{S}(x_1,y,0) \rangle \cdot g_x(|x_1-x_2|) \cdot \langle \vec{S}(x_2,y,0) \cdot \vec{S}(x_2,y,z) \rangle$, with $g_x(|x_1-x_2|)$ the in-plane correlation along the $x$-direction. For $K \ge K_2$, each factor $\langle \vec{S}(x_i,y,z) \cdot \vec{S}(x_i,y,0) \rangle$ measures how a P-plane spin responds to the ordered intersection line. This response inherits the critical fluctuations of the P plane: if the intra-plane correlation scales as $|z|^{-\eta}$, then the factor scales as $|z|^{-\eta/2}$. The $x$-direction correlation $g_x(|x_1-x_2|)$ is most likely to follow a logarithmic form and merely contributes a correction to the power-law decay. Combining these factors yields the leading scaling $\langle \vec{S}(x_1,y,z) \cdot \vec{S}(x_1,y,0) \rangle \cdot g_x(|x_1-x_2|) \cdot \langle \vec{S}(x_2,y,0) \cdot \vec{S}(x_2,y,z) \rangle \sim |z|^{-\eta}$. According to our definition of $G_{\rm P}$ with $z = L/2$, one has $G_{\rm P} \sim L^{-\eta}$.

\headline{Comparison among q2D Systems}
The significance of our results becomes evident in relation to interface and surface critical phenomena in the critical 3D XY model~\cite{Metlitski,Hu2021,ToldinMetlitski22,Krishnan2023,Sun2023}. Figure~\ref{Fig7} shows a schematic comparison. The features of effective interactions mediated via the critical 3D bulk are illustrated. Despite extensive efforts, conclusive evidence for LR order on such interfaces and surfaces has not been established for any finite coupling. Instead, a highly nontrivial critical phase---the E-Log phase---has been observed under strong interface or surface couplings, characterized by the logarithmic decay of correlations and thus the absence of LR order. We note that, in contrast to the 3D bulk environment of interface and open surfaces, the P planes in our setup exhibit genuinely 2D characteristics. This result highlights the crucial role of dimensionality: a 2D critical environment can induce order more effectively than a 3D one---a counterintuitive yet central insight based on critical fluctuations. 

A crucial point is that the critical behavior in the P planes can be tuned continuously via $K$, whereas the 3D XY model with interface or open surfaces exhibits bulk criticality only at a single point, $K=K_c$ (see Fig.~\ref{Fig7}). Consider the two-point correlation function $g(r)$. In the critical 3D bulk, it decays as $\sim r^{-2X_{\rm 3D}}$ with $X_{\rm 3D} \approx 0.52$~\cite{Xu2019}, while in the low-temperature phase of the 2D XY model, it decays as $\sim r^{-2X_{\rm 2D}}$ with $X_{\rm 2D} \leq 1/8$~\cite{Kosterlitz17}. That is, in the former case, the correlation function decays much more rapidly with $r$. This difference suggests a possible explanation for the counterintuitive finding: the 2D critical P planes may mediate more pronounced effective LR interactions than the 3D critical bulk does. This schematic analysis would serve as a starting point for a rigorous analytical derivation based on field theories.

\headline{Outlook}
In addition to a field-theoretical investigation, the present study can be extended to various contexts. First, our setup can be utilized to investigate classical and quantum critical systems, such as Ising, XY and Heisenberg models. It would be interesting to examine whether---and if so, how---the ordering of the V plane can be enhanced. Second, the results of LR order motivate a future study on impurity problems in q1D quantum systems. Specifically, it would be intriguing to study how an impurity on the spine of a comb lattice behaves when the surrounding teeth are tuned to a critical phase. In a conventional, non-critical environment, an impurity may disrupt the 1D order. However, a critical bath of teeth could fundamentally alter the impurity physics, potentially leading to novel screening or delocalization phenomena that are inaccessible in standard, non-critical settings. These effects might be realized with the Bose-Hubbard Hamiltonian. Third, our study highlights ``exotic phenomena induced by fluctuations" as a more or less unifying theoretical framework by providing a missing piece, namely the realization of LR order by critical fluctuations, supplementing the Casimir effect~\cite{Dantchev2023}. Thus, the present results are relevant to a large family of many-body effects. Finally, an experimental scheme, using optical lattice quantum emulator, is formulated in SM, where Refs.~\cite{andrews1997observation,bakr2009quantum,gauthier2016direct} are included.

\end{document}


\title{\fontsize{11.4}{15}\selectfont Supplemental Material for \\
``Long-Range Order in a Strictly Short-Range Quasi-2D XY Model: When Critical Fluctuations Matter"}

\maketitle

\section{Monte Carlo simulations}
Wolff's cluster algorithm is employed to perform Monte Carlo simulations. In each Markov chain simulation, the first one sixth of Monte Carlo steps are used for thermalization, after which samplings are performed once every $\delta N$ Wolff clusters are generated. Thus, correlations between sampled configurations are suppressed. For fixed $W$ and $L$, the value of $\delta N$ is typically set to decrease as $K$ increases. For $W=0.4$, $0.8$ and $1.5$, the ranges of cluster numbers are summarized in Table~\ref{Tab_steps}. 

\begin{table}
	\caption{The ranges of Wolff cluster numbers with $W=0.4$, $0.8$ and $1.5$. $N_{\rm min}$ and $N_{\rm max}$ denote the minimum and maximum numbers of created Wolff clusters respectively, while ``-" implies $N_{\rm max}=N_{\rm min}$.}~\label{Tab_steps}
	\centering
	\scalebox{1.0}{
		\tabcolsep 0.1in
		\begin{tabular}{c|c|c}
			\hline
			\hline      
			$L$   &  $N_{\rm min}$ & $N_{\rm max}$      \\
			\hline						
			8   & $2.2 \times 10^8$ & $2.9 \times 10^8$ \\
			16  & $2.2 \times 10^8$ & $5.8 \times 10^8$ \\
			24  & $8.6 \times 10^8$ & - \\
			32  & $7.2 \times 10^8$ & $3.5 \times 10^{10}$ \\
			48  & $2.2 \times 10^8$ & $1.7 \times 10^{10}$ \\
			64  & $5.8 \times 10^7$ & $2.8 \times 10^{10}$ \\
			80  & $7.2 \times 10^7$ & - \\
			96  & $2.9 \times 10^7$ & $1.1 \times 10^8$ \\
			128 & $3.6 \times 10^6$ & $4.6 \times 10^9$ \\
			192 & $3.6 \times 10^6$ & $1.8 \times 10^8$ \\
			256 & $1.4 \times 10^6$ & $3.1 \times 10^8$ \\
			384 & $2.9 \times 10^6$ & $1.2 \times 10^7$ \\
			\hline
			\hline
		\end{tabular}
	}
\end{table}

\section{Critical phenomena in the V plane}

\subsection{The phase transition points for $W=0.8$}
To quantify the variation of $\xi_y/L$ with $K$, we compute the derivative ${\rm d}(\xi_y/L)/{\rm d}K$. To this end, we first apply a linear method to perform interpolation of $\xi_y/L$, and then employ a differential module without any smoothing. Such a procedure is utilized to obtain all numerical differentials throughout this Letter. When $L$ is sufficiently large, we find a pair of peaks $K_1(L)$ and $K_2(L)$ [$K_1(L)<K_2(L)$] from the ${\rm d}(\xi_y/L)/{\rm d}K$-$K$ curve and a peak $K_1(L)$ from the ${\rm d}G_y/{\rm d}K$-$K$ curve. 
The error bars of $K_1(L)$ and $K_2(L)$ are non-statistical and approximated using bounds introduced within the interpolation processes. For $K_1(L)$, we perform least-squares fits according to the scaling formula
\begin{equation}~\label{Equ_Kn}
	K_n(L)=K_n+a[{\rm ln}(L/l_0)]^{-2}
\end{equation}
with $n=1$, where $a$ denotes a non-universal constant and $l_0$ represents a reference length. Similarly, we estimate $K_2$ according to Eq.~(\ref{Equ_Kn}) with $n=2$. The results of fits are summarized in Table~\ref{Tab_Kn8}.

Similar to what is done for the $y$ lines, we perform least-squares fits of $K_1(L)$, extracted from ${\rm d}G_x/{\rm d}K$ and ${\rm d}(\xi_x/L)/{\rm d}K$, according to Eq.~(\ref{Equ_Kn}). Details of fits are given in Table~\ref{Tab_Kn8}. We also perform fits of $K_2(L)$, which is the value of $K$ satisfying ${\rm d}G_x/{\rm d}K=f_0$, obtained via a linear method. The error bar of $K_2(L)$ is also approximated using a bound introduced within the interpolation process. The cases with $f_0=0.3$, $0.4$, $0.5$, $0.7$ and $1$ are considered. As $L$ increases, the difference between estimated $K_2(L)$ becomes negligible. The results of fits are summarized in Table~\ref{B_Tab_7}. It is found that the results of $K_2$ are all consistent with $K_2 \approx 1.11996$, although a larger value of $f_0$ typically causes more sizable uncertainties.

\subsection{The anomalous dimension at $K_1$}
\begin{figure}[t]
	\centering
	\includegraphics[height=8cm,width=8cm]{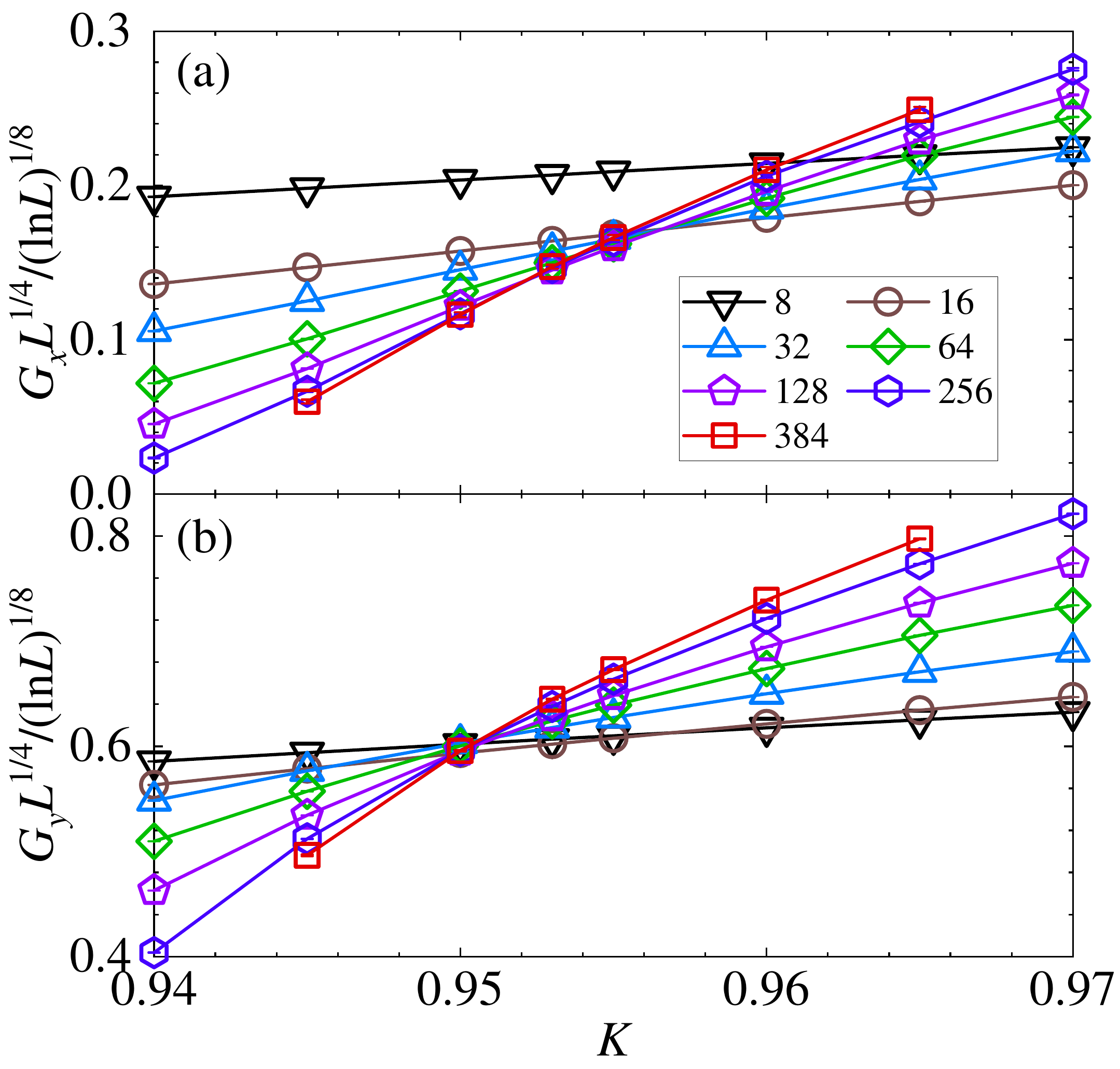}	
	\caption{Scaled large-distance spin-spin correlations $G_xL^{1/4}/({\rm ln}L)^{1/8}$ and $G_yL^{1/4}/({\rm ln}L)^{1/8}$ versus $K$ for $W=0.4$.}\label{SM_G}
\end{figure}

We conduct a detailed analysis of $G_x$ and $G_y$ near $K_1$ for $W=0.4$ and $0.8$. For a set of Monte Carlo data around $K_1$, we perform fits according to the FSS formula
\begin{equation}~\label{K1}	
	\begin{split}
		G_{\alpha}&=L^{-\eta}({\rm ln}L+C_1)^{1/8}[a_0+\sum_{k=1}^3a_k \epsilon ^k ({\rm ln}L+C_2)^{2k} \\
		&+ d_1L^{-1} + d_2L^{-2} + n_0\epsilon + n_1\epsilon^2({\rm ln}L+C_2)^2]
	\end{split}
\end{equation}
with $\alpha \in \{x,y\}$ and $\epsilon=K_1-K$. The term with $a_k$ (for $k=0,1,2,3$) originates from the Taylor expansion of the scaled correlation. This expansion is taken with respect to the scaling variable  $\epsilon[{\rm ln}(L/l_0)]^2 = \epsilon ({\rm ln}L+C_2)^2$ around the critical point. The exponent $\eta$ denotes the anomalous dimension, while $C_1$, $C_2$, $d_1$, $d_2$, $n_0$ and $n_1$ are non-universal. The term $({\rm ln}L+C_1)^{1/8}$ represents a multiplicative logarithmic correction. The terms $d_1L^{-1}$ and $d_2L^{-2}$ account for additive finite-size corrections. The term $n_0\epsilon$ describes the asymmetry dependence of the scaling function, while the term $n_1\epsilon^2({\rm ln}L+C_2)^2$ originates from the nonlinearity of the renormalization group invariant function. We perform a series of fits, and the preferred ones are those with $C_1=C_2=0$. In the case $W=0.4$, from $G_x$, we obtain $K_1=0.950(2)$, $\eta=0.30(3)$, $a_3 \approx 0$ and $d_1 \approx 0$ as well as $\chi^2/{\rm DOF} \approx 0.7$, with $L_{\rm min}=64$. Setting $a_3=d_1=0$ gives $K_1=0.950(1)$ and $\eta=0.29(3)$ as well as $\chi^2/{\rm DOF} \approx 0.6$, with $L_{\rm min}=64$. Fixing $\eta=1/4$ results in $K_1=0.9515(2)$ and $\chi^2/{\rm DOF} \approx 0.7$ with $L_{\rm min}=64$. From $G_y$, we obtain $K_1=0.9508(3)$ and $\eta=0.249(3)$ as well as $\chi^2/{\rm DOF} \approx 2.0$, with $L_{\rm min}=64$. Fixing $\eta=1/4$ results in $K_1 = 0.9507(3)$ and $\chi^2/{\rm DOF} \approx 1.7$ with $L_{\rm min}=64$. Details of fits are summarized in Table~\ref{Tab_1}. For $W = 0.4$, the final estimate is $K_1 \approx 0.951$. The scaled correlations $G_xL^{1/4}/({\rm ln}L)^{1/8}$ and $G_yL^{1/4}/({\rm ln}L)^{1/8}$ are shown in Figs.~\ref{SM_G}(a) and~\ref{SM_G}(b), respectively. 
In the case $W=0.8$, the analysis of $G_x$ gives $K_1=0.76(1)$, $\eta=0.31(4)$, $a_3 \approx 0$ and $d_1 \approx 0$ as well as $\chi^2/{\rm DOF} \approx 1.3$, with $L_{\rm min}=16$. Setting $a_3=d_1=0$, we obtain $K_1=0.77(1)$ and $\eta=0.26(3)$ as well as $\chi^2/{\rm DOF} \approx 1.2$, with $L_{\rm min}=64$. Fixing $\eta=1/4$ results in $K_1=0.7727(10)$, $\chi^2/{\rm DOF} \approx 1.7$ with $L_{\rm min}=32$. The analysis of $G_y$ gives $K_1=0.769(2)$, $\eta=0.258(7)$ and $a_3 \approx 0$ as well as $\chi^2/{\rm DOF} \approx 1.3$, with $L_{\rm min}=32$. Setting $a_3=0$ gives $K_1=0.768(3)$ and $\eta=0.258(9)$ as well as $\chi^2/{\rm DOF} \approx 1.2$, with $L_{\rm min}=32$. Fixing $\eta=1/4$ results in $K_1 = 0.7714(5)$ and $\chi^2/{\rm DOF} \approx 1.3$ with $L_{\rm min}=32$. Details of fits are summarized in Table~\ref{Tab_1}. For $W = 0.8$, the final estimate is $K_1 \approx 0.772$. For $W = 0.4$ and $0.8$, the values of $\eta$ are consistent with $1/4$. The values from $G_x$ and $G_y$ are in agreement with each other.

\begin{figure}[t]
	\centering
	\includegraphics[height=6cm,width=8cm]{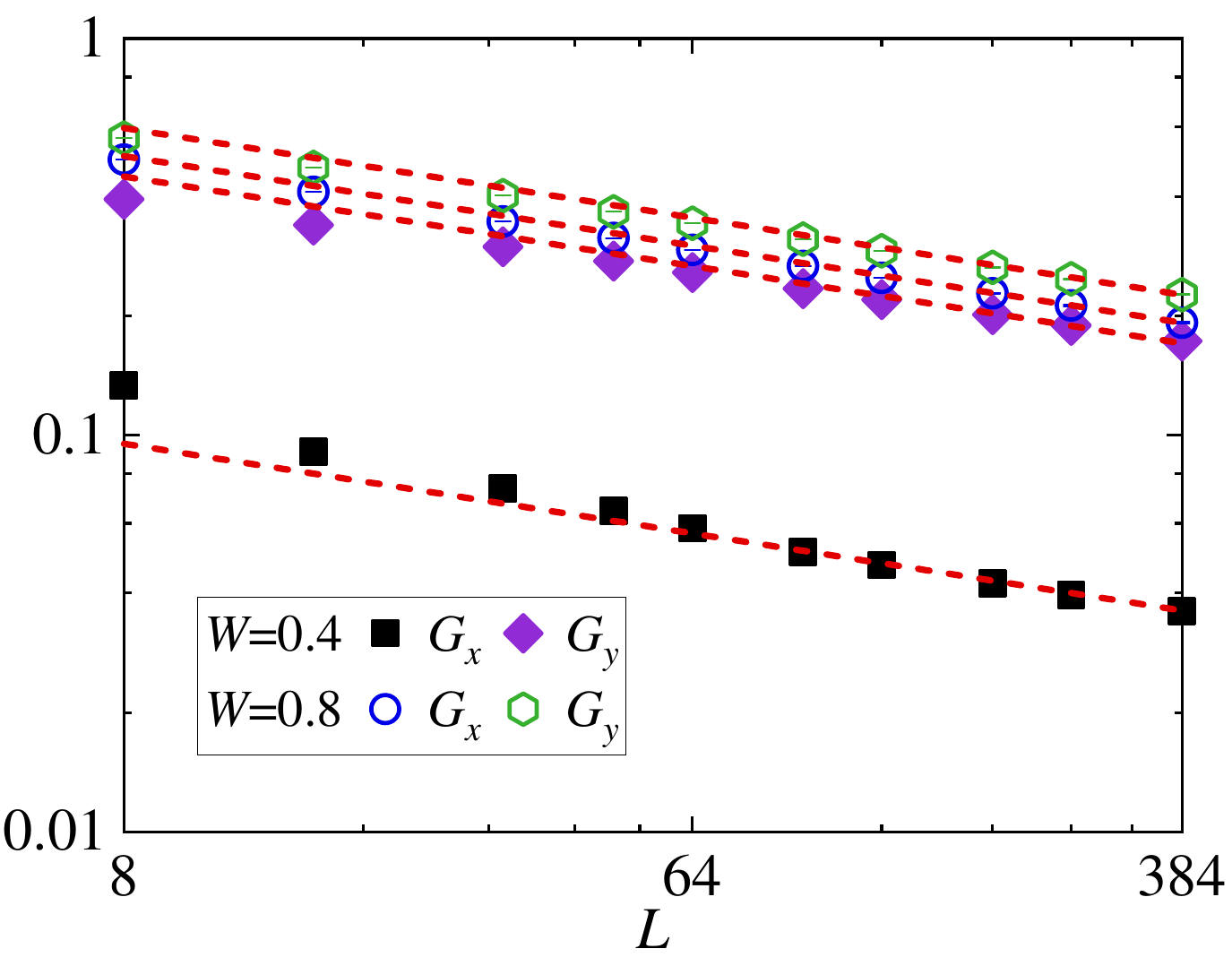}	
	\caption{Log-log plot of the large-distance spin-spin correlations $G_x$ and $G_y$ versus $L$ at $K_1$ for $W=0.4$ and $0.8$. The slope of dashed lines represents $-\eta=-1/4$.}\label{SM_GL}
\end{figure}

\begin{figure}[t]
	\centering
	\includegraphics[height=8cm,width=8cm]{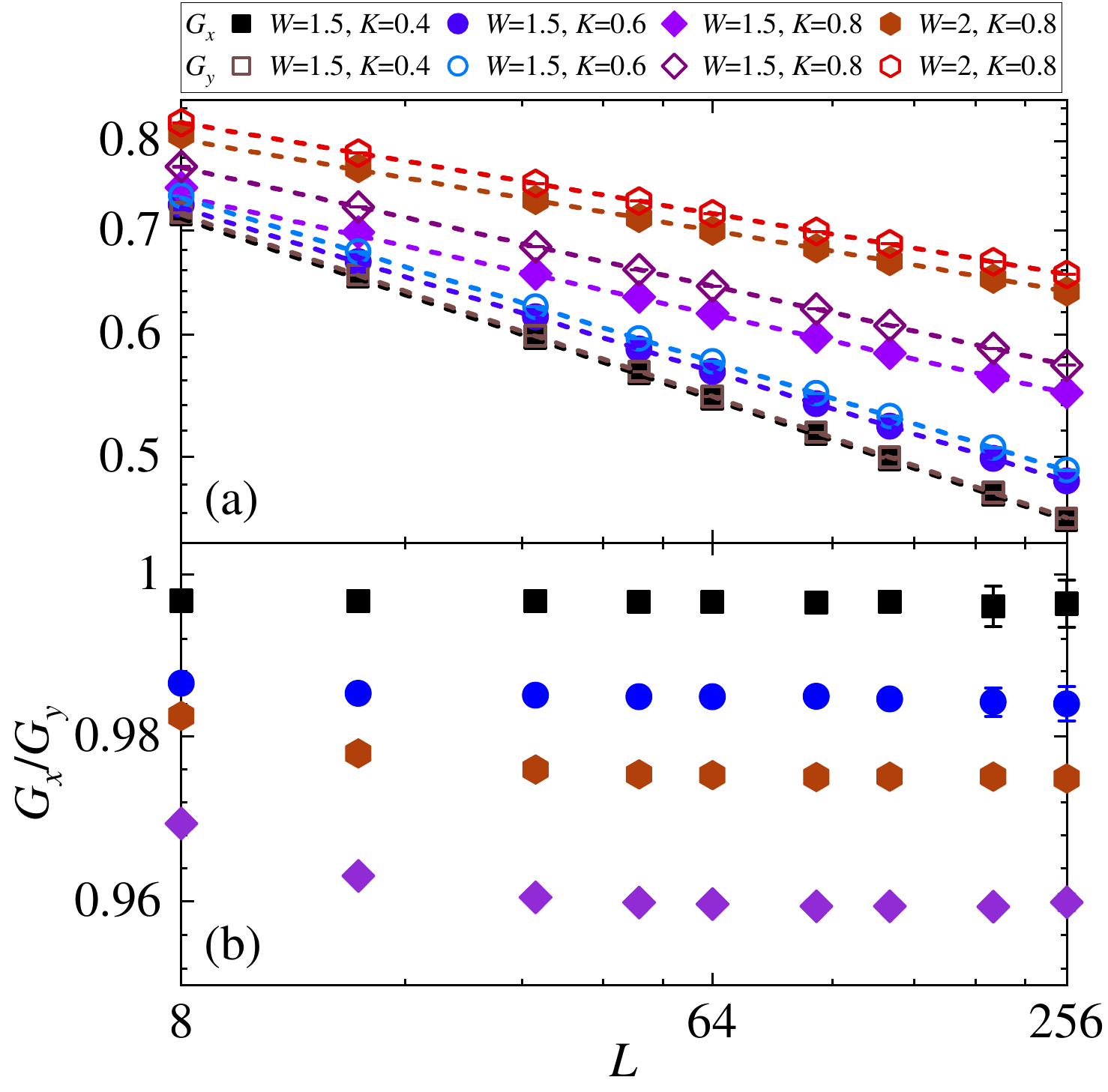}	
	\caption{(a) Log-log plot of the large-distance spin-spin correlations $G_x$ and $G_y$ versus $L$ for $(W=1.5, K=0.4)$, $(W=1.5, K=0.6)$, $(W=1.5, K=0.8)$ and $(W=2, K=0.8)$. The slopes of dashed lines represent $-\eta$. (b) Log-log plot of the ratio $G_x/G_y$ versus $L$.}\label{SM_QLR}
\end{figure}

\begin{figure}[t]
	\includegraphics[height=12.7cm,width=8cm]{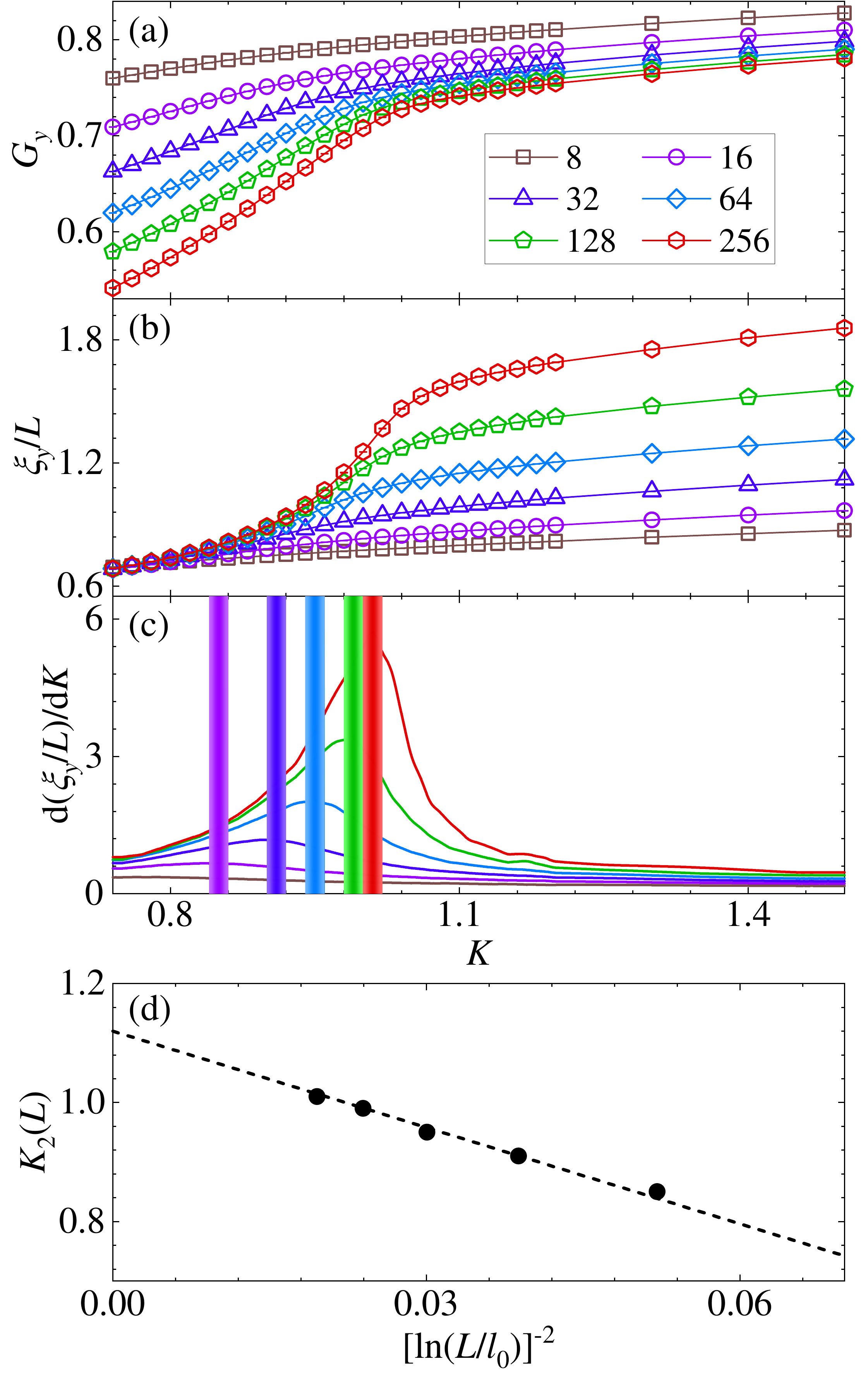}
	\caption{Quantities of a $y$ line in the V plane with $W=1.5$. (a) The large-distance spin-spin correlation $G_y$ versus $K$. (b) The reduced correlation length $\xi_y/L$ versus $K$. (c) The derivative ${\rm d}(\xi_y/L)/{\rm d}K$ versus $K$. The shadowed areas, whose widths represent two error bars, denote $K_2(L)$ (the $K$ coordinate of the peak). (d) $K_2(L)$ from ${\rm d}(\xi_y/L)/{\rm d}K$ versus $[{\rm ln}(L/l_0)]^{-2}$, where $l_0=0.2$ comes from a preferred fit. The intercept $K_2=1.11996$ represents the thermodynamic limit of $K_2(L)$.}~\label{SM_Fig1}
\end{figure}

\begin{figure}[t]
	\includegraphics[height=12.7cm,width=8cm]{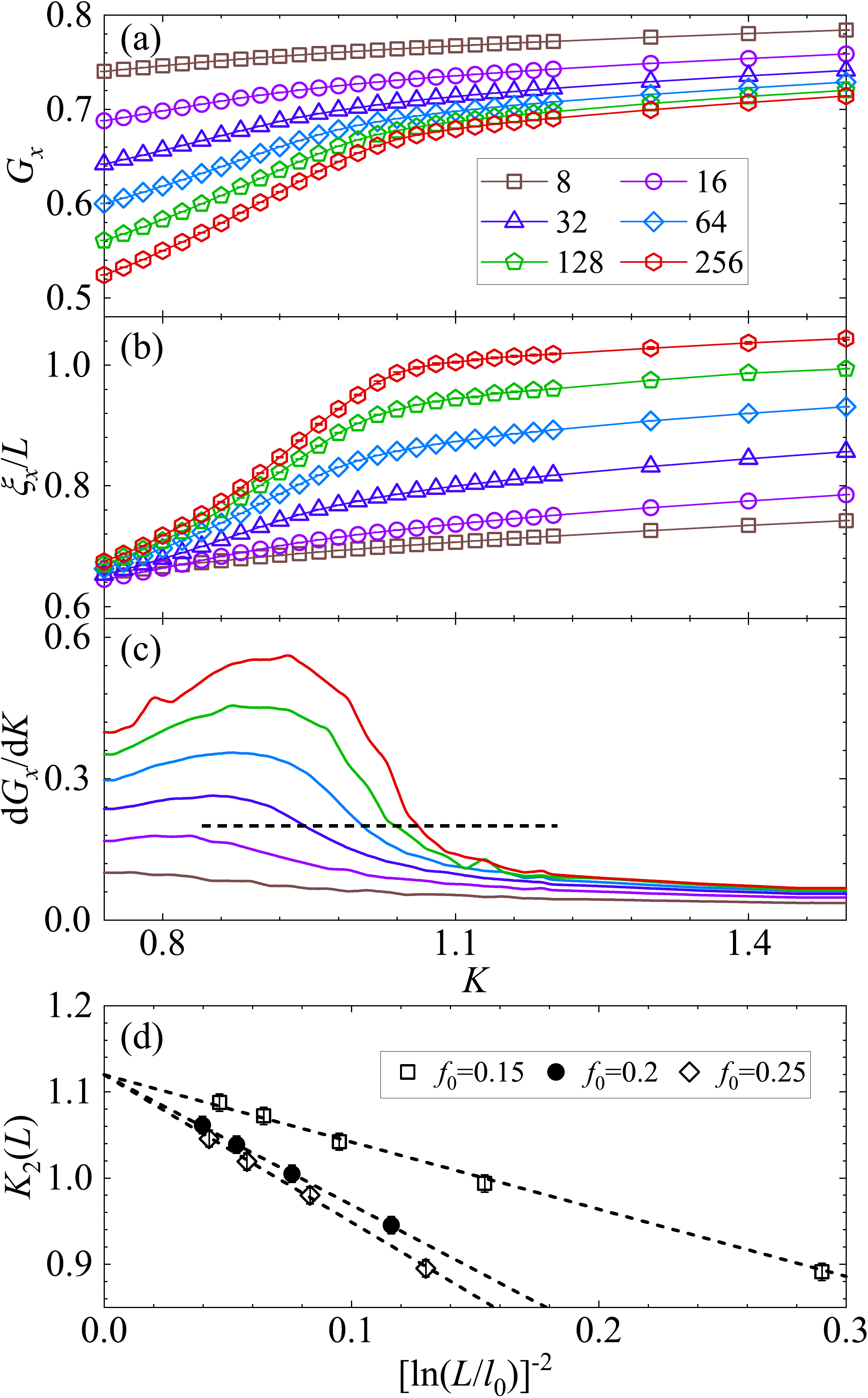}
	\caption{Quantities of an $x$ line in the V plane with $W=1.5$. (a)-(b) Same as Figs.~\ref{SM_Fig1}(a) and~\ref{SM_Fig1}(b) but for the $x$ line. (c) The derivative ${\rm d}G_x/{\rm d}K$ versus $K$. The dashed line represents $f_0=0.2$. (d) $K_2(L)$ from ${\rm d}G_x/{\rm d}K$ versus $[{\rm ln}(L/l_0)]^{-2}$, where $l_0=2.5$ ($f_0=0.15$), $1.7$ ($f_0=0.2$) and $2.0$ ($f_0=0.25$) come from preferred fits. The intercept $K_2=1.11996$ represents the thermodynamic limit of $K_2(L)$.}~\label{SM_Fig2}
\end{figure}

To suppress the uncertainties in the fits, we perform FSS analyses without off-critical terms. To this end, we perform simulations at $K_1=0.951$ and $0.772$ for $W=0.4$ and $0.8$, respectively. According to Eq.~(\ref{K1}), the FSS form at $K_1$ reads
\begin{equation}~\label{eq2}	
	G_{\alpha}=L^{-\eta}({\rm ln}L+C_1)^{1/8}(a_0+d_1L^{-1}+d_2L^{-2}).
\end{equation}
We fit $G_x$ and $G_y$ to Eq.~(\ref{eq2}). In the case $W=0.4$, from $G_x$, we obtain $\eta=0.30(4)$, $d_1 \approx 0$ and $\chi^2/{\rm DOF} \approx 1.1$, with $L_{\rm min}=64$. Setting $d_1=0$ gives $\eta=0.247(7)$ and $\chi^2/{\rm DOF} \approx 1.3$ with $L_{\rm min}=64$. Fixing $\eta=1/4$ results in $\chi^2/{\rm DOF} \approx 1.0$ with $L_{\rm min}=64$. From $G_y$, we obtain $\eta=0.259(8)$ and $\chi^2/{\rm DOF} \approx 0.4$ with $L_{\rm min}=96$. Fixing $\eta=1/4$ results in $\chi^2/{\rm DOF} \approx 0.9$ with $L_{\rm min}=128$. In the case $W=0.8$, from $G_x$, we obtain $\eta=0.251(5)$ and $\chi^2/{\rm DOF} \approx 1.0$ with $L_{\rm min}=32$. Fixing $\eta=1/4$ results in $\chi^2/{\rm DOF} \approx 1.0$ with $L_{\rm min}=48$. From $G_y$, we obtain $\eta=0.24(1)$ and $\chi^2/{\rm DOF} \approx 0.2$, with $L_{\rm min}=64$. Fixing $\eta=1/4$ results in $\chi^2/{\rm DOF} \approx 1.1$ with $L_{\rm min}=48$. All the values of $\eta$ obtained from the free fits are close to $\eta=1/4$. Moreover, the fits with fixed $\eta=1/4$ support the consistency with the BKT value. Details of fits are given in Table~\ref{Tab_2}. The FSS analyses of $G_x$ and $G_y$ are displayed in Fig.~\ref{SM_GL}.

Summarizing, on the $K_1$ transition line, the exponent $\eta$ takes a universal value of $1/4$, which characterizes the spin-spin correlations in both $x$- and $y$-directions, and corresponds to the BKT universality class.

\subsection{Isotropic critical phase}
In the region between $K_1$ and $K_2$ lines, we perform fits according to the form
\begin{equation}~\label{QLR}	
	G_{\alpha}=L^{-\eta}(a+bL^{-\omega})
\end{equation}
with $\alpha \in \{x,y\}$, where $\eta$ denotes the anomalous dimension, $\omega$ is a correction exponent, while $a$ and $b$ are non-universal. At $W=1.5$, from $G_x$, we obtain $\eta=0.1297(1)$, $0.11729(6)$ and $0.0845(2)$ as well as $\chi^2/{\rm DOF} \approx 0.8$, $0.6$ and $1.1$, with $L_{\rm min}=8$, $8$ and $32$, for $K=0.4$, $0.6$ and $0.8$, respectively. From $G_y$, we find $\eta=0.1302(4)$, $0.1171(1)$ and $0.08483(4)$ as well as $\chi^2/{\rm DOF} \approx 0.9$, $1.0$ and $1.7$, with $L_{\rm min}=96$, $8$ and $48$, for $K=0.4$, $0.6$ and $0.8$, respectively. Additionally, for the parameter set $(W=2, K=0.8)$, preferred fits yield $\eta=0.06511(5)$ and $0.06501(4)$ as well as $\chi^2/{\rm DOF} \approx 1.3$ and $0.7$, with $L_{\rm min}=16$ and $48$, from $G_x$ and $G_y$, respectively. Details of fits are given in Tables~\ref{Tab_QLR1},~\ref{Tab_QLR2} and~\ref{Tab_QLR3}. The dependence of $G_x$ and $G_y$ on $L$ are shown in Fig.~\ref{SM_QLR}(a). These results indicate that the values of $\eta$ remain consistent between the $x$- and $y$-directions and exhibit a decrease with increasing $W$ or $K$. Figure~\ref{SM_QLR}(b) illustrates that the ratio $G_x/G_y$ quickly saturates as $L$ increases.

\begin{figure}[t]
	\centering
	\includegraphics[height=6.5cm,width=8.5cm]{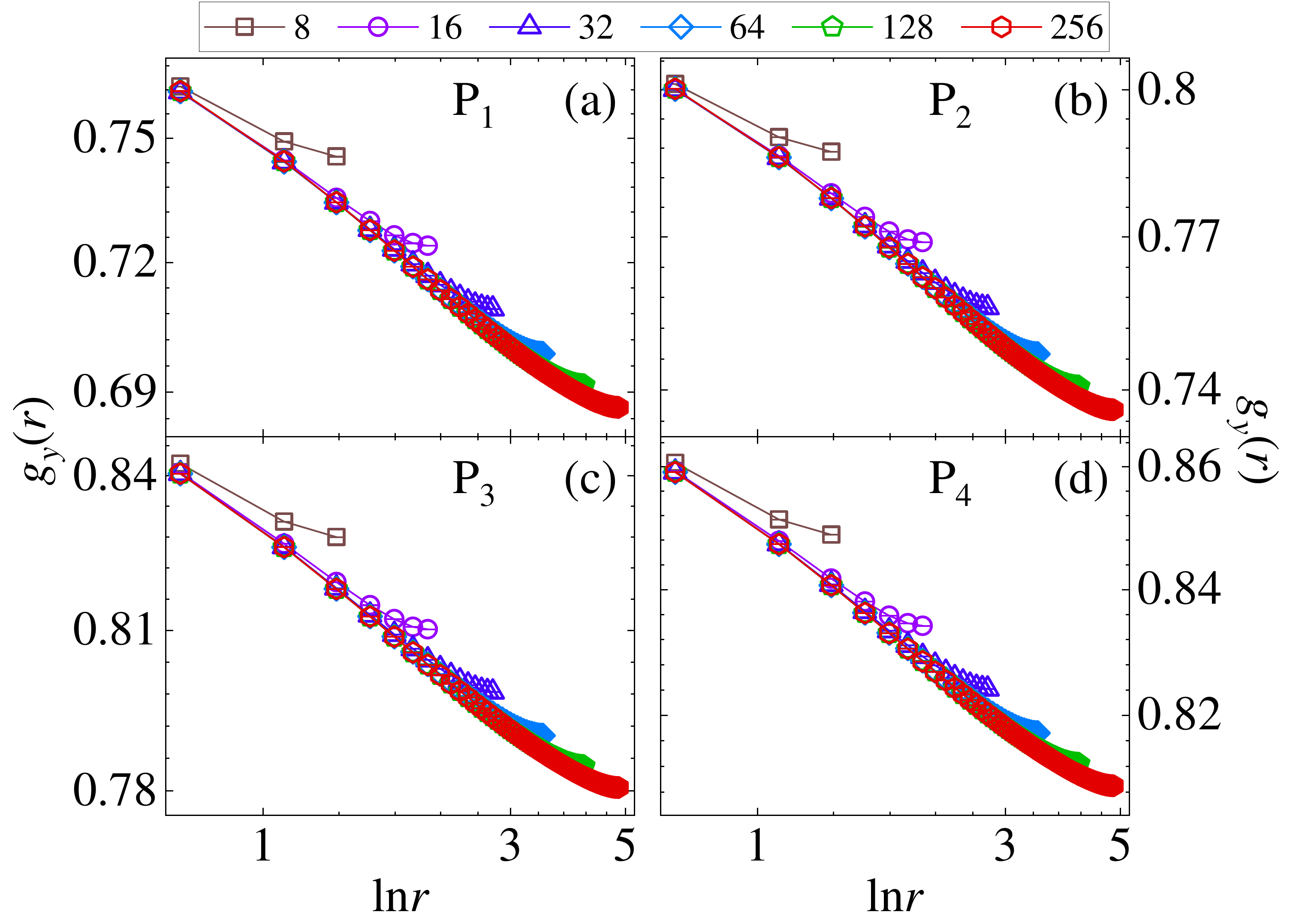}
	\caption{Log-log plots of the spin-spin correlation function $g_y(r)$ versus ${\rm ln}r$ for ${\rm P_1}$, ${\rm P_2}$, ${\rm P_3}$ and ${\rm P_4}$.}\label{SM_gy}
\end{figure}

\begin{figure}[t]
	\includegraphics[height=10.0cm,width=8cm]{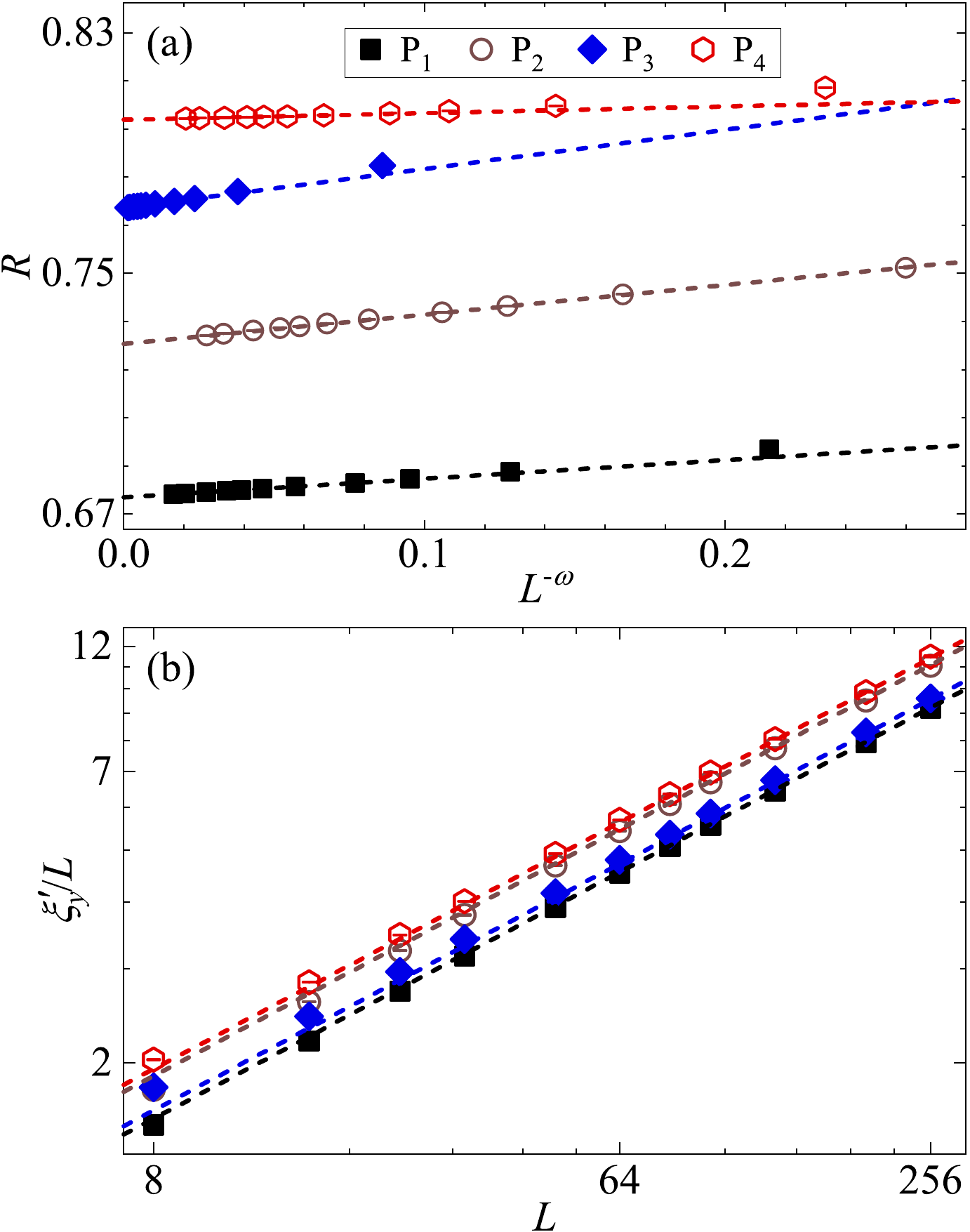}
	\caption{Additional quantities for the LR ordered phase of a $y$ line in the V plane. (a) The residual magnetic fluctuations $R$ versus $L^{-\omega}$. For ${\rm P_1}$, ${\rm P_2}$, ${\rm P_3}$ and ${\rm P_4}$, the values $\omega=0.74$, $0.648$, $1.18$ and $0.7$ are from preferred fits, respectively. (b) Log-log plot of the reduced correlation length $\xi'_y/L$ versus $L$. The slope $0.51$ of dashed lines stands for $q'$.}~\label{SM_Fig3}
\end{figure}

\subsection{The phase transition point for $W=1.5$}
From Figs.~\ref{SM_Fig1}(a) and~\ref{SM_Fig1}(b), we infer a singularity at $K_2$ from $G_y$ and $\xi_y/L$, respectively. This result indicates a phase transition at $K_2$ for a $y$ line. Figure~\ref{SM_Fig1}(c) shows that, for each $L$, the derivative ${\rm d}(\xi_y/L)/{\rm d}K$ is peaked at $K_2(L)$. We perform fits of $K_2(L)$ according to Eq.~(\ref{Equ_Kn}), which yield $K_2=1.2(2)$ and $\chi^2/{\rm DOF} \approx 0.3$ with $L_{\rm min}=32$. The estimate of $K_2$ is consistent with $K_2 \approx 1.11996$. By fixing $K_2=1.11996$, we obtain $l_0=0.2(1)$ and $\chi^2/{\rm DOF} \approx 0.2$ with $L_{\rm min}=32$. Details of fits are presented in Table~\ref{Tab_Kn15}. The dependence of $K_2(L)$ on $[{\rm ln}(L/l_0)]^{-2}$ is illustrated in Fig.~\ref{SM_Fig1}(d).

For an $x$ line, we observe that $G_x$ exhibits a kink at $K_2$ [Fig.~\ref{SM_Fig2}(a)]. For $K \ge K_2$, Fig.~\ref{SM_Fig2}(b) implies that $\xi_x/L$ converges roughly as $L \rightarrow \infty$, indicating the possibility of a critical phase. The sharp reduction in ${\rm d}G_x/{\rm d}K$ at $K_2$ provides complementary evidence of a phase transition [Fig.~\ref{SM_Fig2}(c)]. We fit $K_2(L)$ to Eq.~(\ref{Equ_Kn}), where $K_2(L)$ is defined with ${\rm d}G_x/{\rm d}K=f_0$. For $f_0=0.2$, we obtain $K_2=1.13(7)$ with $L_{\rm min}=32$. The stability of result is checked by varying $f_0$. Details of fits are given in Table~\ref{B_Tab_8}. The estimates of $K_2$ again agree well with $K_2 \approx 1.11996$. The dependence of $K_2(L)$ on $[{\rm ln}(L/l_0)]^{-2}$ is shown in Fig.~\ref{SM_Fig2}(d).
\bigskip
\section{The phases at $K \ge K_2$}

\subsection{$y$ lines}

As discussed in the main text, the $y$ lines may enter a LR ordered phase with $K \ge K_2$, which can be related to the Goldstone mode. It follows that the large-distance spin-spin correlation $G_y$, the magnetic fluctuations $\langle M^2_y \rangle$ at zero mode, the magnetic fluctuations $\langle M^2_{yk} \rangle$ at smallest non-zero mode and the correlation length $\xi_y$ obey the FSS formulae
\begin{equation}~\label{Equ_Gy}
	G_y=a+ b L^{-q},
\end{equation}
\begin{equation}~\label{Equ_f0}
	\langle M^2_y \rangle=a+b L^{-q},
\end{equation}
\begin{equation}~\label{Equ_f1}
	\langle M^2_{yk} \rangle=L^{-q}(a+b L^{-\omega})
\end{equation}
and
\begin{equation}~\label{Equ_xi}
	(\xi_y/L)^2=L^{q}(a+b L^{-\omega})+c
\end{equation}
respectively, where the exponent $q$ controls the leading FSS behavior, $\omega$ is a correction exponent, while $a$, $b$ and $c$ are non-universal constants. We fit the Monte Carlo data of $G_y$, $\langle M^2_y \rangle$, $\langle M^2_{yk} \rangle$ and $\xi_y$ to Eqs.~(\ref{Equ_Gy}),~(\ref{Equ_f0}),~(\ref{Equ_f1}) and (\ref{Equ_xi}), respectively. The results are given in Tables~\ref{Tab_Gy},~\ref{Tab_f0},~\ref{Tab_f1} and~\ref{Tab_xi}. The results of FSS analyses indicate a LR ordered phase. Figure~\ref{SM_gy} demonstrates the spin-spin correlation function $g_y(r)=\langle \vec{S}_{\bf r} \cdot \vec{S}_{{\bf r}+(0,r)} \rangle$, where ${\bf r}$ denotes a site in the V plane and $r$ is the spatial distance, with ${\rm P_1}$, ${\rm P_2}$, ${\rm P_3}$ and ${\rm P_4}$. For each parameter set, except in the small-$r$ regime, the slope of $g_y(r)$ versus ${\rm ln}r$ in the log-log plot does not increase in magnitude as ${\rm ln}r$ increases. Consequently, a decay faster than the logarithmic form $g_y(r) \sim ({\rm ln}r)^{-\hat{q}'}$ (where $\hat{q}'$ is an unspecified exponent) is not observed, because a purely logarithmic decay would be represented by a straight line in such a plot.

Having confirmed scaling formulae (\ref{Equ_f0}) and (\ref{Equ_f1}), we analyze the residual magnetic fluctuations $R=\langle M^2_y \rangle-f\langle M^2_{yk} \rangle$, where the parameter $f$ is set, according to Eqs.~(\ref{Equ_f0}) and (\ref{Equ_f1}), to cancel the term with $L^{-q}$. Therefore, $R$ features reduced finite-size corrections and obeys
\begin{equation}~\label{Equ_fr}
	R=a+b L^{-\omega}.
\end{equation}
Using Eq.~(\ref{Equ_fr}), we obtain $a=0.67556(9)$, $0.72676(5)$, $0.77175(3)$ and $0.8011(1)$ as well as $\chi^2/{\rm DOF} \approx 0.7$, $1.5$, $0.7$ and $0.8$, with $L_{\rm min}=48$, $32$, $32$ and $64$, for ${\rm P_1}$, ${\rm P_2}$, ${\rm P_3}$ and ${\rm P_4}$, respectively. Details of fits are given in Table~\ref{Tab_fr}. Figure~\ref{SM_Fig3}(a) displays $R$ versus $L^{-\omega}$. Indeed, if compared to $\langle M^2_y \rangle$ (main text), $R$ converges faster as $L$ increases.

Finally, we provide additional evidence of LR ordered phase by using an alternative definition of correlation length, namely $\xi'_y=(2{\rm sin}\frac{\pi}{L})^{-1}\sqrt{\frac{\Gamma_{\rm V}(0,0)}{\Gamma_{\rm V}(0,\frac{2\pi}{L})}-1}$. The power-law dependence of $\xi'_y/L$ on $L$ is indicated by Fig.~\ref{SM_Fig3}(b). By fitting $\xi'_y/L$ to
\begin{equation}~\label{Equ_xiV}
	\xi'_y/L=aL^{q'}+b,
\end{equation}
we obtain $q'=0.516(6)$, $0.50(1)$, $0.514(2)$ and $0.504(2)$ as well as $\chi^2/{\rm DOF} \approx 1.1$, $0.9$, $1.7$ and $2.5$, with $L_{\rm min}=64$, $80$, $24$ and $24$, for ${\rm P_1}$, ${\rm P_2}$, ${\rm P_3}$ and ${\rm P_4}$, respectively. It is found that, for ${\rm P_1}$, ${\rm P_2}$, ${\rm P_3}$ and ${\rm P_4}$, the estimates of $q'$ are close to each other. The fits are detailed in Table~\ref{Tab_xiV}.

\begin{figure}[t]
	\centering
	\includegraphics[height=10cm,width=8.5cm]{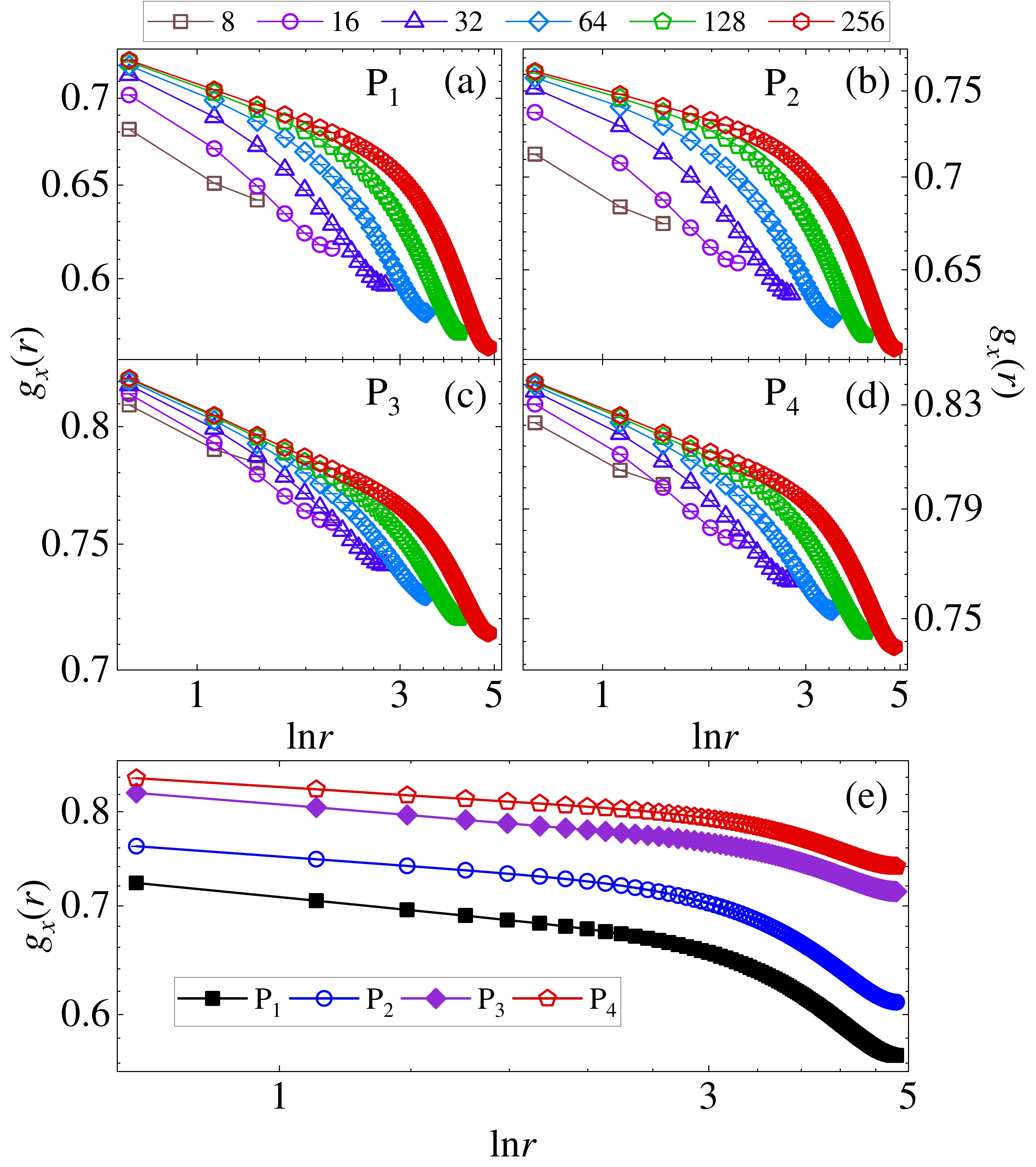}	
	\caption{(a)-(d) Log-log plots of the spin-spin correlation function $g_x(r)$ versus ${\rm ln}r$ for ${\rm P_1}$, ${\rm P_2}$, ${\rm P_3}$ and ${\rm P_4}$. (e) Log-log plot of the spin-spin correlation function $g_x(r)$ versus ${\rm ln}r$ with ${\rm P_1}$, ${\rm P_2}$, ${\rm P_3}$ and ${\rm P_4}$, for $L=256$.}\label{SM_gx}
\end{figure}

\begin{figure}[t]
	\centering
	\includegraphics[height=4.5cm,width=8.5cm]{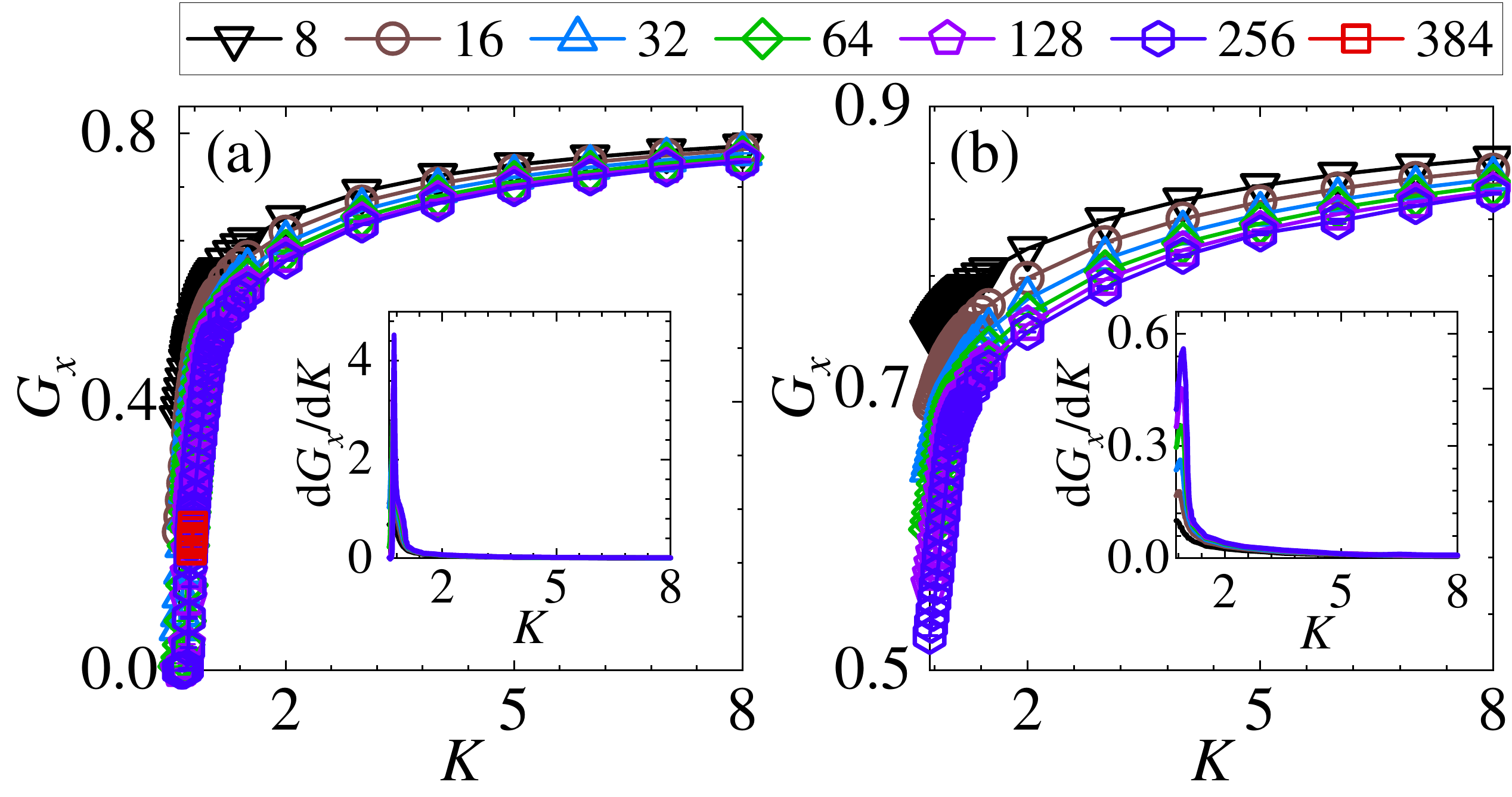}	
	\caption{The large-distance spin-spin correlation $G_x$ versus $K$ for $W=0.8$ (a) and $1.5$ (b). Inset: ${\rm d}G_x/{\rm d}K$ versus $K$.}\label{SM_dG}
\end{figure}

\begin{figure}[t]
	\centering
	\includegraphics[height=6cm,width=8cm]{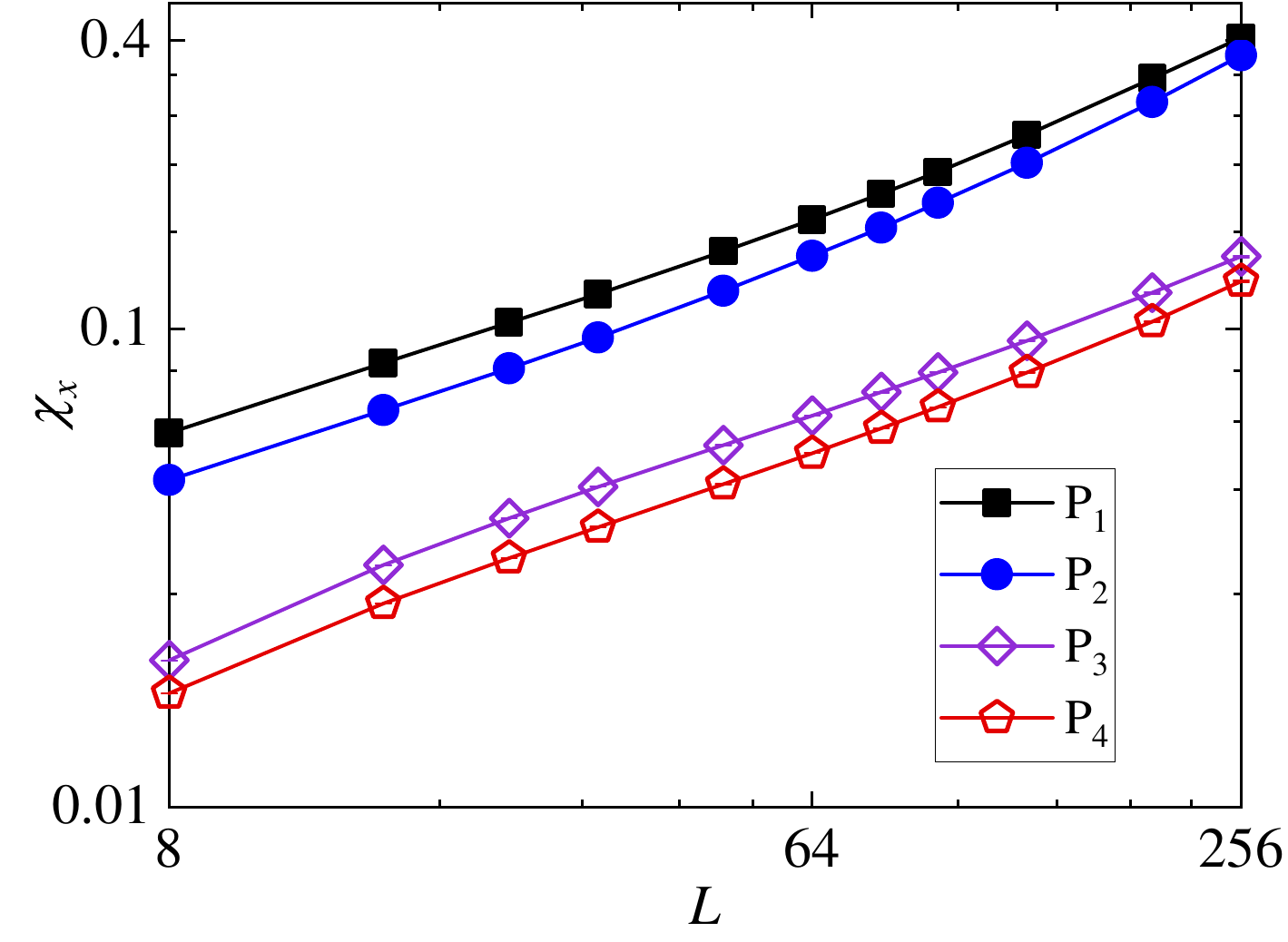}	
	\caption{Log-log plot of the susceptibility $\chi_x$ versus $L$ for ${\rm P_1}$, ${\rm P_2}$, ${\rm P_3}$ and ${\rm P_4}$.}\label{SM_Xx}
\end{figure}

\subsection{$x$ lines}
Assuming the $x$ lines are critical, we fit $\xi_x/L$ to the scaling formula
\begin{equation}~\label{Equ_xix1}
	\xi_x/L=a+bL^{-\omega},
\end{equation}
where $a$ represents the non-zero value of $\xi_x/L$ in the thermodynamic limit and $bL^{-\omega}$ is a power-law correction term. We obtain $a=0.770(8)$, $0.788(9)$, $1.24(2)$ and $1.19(1)$ as well as $\chi^2/{\rm DOF} \approx 0.3$, $1.0$, $1.2$ and $1.0$, with $L_{\rm min}=80$, $80$, $48$ and $48$, for ${\rm P_1}$, ${\rm P_2}$, ${\rm P_3}$ and ${\rm P_4}$, respectively. The results are summarized in Table~\ref{Tab_xix1}.  We also fit $\xi_x/L$ to the formula
\begin{equation}~\label{Equ_xix2}
	\xi_x/L=a+b({\rm ln}L)^{-\omega}
\end{equation}
with $b({\rm ln}L)^{-\omega}$ a logarithmic correction term. We obtain $a=0.80(2)$, $0.82(1)$, $1.5(1)$ and $1.43(5)$ as well as $\chi^2/{\rm DOF} \approx 0.4$, $0.8$, $0.3$ and $1.1$, with $L_{\rm min}=80$, $64$, $64$ and $48$, for ${\rm P_1}$, ${\rm P_2}$, ${\rm P_3}$ and ${\rm P_4}$, respectively. The results are summarized in Table~\ref{Tab_xix2}. Hence, it is practically difficult to differentiate scaling formulae (\ref{Equ_xix1}) and (\ref{Equ_xix2}). However, both formulae imply that $\xi_x/L$ is finite in the thermodynamic limit.

The large-distance spin-spin correlation $G_x$ is fitted according to the logarithmic FSS formula
\begin{equation}~\label{Equ_Gx1}
	G_x=a[{\rm ln}(L/l_0)]^{-\hat{q}}
\end{equation}
with $\hat{q}$ a critical exponent. The results are presented in Table~\ref{Tab_Gx1}. We also perform fits according to the power-law FSS formula 
\begin{equation}~\label{Equ_Gx2}
	G_x=aL^{-\tilde{q}}
\end{equation}
with $\tilde{q}$ a critical exponent, but find that $\chi^2/{\rm DOF}$ is generally huge. The fits are detailed in Table~\ref{Tab_Gx2}. Figures~\ref{SM_gx}(a)-\ref{SM_gx}(d) show that the decay of the spin-spin correlation function $g_x(r)=\langle \vec{S}_{\bf r} \cdot \vec{S}_{{\bf r}+(r,0)} \rangle$ is faster at larger ${\rm ln}r$. This behavior stands in sharp contrast to that of $g_y(r)$, shown in Fig.~\ref{SM_gy}. In Fig.~\ref{SM_gx}(e), as $W$ or $K$ increases, local correlation grows, while the decay of $g_x(r)$ at large ${\rm ln}r$ becomes slow. These results indicate LR order along $x$-direction in the limit $W \to \infty$ and $K \to \infty$.

Figure~\ref{SM_dG} demonstrates that, for $K > K_2$, $G_x$ increases more and more slowly with increasing $K$ and the derivative ${\rm d}G_x/{\rm d}K$ does not exhibit any singularity. Hence, phase transition is absent along the $x$ lines for finite $K>K_2$.

We compute the susceptibility $\chi_{x}=L(\langle M_{x}^2 \rangle - \langle M_{x} \rangle^2)$. For ${\rm P_1}$, ${\rm P_2}$, ${\rm P_3}$ and ${\rm P_4}$, Fig.~\ref{SM_Xx} demonstrates that $\chi_x$ diverges with increasing $L$. This behavior is reminiscent of the divergent compressibility found in iTQF.

\begin{figure}[t]
	\includegraphics[height=10.2cm,width=8cm]{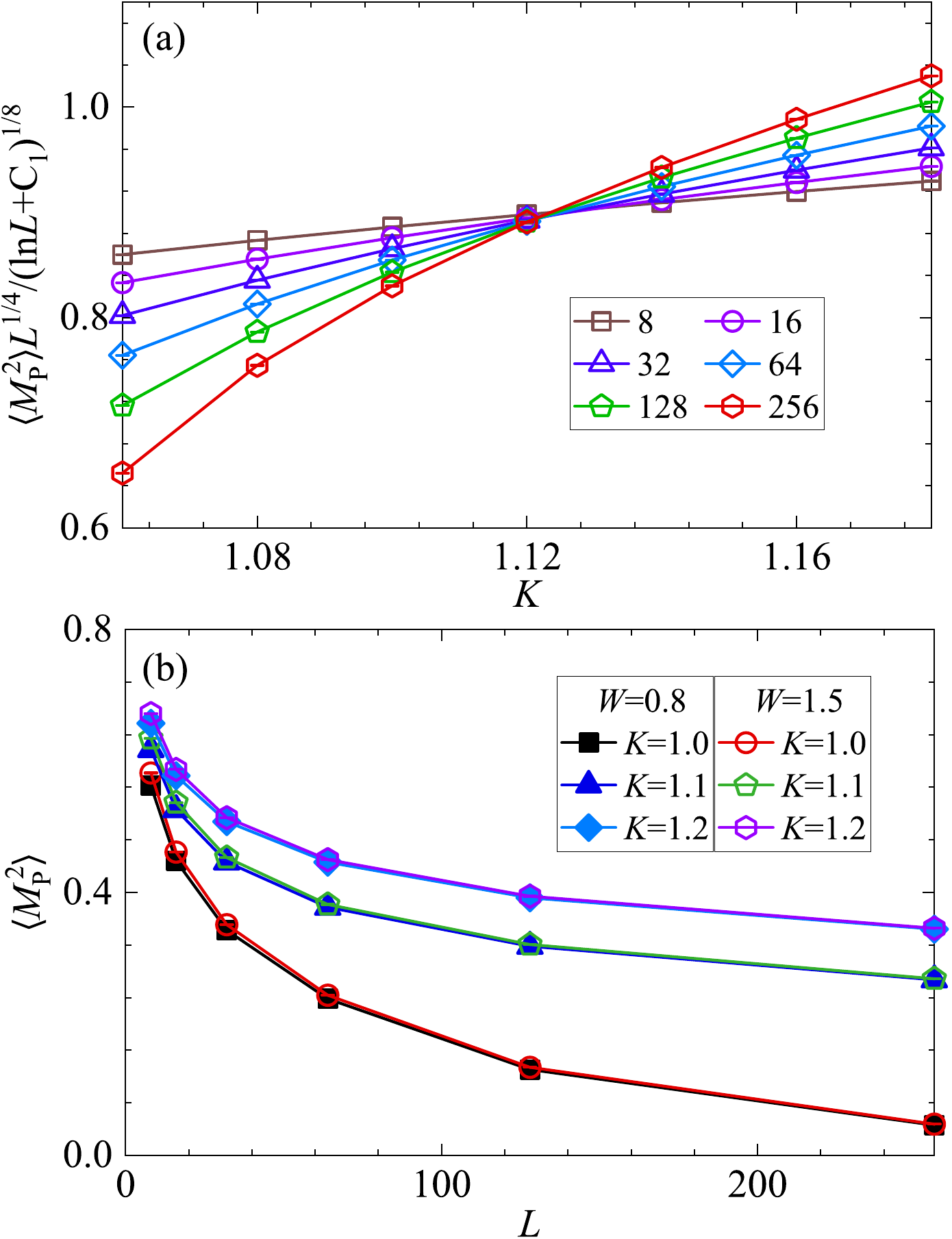}
	\caption{(a) Scaled magnetic fluctuations $\langle M^2_{\rm P} \rangle L^{1/4}/({\rm ln}L+C_1)^{1/8}$ versus $K$ for $W=1.5$. The parameter $C_1=2.32$ comes from a preferred fit. (b) $\langle M^2_{\rm P} \rangle$ versus $L$ for different $W$ and $K$.}~\label{SM_Fig4}
\end{figure}

\section{The P planes}
It is inferred that the P planes experience a BKT transition at $K=J_{\rm BKT}$ independent of $W$, which separates a disordered phase from the QLR ordered phase. To verify this point, for a P plane, we sample the magnetic fluctuations $\langle M^2_{\rm P} \rangle$ with $M_{\rm P}=L^{-2} |\sum_{\bf_{r \in {\rm P}}} \vec{S}_{\bf r}|$, where the summation runs over sites in the P plane. For $W=1.5$, we fit $\langle M^2_{\rm P} \rangle$ to the scaling form
\begin{equation}~\label{XP_K1}
	\begin{split}
		\langle M^2_{\rm P} \rangle&=L^{-1/4}({\rm ln}L+C_1)^{1/8}[a_0+\sum_{k=1}^3a_k \epsilon ^k ({\rm ln}L+C_2)^{2k} \\
		&+ d_1L^{-1} + d_2L^{-2} + n_0\epsilon + n_1\epsilon^2({\rm ln}L+C_2)^2]
	\end{split}
\end{equation}	
with $\epsilon = 1.11996 - K$. An initial fit yields $n_1 \approx 0$ and $\chi^2/{\rm DOF} \approx 0.2$ with $L_{\rm min}=32$. Subsequently fixing $n_1 = 0$ results in $\chi^2/{\rm DOF} \approx 0.5$ and $L_{\rm min}=32$. Details of fits are given in Table~\ref{Tab_Xp}. We confirm the existence of the BKT transition [Fig.~\ref{SM_Fig4}(a)]. Furthermore, Fig.~\ref{SM_Fig4}(b) illustrates that the Monte Carlo data of $\langle M^2_{\rm P} \rangle$ at $W=0.8$ and $1.5$ become more indistinguishable as $L$ increases. This observation is consistent with the expectation that all macroscopic quantities of a P plane are independent of $W$ in the $L \rightarrow \infty$ limit. 

\begin{figure}[t]
	\centering
	\includegraphics[height=6.5cm,width=8.5cm]{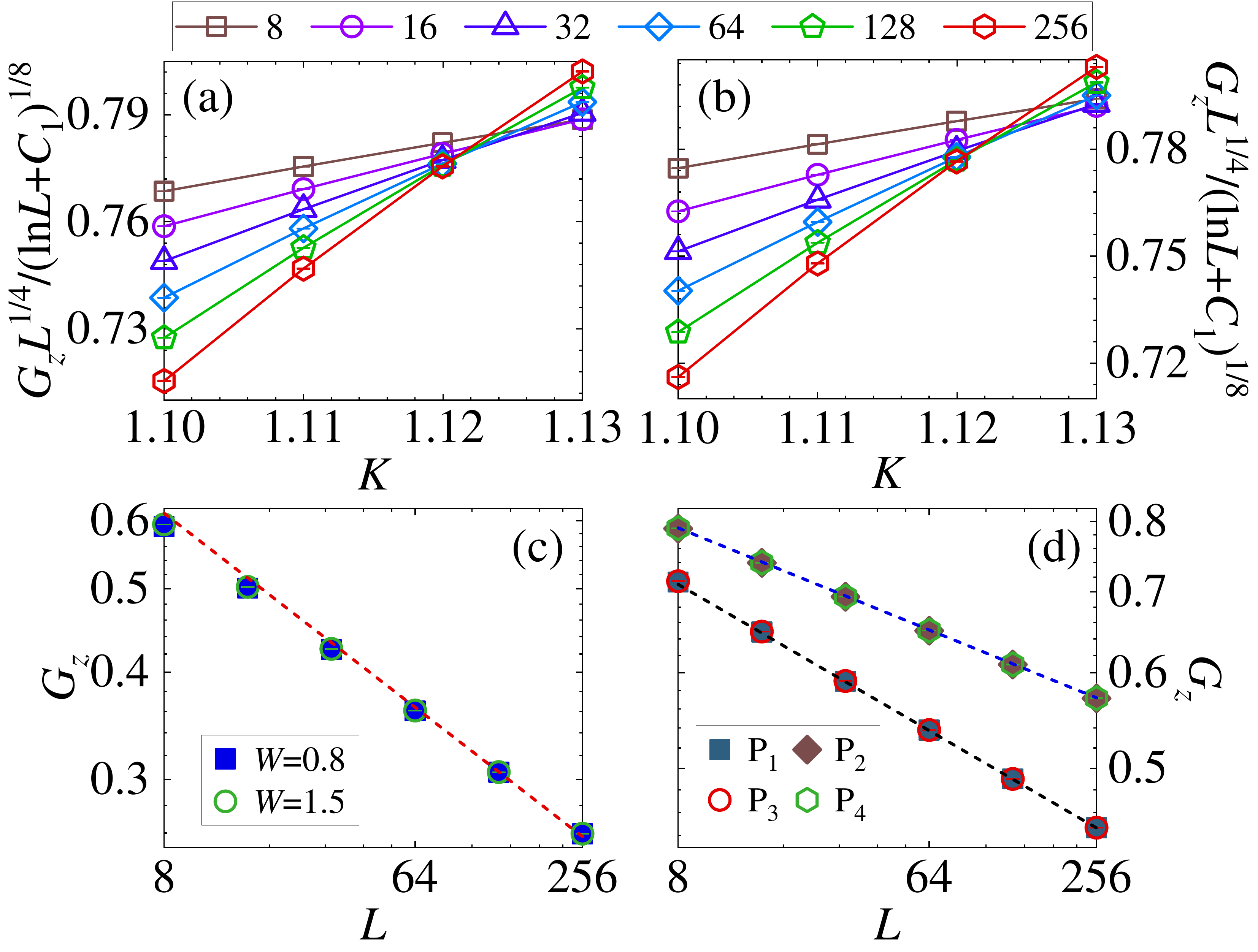}	
	\caption{(a)-(b) Scaled large-distance spin-spin correlation $G_zL^{1/4}/({\rm ln}L+C_1)^{1/8}$ versus $K$ for $W=0.8$ and $1.5$. The parameter $C_1$ comes from preferred fits (for $W=0.8$, $C_1=4.72$; for $W=1.5$, $C_1=4.65$). (c) Log-log plot of the large-distance spin-spin correlation $G_z$ versus $L$ at $K_2$ for $W=0.8$ and $1.5$. The slope of dashed line represents $-\eta=-1/4$. (d) Log-log plot of the large-distance spin-spin correlation $G_z$ versus $L$ for ${\rm P_1}$, ${\rm P_2}$, ${\rm P_3}$ and ${\rm P_4}$. The slopes of black and blue lines represent $-\eta=-0.1339$ and $-0.0934$, respectively.}\label{SM_Gz}
\end{figure}

Along the $z$-direction in a P plane, we sample the large-distance spin-spin correlation $G_z = \langle \vec{S}_{{\bf r}+(0,0,L/4)} \cdot \vec{S}_{{\bf r}-(0,0,L/4)} \rangle$, where {\bf r} is a site in the V plane, with $W=0.8$ and $1.5$. We fit $G_z$ to Eq.~(\ref{K1}) with $\eta = 1/4$ and $\epsilon = 1.11996 - K$. For $W=0.8$, an initial fit yields $n_1 \approx 0$ and $\chi^2/{\rm DOF} \approx 1.4$ with $L_{\rm min}=8$. Subsequently fixing $n_1 = 0$ results in $\chi^2/{\rm DOF} \approx 1.3$ and $L_{\rm min}=8$. For $W=1.5$, a preferred fit is obtained with $\chi^2/{\rm DOF} \approx 1.5$ and $L_{\rm min}=8$. Details of fits are given in Table~\ref{Tab_3}. Figures~\ref{SM_Gz}(a) and~\ref{SM_Gz}(b) show $G_zL^{1/4}$ versus $K$ with multiplicative finite-size corrections, demonstrating a scale-invariant point at $K_2$, for $W=0.8$ and $1.5$, respectively. These results confirm that $G_z$ is well described by Eq.~(\ref{K1}), supporting the presence of a BKT transition in a P plane. At $K=K_2$, we perform fits to Eq.~(\ref{eq2}). For $W=0.8$, the free fit gives $\eta=0.250(2)$ and $\chi^2/{\rm DOF} \approx 0.1$ with $L_{\rm min}=8$. Fixing $\eta=1/4$ also yields a stable result with $\chi^2/{\rm DOF} \approx 0.1$ and $L_{\rm min}=16$. For $W=1.5$, the free fit gives $\eta=0.249(2)$ and $\chi^2/{\rm DOF} \approx 0.1$ with $L_{\rm min}=8$, while fixing $\eta=1/4$ gives $\chi^2/{\rm DOF} \approx 0.1$ with $L_{\rm min}=16$. Details of fits are given in Table~\ref{Tab_4}. The power-law decay of $G_z$ upon increasing $L$ is shown in Fig.~\ref{SM_Gz}(c). The estimates of $\eta$ are consistent with the BKT value. For $K>K_2$, as shown in Fig.~\ref{SM_Gz}(d), $G_z$ also exhibits a power-law decay with $L$. We fit $G_z$ to Eq.~(\ref{QLR}). Preferred fits yield $\eta=0.13396(5)$, $0.09293(3)$, $0.13401(5)$ and $0.09292(3)$ as well as $\chi^2/{\rm DOF} \approx 1.6$, $0.8$, $0.7$ and $0.7$ with $L_{\rm min}=8$, for ${\rm P_1}$, ${\rm P_2}$, ${\rm P_3}$ and ${\rm P_4}$, respectively. Details of fits are presented in Table~\ref{Tab_5}. 

\begin{figure}[t]
	\centering
	\includegraphics[height=3.8cm,width=8.5cm]{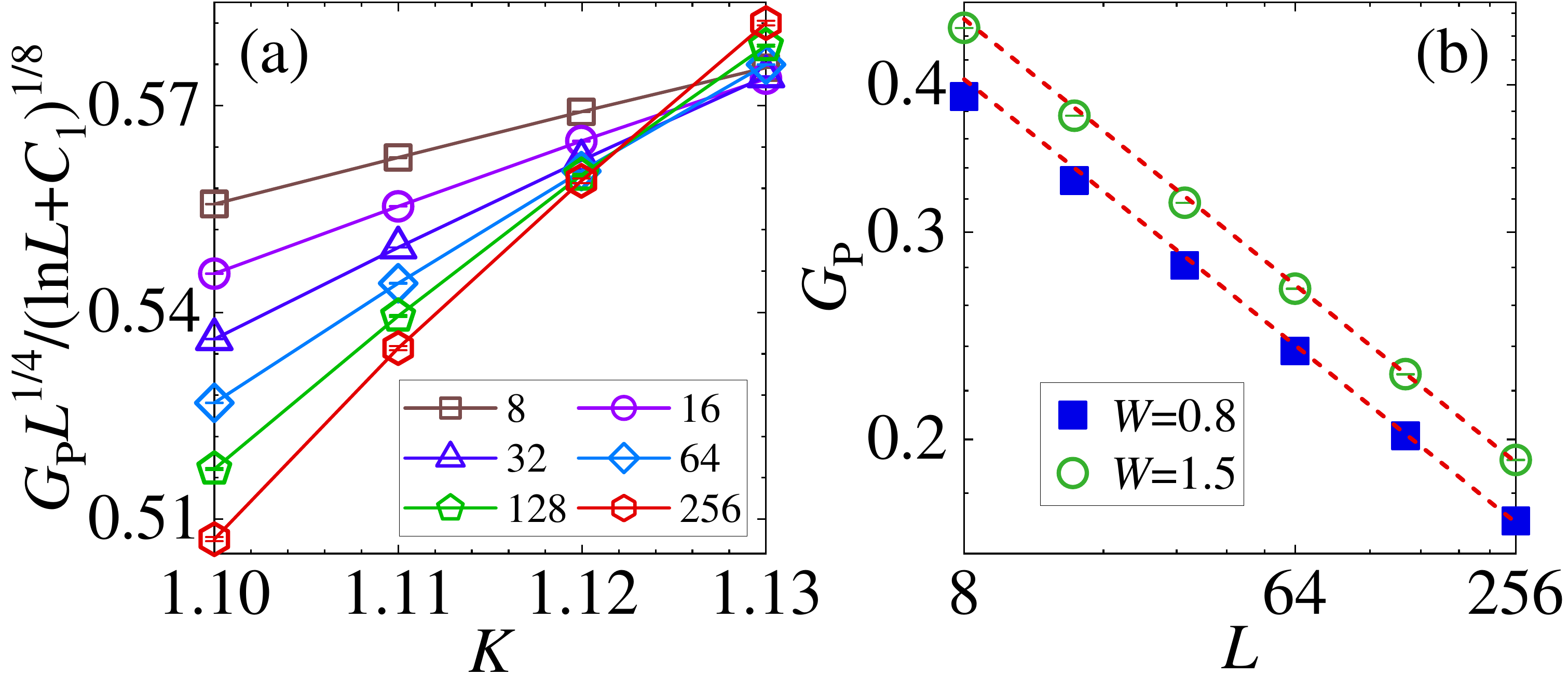}	
	\caption{(a) Scaled large-distance spin-spin correlation $G_{\rm P}L^{1/4}/({\rm ln}L+C_1)^{1/8}$ versus $K$ for $W=1.5$. The parameter $C_1=7.2$ comes from a preferred fit. (b) Log-log plot of the large-distance spin-spin correlation $G_{\rm P}$ versus $L$ at $K_2$ for $W=0.8$ and $1.5$. The slope of dashed lines represents $-\eta=-1/4$.}\label{SM_GP}
\end{figure}

Near $K_2$, we fit $G_{\rm P}$ to Eq.~(\ref{K1}) with $\eta = 1/4$ and $\epsilon = 1.11996 - K$. For $W=0.8$, we obtain $n_1 \approx 0$ and $\chi^2/{\rm DOF} \approx 1.9$ with $L_{\rm min}=8$. Fixing $n_1 = 0$ then gives $\chi^2/{\rm DOF} \approx 1.8$ with $L_{\rm min}=8$. For $W=1.5$, the initial fit gives $a_3 \approx 0$ and $\chi^2/{\rm DOF} \approx 1.1$ with $L_{\rm min}=8$. Setting $a_3 = 0$ subsequently results in $\chi^2/{\rm DOF} \approx 1.1$ with $L_{\rm min}=8$. Details of fits are given in Table~\ref{Tab_6}. Figure~\ref{SM_GP}(a) shows $G_{\rm P}L^{1/4}$ versus $K$ with multiplicative finite-size corrections ($W=1.5$). The scaling behavior for $W=0.8$ has been presented in End Matter. At $K=K_2$, we fit $G_{\rm P}$ to Eq.~(\ref{eq2}). For $W=0.8$, the free fit gives $\eta=0.2676(8)$ and $\chi^2/{\rm DOF} \approx 0.2$ with $L_{\rm min}=16$, and fixing $\eta=1/4$ gives $\chi^2/{\rm DOF} \approx 0.8$ with $L_{\rm min}=32$. For $W=1.5$, the free fit gives $\eta=0.255(6)$ and $\chi^2/{\rm DOF} \approx 1.9$ with $L_{\rm min}=8$, and fixing $\eta=1/4$ gives $\chi^2/{\rm DOF} \approx 1.2$ with $L_{\rm min}=8$. Details of fits are given in Table~\ref{Tab_7}. The estimates of $\eta$ are again close to $1/4$. We plot $G_{\rm P}$ versus $L$ in Fig.~\ref{SM_GP}(b). For $K>K_2$, by fitting $G_{\rm P}$ to Eq.~(\ref{QLR}), we obtain $\eta=0.1342(3)$, $0.0939(1)$, $0.1336(1)$ and $0.0929(2)$ as well as $\chi^2/{\rm DOF} \approx 1.1$, $0.8$, $0.4$ and $0.5$ with $L_{\rm min}=16$, $8$, $8$ and $64$, for ${\rm P_1}$, ${\rm P_2}$, ${\rm P_3}$ and ${\rm P_4}$, respectively. Details of fits are given in Table~\ref{Tab_8}. The power-law decay has been shown in End Matter. The value of $\eta$ depends on $K$, while the results of $\eta$ obtained from $G_z$ and $G_{\rm P}$ are close to each other. Our final estimates are $\eta \approx 0.1339$ and $0.0934$ for $K=1.5$ and $2$, respectively.

\section{Possible experimental accessibility}
We discuss the possible experimental accessibility. The realization of the orthogonal geometry, where the V plane intersects a stack of P planes, requires optical potential engineering beyond a standard 3D lattice. This can be achieved by superimposing two independent optical potentials onto an ultracold atomic cloud. First, a 1D optical lattice created by counter-propagating lasers along the $x$ axis confines the atoms into a stack of decoupled 2D planes (the P planes). Second, the V plane is formed by a separate, sheet-like potential, generated by a rapidly scanned or spatially shaped laser beam~\cite{gauthier2016direct}, which provides strong confinement in the $z$-direction while leaving the atoms free in the $xy$ plane. When these two potentials are combined, the V plane orthogonally intersects each of the P planes along a 1D line, defining the unique ``T-junction" geometry. The coupling between the V plane and the P planes, a crucial parameter, is then determined by the potential barrier at these intersections and can be tuned by adjusting the relative intensities of the lattice and the light sheet~\cite{bakr2009quantum}, allowing for the exploration of dimensional crossover and LR order in this complex heterostructure. 
	
The detection of the LR and QLR orders in this heterostructure requires a probe capable of resolving phase correlations within and between distinct planes. This can be achieved through high-resolution in situ imaging, such as with a quantum gas microscope~\cite{bakr2009quantum}, followed by time-of-flight expansion within the plane of interest---a technique that converts the in-trap momentum distribution into a detectable spatial density~\cite{andrews1997observation}. To distinguish between the LR and QLR orders, one can analyze the visibility of interference pattern as a function of the observed system size: a stable, high-contrast pattern indicates LR order, while a decaying contrast signifies QLR order.

\newpage

\begin{table*}
	\caption{Fits of $K_n(L)$ ($n=1$,$2$) to Eq.~(\ref{Equ_Kn}) for $y$ and $x$ lines with $W=0.8$.}~\label{Tab_Kn8}
	\centering
	\scalebox{1.0}{
		\tabcolsep 0.1in
\begin{tabular}{c|c|c|c|c|c|c}
			\hline
			\hline      
			Quantity  &$n$  & $L_{\rm min}$ & $\chi^2/{\rm DOF}$ & $K_n$         & $a$     &$l_0$      \\
			\hline									
			${\rm d}(\xi_y/L)/{\rm d}K$ & 1 & 32  & 0.97/2 & 0.75(1)   & -0.7(3)   & 1       \\
			&   & 64  & 0.97/1 & 0.75(3)   & -0.7(6)   & 1       \\
			${\rm d}G_y/{\rm d}K$  & 1 & 16  & 3.05/3 & 0.750(10) & -0.8(1)   & 1       \\
			&   & 32  & 1.53/2 & 0.76(1)   & -1.1(3)   & 1       \\
			&   & 64  & 0.97/1 & 0.75(3)   & -0.7(6)   & 1       \\
			${\rm d}(\xi_y/L)/{\rm d}K$ & 2 & 16  & 0.14/2 & 1.07(2)   & -1.4(7)   & 1.9(9)  \\
			&   & 32  & 0.08/1 & 1.08(7)   & -2.1(35)  & 1.2(26) \\
			&   & 16  & 2.78/3 & 1.11996   & -3.3(4)   & 0.8(2)  \\
			&   & 32  & 0.26/2 & 1.11996   & -4.5(11)  & 0.4(2)  \\
			&   & 64  & 0.17/1 & 1.11996   & -5.1(25)  & 0.3(4)  \\
			&   & 16  & 2.80/4 & 1.11996   & -3.28(7)  & 0.8     \\
			&   & 32  & 2.44/3 & 1.11996   & -3.24(10) & 0.8     \\
			&   & 64  & 1.06/2 & 1.11996   & -3.4(1)   & 0.8     \\
			&   & 128 & 0.55/1 & 1.11996   & -3.5(2)   & 0.8     \\
			${\rm d}G_x/{\rm d}K$  & 1 & 16  & 1.15/3 & 0.751(10) & -0.7(1)   & 1       \\
			&   & 32  & 0.41/2 & 0.76(1)   & -0.9(3)   & 1       \\
			&   & 64  & 0.41/1 & 0.76(3)   & -0.8(6)   & 1       \\
			${\rm d}(\xi_x/L)/{\rm d}K$ & 1 & 32  & 0.97/2 & 0.75(1)   & -0.7(3)   & 1       \\
			&   & 64  & 0.97/1 & 0.75(3)   & -0.7(6)   & 1       \\					
			\hline
			\hline
		\end{tabular}
	}
\end{table*}

\begin{table*}
	\caption{Fits of $K_2(L)$, obtained by using different $f_0$, to Eq.~(\ref{Equ_Kn}) for an $x$ line with $W=0.8$.}~\label{B_Tab_7}
	\centering
	\scalebox{1.0}{
		\tabcolsep 0.1in
		\begin{tabular}{c|c|c|c|c|c}
			\hline
			\hline      
			$f_0$  & $L_{\rm min}$ & $\chi^2/{\rm DOF}$ & $K_2$         & $a$     & $l_0$     \\
			\hline			
			0.3   & 16  & 0.05/2 & 1.13(6) & -2.9(58)    & 0.2(7)    \\
			& 32  & 0.05/1 & 1.1(2)  & -3.5(187)   & 0.1(15)   \\
			& 16  & 0.11/3 & 1.11996 & -1.8(6)     & 0.4(3)    \\
			& 32  & 0.07/2 & 1.11996 & -1.7(9)     & 0.5(6)    \\
			& 64  & 0.02/1 & 1.11996 & -1.4(13)    & 0.9(19)   \\
			& 16  & 0.11/4 & 1.11996 & -1.85(9)    & 0.4       \\
			& 32  & 0.11/3 & 1.11996 & -1.9(1)     & 0.4       \\
\hline
			\hline
		\end{tabular}
	}
\end{table*}

\begin{table*}[h]
	\caption{Fits of $G_x$ and $G_y$ to Eq.~(\ref{K1}) ($C_1=C_2=0$) near $K_1$ for $W=0.4$ and $0.8$. ``-" means that the corresponding scaling term is not considered in the least-squares fit.}~\label{Tab_1}
	\centering
	\scalebox{0.8}{
		\tabcolsep 0.1in
		\begin{tabular}{c|c|c|c|c|c|c|c|c|c|c|c|c|c}
			\hline
			\hline 
			$W$   &Quantity     & $L_{\rm min}$ & $\chi^2/{\rm DOF}$  & $\eta$  & $a_0$   & $a_1$  & $K_1$   & $a_2$          & $a_3$         & $d_1$        & $d_2$        & $n_0$      & $n_1$        \\
			\hline
			0.4 & $G_x$ & 32  & 34.25/10  & 0.42(2)  & 0.102(9) & -1.27(9)  & 0.9436(6)  & -1.0(3)   & -0.5(2)  & 6.4(6)   & -88.4(96)    & 7.7(7)    & 9.5(23)  \\
			&    & 64  & 4.30/6    & 0.30(3)  & 0.16(4)  & -0.4(1)   & 0.950(2)   & -0.3(2)   & -0.1(4)  & -1.3(25) & 84.8(803)    & -0.6(9)   & 3.9(27)  \\
			&    & 128 & 0.48/2    & -0.2(18) & -0.0(2)  & -0.1(10)  & 0.94(3)    & -0.1(13)  & -0.0(4)  & 2.5(327) & -78.1(14114) & -0.2(84)  & 1.2(157) \\
			&    & 64  & 4.63/8    & 0.29(3)  & 0.142(9) & -0.38(9)  & 0.950(1)   & -0.26(5)  & -       & -       & 44.0(51)     & -0.7(7)   & 3.1(11)  \\
			&    & 64  & 6.13/7    & 1/4     & 0.12(2)  & -0.28(3)  & 0.951(2)   & -0.2(2)   & 0.0(3)   & 0.7(14)  & 19.1(417)    & -1.3(4)   & 2.9(20)  \\
			&    & 64  & 6.42/9    & 1/4     & 0.131(2) & -0.272(9) & 0.9515(2)  & -0.23(5)  & -       & -       & 39.5(20)     & -1.4(2)   & 3.0(10)  \\
			&    & 128 & 3.81/5    & 1/4     & 0.14(1)  & -0.25(3)  & 0.952(1)   & -0.2(1)   & -       & -       & 17.1(452)    & -1.9(6)   & 2.7(26)  \\
\hline
			\hline
		\end{tabular}
	}
\end{table*}

\begin{table*}
	\caption{Fits of $G_x$ and $G_y$ to Eq.~(\ref{eq2}) ($C_1=0$) at $K_1$ for $W=0.4$ and $0.8$.}~\label{Tab_2}
	\centering
	\scalebox{1.0}{
		\tabcolsep 0.1in
		\begin{tabular}{c|c|c|c|c|c|c|c}
			\hline
			\hline 					
			$W$ & Quantity & $L_{\rm min}$ & $\chi^2/{\rm DOF}$ & $\eta$        & $a_0$        & $d_1$       & $d_2$       \\
			\hline
			0.4 & $G_x$ & 8   & 828.77/6 & 0.399(2)  & 0.274(2)   & -1.06(3) & 8.8(1)       \\
			&    & 16  & 245.45/5 & 0.297(5)  & 0.157(4)   & 0.78(5)  & -6.3(4)      \\
			&    & 32  & 22.27/4  & 0.09(2)   & 0.047(6)   & 1.99(3)  & -22.2(7)     \\
			&    & 48  & 12.26/3  & 0.16(3)   & 0.07(1)    & 1.7(2)   & -13.9(41)    \\
			&    & 64  & 2.24/2   & 0.30(4)   & 0.18(5)    & -2.1(21) & 113.6(662)   \\
			&    & 96  & 0.37/1   & 0.1(2)    & 0.06(6)    & 2.6(11)  & -72.7(425)   \\
			&    & 64  & 3.75/3   & 0.247(7)  & 0.125(5)   & -       & 45.2(25)     \\
\hline
			\hline
		\end{tabular}
	}
\end{table*}

\begin{table*}
	\caption{Fits of $G_x$ to Eq.~(\ref{QLR}) at $K=0.4$, $0.6$ and $0.8$ for $W=1.5$.}~\label{Tab_QLR1}
	\centering
	\scalebox{1.0}{
		\tabcolsep 0.1in
		\begin{tabular}{c|c|c|c|c|c|c}
			\hline
			\hline 					
			$K$  & $L_{\rm min}$ & $\chi^2/{\rm DOF}$ & $\eta$        & $a$        & $b$       & $\omega$       \\
			\hline
			0.4 & 8   & 1.82/5 & 0.1301(2)   & 0.936(1)  & 0.08(4)   & 1.5(3)  \\
			& 16  & 1.79/4 & 0.1302(4)   & 0.936(2)  & 0.1(4)    & 1.7(15) \\
			& 8   & 4.65/6 & 0.1297(1)   & 0.9338(5) & 0.038(2)  & 1       \\
			& 16  & 2.18/5 & 0.1299(2)   & 0.9348(8) & 0.029(6)  & 1       \\
			& 32  & 2.03/4 & 0.1300(3)   & 0.935(2)  & 0.02(2)   & 1       \\
			& 48  & 0.47/3 & 0.1307(6)   & 0.939(3)  & -0.03(5)  & 1       \\
			& 64  & 0.23/2 & 0.131(1)    & 0.941(6)  & -0.07(9)  & 1       \\
\hline
			\hline
		\end{tabular}
	}
\end{table*}

\begin{table*}
	\caption{Fits of $G_y$ to Eq.~(\ref{QLR}) at $K=0.4$, $0.6$ and $0.8$ for $W=1.5$.}~\label{Tab_QLR2}
	\centering
	\scalebox{1.0}{
		\tabcolsep 0.1in
		\begin{tabular}{c|c|c|c|c|c|c}
			\hline
			\hline 					
			$K$  & $L_{\rm min}$ & $\chi^2/{\rm DOF}$ & $\eta$        & $a$        & $b$       & $\omega$       \\
			\hline
			0.4 & 8   & 2.50/5  & 0.1300(2)  & 0.939(1)   & 0.07(4)   & 1.4(3)  \\
			& 16  & 2.42/4  & 0.1301(3)  & 0.939(2)   & 0.2(6)    & 1.8(14) \\
			& 8   & 5.03/6  & 0.1296(1)  & 0.9365(5)  & 0.039(2)  & 1       \\
			& 16  & 2.92/5  & 0.1298(2)  & 0.9374(8)  & 0.031(6)  & 1       \\
			& 32  & 2.57/4  & 0.1299(3)  & 0.938(2)   & 0.02(2)   & 1       \\
			& 48  & 1.82/3  & 0.1304(6)  & 0.941(3)   & -0.02(5)  & 1       \\
			& 64  & 1.82/2  & 0.130(1)   & 0.940(5)   & -0.01(9)  & 1       \\
\hline
			\hline
		\end{tabular}
	}
\end{table*}

\begin{table*}
	\caption{Fits of $G_x$ and $G_y$ to Eq.~(\ref{QLR}) at $W=2$ and $K=0.8$.}~\label{Tab_QLR3}
	\centering
	\scalebox{1.0}{
		\tabcolsep 0.1in
		\begin{tabular}{c|c|c|c|c|c|c}
			\hline
			\hline 					
			Quantity  & $L_{\rm min}$ & $\chi^2/{\rm DOF}$ & $\eta$        & $a$        & $b$       & $\omega$       \\
			\hline
			$G_x$ & 16  & 6.54/4  & 0.0652(1)  & 0.9181(5)  & 0.8(6)   & 2.1(3)  \\
			& 32  & 7.84/4  & 0.0648(1)  & 0.9164(7)  & 0.038(7) & 1       \\
			& 48  & 6.10/3  & 0.0651(3)  & 0.918(1)   & 0.02(2)  & 1       \\
			& 64  & 5.92/2  & 0.0650(4)  & 0.917(2)   & 0.03(3)  & 1       \\
			& 96  & 0.09/1  & 0.0669(9)  & 0.928(5)   & -0.2(1)  & 1       \\
			& 16  & 6.68/5  & 0.06511(5) & 0.9179(2)  & 0.60(2)  & 2       \\
			& 32  & 6.60/4  & 0.06513(8) & 0.9180(4)  & 0.6(1)   & 2       \\
			& 48  & 6.08/3  & 0.0652(1)  & 0.9184(7)  & 0.3(4)   & 2       \\
			& 64  & 5.56/2  & 0.0651(2)  & 0.918(1)   & 0.9(9)   & 2       \\
			& 48  & 6.76/4  & 0.06533(5) & 0.9189(2)  & -       & -      \\
			& 64  & 6.56/3  & 0.06531(7) & 0.9188(3)  & -       & -      \\
			& 96  & 4.20/2  & 0.0651(1)  & 0.9179(6)  & -       & -      \\
			& 128 & 0.46/1  & 0.0655(2)  & 0.920(1)   & -       & -\\
			$G_y$ & 32  & 0.34/4  & 0.0653(1)  & 0.9426(5)  & -0.026(6) & 1        \\
			& 48  & 0.28/3  & 0.0653(2)  & 0.9424(10) & -0.02(1)  & 1        \\
			& 64  & 0.17/2  & 0.0654(3)  & 0.943(2)   & -0.03(3)  & 1        \\
			& 96  & 0.00/1  & 0.0656(7)  & 0.944(4)   & -0.06(9)  & 1        \\
			& 32  & 0.56/4  & 0.06514(6) & 0.9415(3)  & -0.37(8)  & 2        \\
			& 48  & 0.52/3  & 0.0652(1)  & 0.9416(5)  & -0.4(3)   & 2        \\
			& 64  & 0.25/2  & 0.0652(2)  & 0.9420(9)  & -0.8(7)   & 2        \\
			& 96  & 0.01/1  & 0.0654(4)  & 0.943(2)   & -2.1(29)  & 2        \\
			& 48  & 2.71/4  & 0.06501(4) & 0.9409(2)  & -        & -       \\
			& 64  & 1.42/3  & 0.06504(5) & 0.9411(2)  & -        & -       \\
			& 96  & 0.55/2  & 0.0651(1)  & 0.9415(5)  & -        & -       \\
			& 128 & 0.04/1  & 0.0652(2)  & 0.9420(8)  & -        & - \\
			\hline
			\hline
		\end{tabular}
	}
\end{table*}

\begin{table*}
	\caption{Fits of $K_2(L)$ to Eq.~(\ref{Equ_Kn}) for a $y$ line with $W=1.5$.}~\label{Tab_Kn15}
	\centering
	\scalebox{1.0}{
		\tabcolsep 0.1in
		\begin{tabular}{c|c|c|c|c}
			\hline
			\hline      
			$L_{\rm min}$ & $\chi^2/{\rm DOF}$ & $K_2$         & $a$     & $l_0$     \\
			\hline			
			16  & 0.28/2 & 1.2(1)  & -11.4(168) & 0.0(1)  \\
			32  & 0.28/1 & 1.2(2)  & -10.9(365) & 0.0(3)  \\
			16  & 0.45/3 & 1.11996 & -6.6(11)   & 0.12(6) \\
			32  & 0.32/2 & 1.11996 & -6.0(17)   & 0.2(1)  \\
			64  & 0.17/1 & 1.11996 & -5.1(25)   & 0.3(4)  \\
			16  & 1.96/4 & 1.11996 & -5.4(1)    & 0.2     \\
			32  & 0.44/3 & 1.11996 & -5.5(2)    & 0.2     \\
			64  & 0.20/2 & 1.11996 & -5.6(2)    & 0.2     \\
			128 & 0.09/1 & 1.11996 & -5.5(3)    & 0.2     \\		
			\hline
			\hline
		\end{tabular}
	}
\end{table*}

\begin{table*}
	\caption{Fits of $K_2(L)$, obtained by using different $f_0$, to Eq.~(\ref{Equ_Kn}) for an $x$ line with $W=1.5$.}~\label{B_Tab_8}
	\centering
	\scalebox{1.0}{
		\tabcolsep 0.1in
		\begin{tabular}{c|c|c|c|c|c}
			\hline
			\hline      
			$f_0$  & $L_{\rm min}$ & $\chi^2/{\rm DOF}$ & $K_2$         & $a$     & $l_0$     \\
			\hline			
			0.15 & 16  & 0.03/2 & 1.14(3) & -1.3(8)   & 1.7(11) \\
			& 32  & 0.03/1 & 1.14(6) & -1.3(27)  & 1.6(39) \\
			& 16  & 0.58/3 & 1.11996 & -0.8(1)   & 2.5(5)  \\
			& 32  & 0.18/2 & 1.11996 & -0.7(2)   & 3.3(15) \\
			& 64  & 0.00/1 & 1.11996 & -0.5(4)   & 5.2(52) \\
			& 16  & 0.58/4 & 1.11996 & -0.79(3)  & 2.5     \\
			& 32  & 0.52/3 & 1.11996 & -0.80(5)  & 2.5     \\
			& 64  & 0.36/2 & 1.11996 & -0.78(8)  & 2.5     \\
			& 128 & 0.03/1 & 1.11996 & -0.7(1)   & 2.5     \\
			0.2    & 32  & 0.00/1 & 1.13(7) & -1.9(33)  & 1.3(28) \\
			& 32  & 0.03/2 & 1.11996 & -1.5(4)   & 1.7(8)  \\
			& 64  & 0.01/1 & 1.11996 & -1.4(8)   & 1.9(20) \\
			& 32  & 0.03/3 & 1.11996 & -1.51(6)  & 1.7     \\
			& 64  & 0.03/2 & 1.11996 & -1.51(10) & 1.7     \\
			& 128 & 0.02/1 & 1.11996 & -1.5(1)   & 1.7     \\			
			0.25  & 32  & 0.14/2 & 1.11996 & -1.7(4)   & 2.0(7)  \\
			& 64  & 0.01/1 & 1.11996 & -2.0(9)   & 1.5(14) \\
			& 32  & 0.15/3 & 1.11996 & -1.72(6)  & 2.0     \\
			& 64  & 0.11/2 & 1.11996 & -1.71(9)  & 2.0     \\
			& 128 & 0.00/1 & 1.11996 & -1.7(1)   & 2.0    \\		
			\hline
			\hline
		\end{tabular}
	}
\end{table*}

\begin{table*}
	\caption{Fits of $G_y$ to Eq.~(\ref{Equ_Gy}) for a $y$ line in the LR ordered phase.}~\label{Tab_Gy}
	\centering
	\scalebox{1.0}{
		\tabcolsep 0.1in
		\begin{tabular}{c|c|c|c|c|c}
			\hline
			\hline      
			{\rm Parameters}   & $L_{\rm min}$ & $\chi^2/{\rm DOF}$ & $a$  & $b$  & $q$   \\
			\hline
			${\rm P_1}$ & 8  & 23.20/8 & 0.67534(3) & 0.2101(2) & 0.5278(5)  \\
			& 16 & 16.61/7 & 0.67526(5) & 0.2094(3) & 0.5262(8)  \\
			& 24 & 6.06/6  & 0.67511(6) & 0.2080(5) & 0.523(1)   \\
			& 32 & 5.91/5  & 0.67513(8) & 0.2082(8) & 0.523(2)   \\
			& 48 & 5.19/4  & 0.6750(1)  & 0.207(2)  & 0.521(4)   \\
			& 64 & 5.18/3  & 0.6750(2)  & 0.207(4)  & 0.521(6)   \\
			& 80 & 1.78/2  & 0.6756(4)  & 0.219(8)  & 0.54(1)    \\
			& 96 & 0.74/1  & 0.6752(6)  & 0.21(1)   & 0.52(2)    \\
			${\rm P_2}$   & 16 & 8.16/7  & 0.72612(4) & 0.1814(3) & 0.5204(8)  \\
			& 24 & 8.02/6  & 0.72613(6) & 0.1816(5) & 0.521(1)   \\
			& 32 & 7.76/5  & 0.72611(7) & 0.1813(7) & 0.520(2)   \\
			& 48 & 5.42/4  & 0.7263(1)  & 0.184(2)  & 0.525(4)   \\
			& 64 & 0.76/3  & 0.7260(2)  & 0.178(3)  & 0.515(6)   \\
			& 80 & 0.74/2  & 0.7259(3)  & 0.178(6)  & 0.51(1)    \\
			& 96 & 0.66/1  & 0.7258(5)  & 0.175(10) & 0.51(2)    \\
			${\rm P_3}$ & 16 & 40.85/7 & 0.77202(4) & 0.1682(3) & 0.535(1)   \\
			& 24 & 6.75/6  & 0.77174(6) & 0.1656(5) & 0.528(2)   \\
			& 32 & 1.13/5  & 0.77161(9) & 0.1641(8) & 0.524(2)   \\
			& 48 & 0.27/4  & 0.7715(2)  & 0.162(2)  & 0.519(5)   \\
			& 64 & 0.27/3  & 0.7715(2)  & 0.162(4)  & 0.519(8)   \\
			& 80 & 0.03/2  & 0.7716(4)  & 0.165(8)  & 0.53(2)    \\
			& 96 & 0.02/1  & 0.7717(6)  & 0.17(1)   & 0.53(3)    \\
			${\rm P_4}$   & 16 & 52.96/7 & 0.80153(4) & 0.1414(2) & 0.5289(10) \\
			& 24 & 30.38/6 & 0.80135(5) & 0.1397(4) & 0.523(2)   \\
			& 32 & 8.04/5  & 0.80113(7) & 0.1374(6) & 0.516(2)   \\
			& 48 & 7.73/4  & 0.8011(1)  & 0.137(2)  & 0.513(5)   \\
			& 64 & 2.59/3  & 0.8007(2)  & 0.131(3)  & 0.499(8)   \\
			& 80 & 2.56/2  & 0.8007(3)  & 0.131(5)  & 0.50(1)    \\
			& 96 & 0.98/1  & 0.8012(5)  & 0.140(10) & 0.52(2)\\
			\hline
			\hline
		\end{tabular}
	}
\end{table*}

\begin{table*}
	\caption{Fits of $\langle M^2_y \rangle$ to Eq.~(\ref{Equ_f0}) for a $y$ line in the LR ordered phase.}~\label{Tab_f0}
	\centering
	\scalebox{1.0}{
		\tabcolsep 0.1in
		\begin{tabular}{c|c|c|c|c|c}
			\hline
			\hline      
			{\rm Parameters}   & $L_{\rm min}$ & $\chi^2/{\rm DOF}$ & $a$  & $b$  & $q$   \\
			\hline
			${\rm P_1}$ & 24 & 48.15/6 & 0.67442(4) & 0.3570(3) & 0.5102(4) \\
			& 32 & 8.56/5  & 0.67462(5) & 0.3591(4) & 0.5127(6) \\
			& 48 & 6.26/4  & 0.67472(9) & 0.361(1)  & 0.514(1)  \\
			& 64 & 6.15/3  & 0.6748(1)  & 0.361(2)  & 0.515(2)  \\
			& 80 & 2.63/2  & 0.6751(2)  & 0.368(4)  & 0.521(4)  \\
			& 96 & 1.62/1  & 0.6748(3)  & 0.362(7)  & 0.516(6)  \\
			${\rm P_2}$   & 32 & 22.21/5 & 0.72532(4) & 0.2991(4) & 0.5003(6) \\
			& 48 & 2.76/4  & 0.72559(7) & 0.3026(9) & 0.505(1)  \\
			& 64 & 2.39/3  & 0.7256(1)  & 0.303(2)  & 0.506(2)  \\
			& 80 & 1.61/2  & 0.7258(2)  & 0.306(3)  & 0.508(4)  \\
			& 96 & 1.47/1  & 0.7258(3)  & 0.307(5)  & 0.510(6)  \\
			${\rm P_3}$ & 24 & 11.66/6 & 0.77102(4) & 0.2727(3) & 0.5059(5) \\
			& 32 & 4.05/5  & 0.77112(5) & 0.2736(4) & 0.5074(8) \\
			& 48 & 0.63/4  & 0.77125(9) & 0.275(1)  & 0.510(2)  \\
			& 64 & 0.63/3  & 0.7713(1)  & 0.275(2)  & 0.510(3)  \\
			& 80 & 0.42/2  & 0.7713(2)  & 0.277(4)  & 0.512(5)  \\
			& 96 & 0.27/1  & 0.7712(3)  & 0.275(7)  & 0.509(8)  \\
			${\rm P_4}$   & 16 & 63.55/7 & 0.80057(2) & 0.2287(1) & 0.4999(3) \\
			& 24 & 12.55/6 & 0.80074(3) & 0.2300(2) & 0.5027(5) \\
			& 32 & 9.48/5  & 0.80078(4) & 0.2305(3) & 0.5035(7) \\
			& 48 & 8.41/4  & 0.80084(7) & 0.2313(8) & 0.505(1)  \\
			& 64 & 2.95/3  & 0.8007(1)  & 0.228(1)  & 0.500(2)  \\
			& 80 & 2.31/2  & 0.8005(2)  & 0.226(3)  & 0.497(4)  \\
			& 96 & 0.38/1  & 0.8008(3)  & 0.232(5)  & 0.505(7)\\
			\hline
			\hline
		\end{tabular}
	}
\end{table*}

\begin{table*}
	\caption{Fits of $\langle M^2_{yk} \rangle$ to Eq.~(\ref{Equ_f1}) for a $y$ line in the LR ordered phase.}~\label{Tab_f1}
	\centering
	\scalebox{1.0}{
		\tabcolsep 0.1in
		\begin{tabular}{c|c|c|c|c|c|c}
			\hline
			\hline      
			{\rm Parameters}   & $L_{\rm min}$ & $\chi^2/{\rm DOF}$   & $q$       & $a$  & $b$   & $\omega$      \\
			\hline
			${\rm P_1}$ & 8  & 7.86/7  & 0.5321(10) & 0.1448(9) & -0.0692(3) & 0.62(2) \\
			& 16 & 1.72/6  & 0.529(2)   & 0.141(1)  & -0.073(2)  & 0.71(4) \\
			& 24 & 1.12/5  & 0.527(2)   & 0.140(2)  & -0.078(7)  & 0.76(8) \\
			& 32 & 0.81/4  & 0.525(4)   & 0.138(3)  & -0.09(3)   & 0.9(2)  \\
			& 48 & 0.53/3  & 0.53(1)    & 0.14(1)   & -0.06(3)   & 0.6(5)  \\
			${\rm P_2}$   & 24 & 11.70/5 & 0.537(7)   & 0.135(8)  & -0.069(5)  & 0.39(7) \\
			& 32 & 10.06/4 & 0.527(7)   & 0.123(7)  & -0.068(4)  & 0.5(1)  \\
			${\rm P_3}$ & 16 & 1.69/6  & 0.516(1)   & 0.1007(6) & -0.147(8)  & 1.08(4) \\
			& 24 & 1.64/5  & 0.516(2)   & 0.1006(9) & -0.15(2)   & 1.10(8) \\
			& 32 & 0.95/4  & 0.513(3)   & 0.099(1)  & -0.2(1)    & 1.3(2)  \\
			& 48 & 0.29/3  & 0.52(1)    & 0.103(8)  & -0.07(8)   & 0.8(6)  \\
			& 64 & 0.18/2  & 0.51(1)    & 0.100(6)  & -0.3(15)   & 1.3(17) \\
			${\rm P_4}$   & 16 & 13.29/6 & 0.514(1)   & 0.0849(6) & -0.088(5)  & 0.98(4) \\
			& 24 & 6.88/5  & 0.511(1)   & 0.0834(7) & -0.12(2)   & 1.16(8) \\
			& 32 & 6.59/4  & 0.512(3)   & 0.084(2)  & -0.10(5)   & 1.1(2) \\
			\hline
			\hline
		\end{tabular}
	}
\end{table*}

\begin{table*}
	\caption{Fits of $\xi_y$ to Eq.~(\ref{Equ_xi}) for a $y$ line in the LR ordered phase.}~\label{Tab_xi}
	\centering
	\scalebox{1.0}{
		\tabcolsep 0.1in
		\begin{tabular}{c|c|c|c|c|c|c|c}
			\hline
			\hline      
			{\rm Parameters}   & $L_{\rm min}$ & $\chi^2/{\rm DOF}$   & $q$   & $a$   &$b$  &$\omega$  &$c$    \\
			\hline
			${\rm P_1}$ & 8  & 1.97/6  & 0.533(2) & 0.117(1) & 0.61(5) & 1.86(6) & 0.093(5) \\
			& 16 & 1.19/5  & 0.530(4) & 0.119(4) & 0.4(2)  & 1.6(3)  & 0.08(2)  \\
			& 24 & 1.12/4  & 0.53(1)  & 0.121(9) & 0.3(3)  & 1.5(8)  & 0.07(5)  \\
			${\rm P_2}$   & 8  & 13.50/6 & 0.534(2) & 0.142(2) & 0.51(7) & 1.9(1)  & 0.157(5) \\
			& 16 & 8.42/5  & 0.52(1)  & 0.16(1)  & 0.20(7) & 0.9(4)  & 0.1(1)   \\
			${\rm P_3}$ & 8  & 1.52/6  & 0.522(3) & 0.185(3) & 1.38(7) & 1.73(5) & 0.10(1)  \\
			& 16 & 1.44/5  & 0.521(6) & 0.187(7) & 1.3(4)  & 1.7(2)  & 0.09(3)  \\
			& 24 & 1.27/4  & 0.52(1)  & 0.19(2)  & 0.9(7)  & 1.5(6)  & 0.1(1)   \\
			${\rm P_4}$   & 8  & 12.37/6 & 0.518(2) & 0.231(4) & 1.26(7) & 1.68(5) & 0.11(1)  \\
			& 16 & 7.47/5  & 0.505(9) & 0.25(2)  & 0.74(9) & 1.2(2)  & -0.01(9) \\
			& 24 & 6.89/4  & 0.51(1)  & 0.24(2)  & 1.4(14) & 1.6(6)  & 0.09(8)\\
			\hline
			\hline
		\end{tabular}
	}
\end{table*}

\begin{table*}
	\caption{Fits of $R$ to Eq.~(\ref{Equ_fr}) for a $y$ line in the LR ordered phase.}~\label{Tab_fr}
	\centering
	\scalebox{1.0}{
		\tabcolsep 0.1in
		\begin{tabular}{c|c|c|c|c|c}
			\hline
			\hline      
			{\rm Parameters}   & $L_{\rm min}$ & $\chi^2/{\rm DOF}$   & $a$       & $b$    & $\omega$      \\
			\hline
			${\rm P_1}$ & 16 & 57.76/7 & 0.67596(2) & 0.092(1)  & 0.876(5) \\
			& 24 & 17.26/6 & 0.67581(3) & 0.081(2)  & 0.830(9) \\
			& 32 & 8.09/5  & 0.67572(5) & 0.075(3)  & 0.80(1)  \\
			& 48 & 2.67/4  & 0.67556(9) & 0.062(5)  & 0.74(3)  \\
			& 64 & 1.91/3  & 0.6755(1)  & 0.055(9)  & 0.71(5)  \\
			& 80 & 1.12/2  & 0.6756(2)  & 0.07(2)   & 0.78(9)  \\
			& 96 & 0.81/1  & 0.6755(4)  & 0.05(3)   & 0.7(2)   \\
			${\rm P_2}$   & 16 & 11.10/7 & 0.72684(3) & 0.0996(5) & 0.656(2) \\
			& 24 & 10.57/6 & 0.72681(4) & 0.0991(8) & 0.654(3) \\
			& 32 & 7.72/5  & 0.72676(5) & 0.098(1)  & 0.648(5) \\
			& 48 & 7.00/4  & 0.72670(9) & 0.095(3)  & 0.64(1)  \\
			& 64 & 0.96/3  & 0.7264(1)  & 0.086(4)  & 0.60(2)  \\
			& 80 & 0.92/2  & 0.7264(3)  & 0.084(9)  & 0.60(3)  \\
			& 96 & 0.91/1  & 0.7264(4)  & 0.09(2)   & 0.60(5)  \\
			${\rm P_3}$ & 16 & 41.96/7 & 0.77188(1) & 0.222(4)  & 1.345(7) \\
			& 24 & 16.34/6 & 0.77182(2) & 0.178(8)  & 1.27(2)  \\
			& 32 & 3.65/5  & 0.77175(3) & 0.13(1)   & 1.18(3)  \\
			& 48 & 0.41/4  & 0.77168(5) & 0.09(2)   & 1.07(7)  \\
			& 64 & 0.14/3  & 0.77165(9) & 0.07(3)   & 1.0(1)   \\
			& 80 & 0.10/2  & 0.7716(1)  & 0.06(6)   & 1.0(2)   \\
			& 96 & 0.01/1  & 0.7716(3)  & 0.04(7)   & 0.9(4)   \\
			${\rm P_4}$   & 16 & 55.36/7 & 0.80148(1) & 0.133(3)  & 1.247(8) \\
			& 24 & 38.21/6 & 0.80143(2) & 0.111(5)  & 1.19(2)  \\
			& 32 & 13.99/5 & 0.80135(3) & 0.079(6)  & 1.08(3)  \\
			& 48 & 12.32/4 & 0.80130(5) & 0.06(1)   & 1.00(6)  \\
			& 64 & 2.45/3  & 0.8011(1)  & 0.022(7)  & 0.7(1)   \\
			& 80 & 0.92/2  & 0.8008(4)  & 0.011(6)  & 0.5(2)   \\
			& 96 & 0.49/1  & 0.8010(3)  & 0.02(2)   & 0.7(3)\\
			\hline
			\hline
		\end{tabular}
	}
\end{table*}

\begin{table*}
	\caption{Fits of $\xi'_y$ to Eq.~(\ref{Equ_xiV}) for the V plane in the LR ordered phase.}~\label{Tab_xiV}
	\centering
	\scalebox{1.0}{
		\tabcolsep 0.1in
		\begin{tabular}{c|c|c|c|c|c}
			\hline
			\hline      
			{\rm Parameters}   & $L_{\rm min}$ & $\chi^2/{\rm DOF}$   & $a$  & $q'$  &$b$    \\
			\hline
			${\rm P_1}$ & 16 & 17.09/7 & 0.595(2) & 0.4961(7) & -0.154(5)  \\
			& 24 & 15.15/6 & 0.600(4) & 0.495(1)  & -0.166(10) \\
			& 32 & 11.97/5 & 0.590(7) & 0.498(2)  & -0.14(2)   \\
			& 48 & 9.80/4  & 0.57(1)  & 0.502(4)  & -0.09(4)   \\
			& 64 & 3.26/3  & 0.52(2)  & 0.516(6)  & 0.06(7)    \\
			& 80 & 0.34/2  & 0.47(4)  & 0.53(1)   & 0.3(1)     \\
			& 96 & 0.17/1  & 0.45(6)  & 0.54(2)   & 0.3(2)     \\
			${\rm P_2}$   & 16 & 8.82/7  & 0.751(3) & 0.4892(7) & -0.313(6)  \\
			& 24 & 8.46/6  & 0.754(5) & 0.489(1)  & -0.32(1)   \\
			& 32 & 6.85/5  & 0.745(9) & 0.491(2)  & -0.30(2)   \\
			& 48 & 1.89/4  & 0.71(2)  & 0.498(4)  & -0.20(5)   \\
			& 64 & 1.87/3  & 0.71(3)  & 0.497(6)  & -0.22(9)   \\
			& 80 & 1.85/2  & 0.71(5)  & 0.50(1)   & -0.2(2)    \\
			& 96 & 1.40/1  & 0.66(8)  & 0.51(2)   & -0.0(3)    \\
			${\rm P_3}$ & 16 & 43.58/7 & 0.520(3) & 0.5215(9) & 0.237(6)   \\
			& 24 & 10.01/6 & 0.544(5) & 0.514(2)  & 0.18(1)    \\
			& 32 & 9.69/5  & 0.548(9) & 0.513(3)  & 0.17(2)    \\
			& 48 & 9.69/4  & 0.55(2)  & 0.513(5)  & 0.17(5)    \\
			& 64 & 0.46/3  & 0.48(3)  & 0.533(8)  & 0.40(9)    \\
			& 80 & 0.44/2  & 0.47(5)  & 0.53(1)   & 0.4(2)     \\
			& 96 & 0.12/1  & 0.44(7)  & 0.55(2)   & 0.6(3)     \\
			${\rm P_4}$   & 16 & 29.01/7 & 0.680(3) & 0.5088(9) & 0.043(7)   \\
			& 24 & 15.01/6 & 0.699(6) & 0.504(2)  & -0.00(1)   \\
			& 32 & 13.37/5 & 0.71(1)  & 0.502(3)  & -0.03(3)   \\
			& 48 & 12.93/4 & 0.70(2)  & 0.504(5)  & 0.00(6)    \\
			& 64 & 12.90/3 & 0.69(4)  & 0.505(8)  & 0.0(1)     \\
			& 80 & 12.54/2 & 0.73(7)  & 0.50(1)   & -0.1(2)    \\
			& 96 & 4.04/1  & 0.50(8)  & 0.55(2)   & 0.8(3)\\
			\hline
			\hline
		\end{tabular}
	}
\end{table*}

\begin{table*}
	\caption{Fits of $\xi_x$ to Eq.~(\ref{Equ_xix1}) for an $x$ line in the critical phase.}~\label{Tab_xix1}
	\centering
	\scalebox{1.0}{
		\tabcolsep 0.1in
		\begin{tabular}{c|c|c|c|c|c}
			\hline
			\hline      
			{\rm Parameters}   & $L_{\rm min}$ & $\chi^2/{\rm DOF}$   & $a$       & $b$    & $\omega$      \\
			\hline
			${\rm P_1}$ & 32 & 12.30/5 & 0.810(4) & -0.748(10) & 0.400(9) \\
			& 48 & 8.33/4  & 0.800(6) & -0.80(3)   & 0.44(2)  \\
			& 64 & 5.59/3  & 0.788(8) & -0.91(8)   & 0.49(4)  \\
			& 80 & 0.60/2  & 0.770(8) & -1.4(3)    & 0.62(7)  \\
			& 96 & 0.19/1  & 0.76(1)  & -1.7(7)    & 0.7(1)   \\
			${\rm P_2}$   & 32 & 17.67/5 & 0.808(3) & -0.74(1)   & 0.47(1)  \\
			& 48 & 6.87/4  & 0.796(4) & -0.87(5)   & 0.54(2)  \\
			& 64 & 2.02/3  & 0.786(5) & -1.1(2)    & 0.62(5)  \\
			& 80 & 1.94/2  & 0.788(9) & -1.0(3)    & 0.60(9)  \\
			& 96 & 1.06/1  & 0.78(1)  & -1.6(9)    & 0.7(2)   \\
			${\rm P_3}$ & 32 & 45.36/5 & 1.36(2)  & -1.121(4)  & 0.230(8) \\
			& 48 & 4.78/4  & 1.24(2)  & -1.20(2)   & 0.32(2)  \\
			& 64 & 1.41/3  & 1.21(2)  & -1.31(8)   & 0.38(3)  \\
			& 80 & 0.18/2  & 1.25(5)  & -1.2(1)    & 0.32(6)  \\
			& 96 & 0.17/1  & 1.24(8)  & -1.2(2)    & 0.3(1)   \\
			${\rm P_4}$   & 32 & 30.99/5 & 1.250(8) & -1.073(8)  & 0.325(9) \\
			& 48 & 4.09/4  & 1.19(1)  & -1.25(5)   & 0.41(2)  \\
			& 64 & 4.09/3  & 1.19(2)  & -1.24(10)  & 0.41(4)  \\
			& 80 & 3.19/2  & 1.22(4)  & -1.1(1)    & 0.36(7)  \\
			& 96 & 0.02/1  & 1.16(3)  & -1.7(6)    & 0.5(1)\\
			\hline
			\hline
		\end{tabular}
	}
\end{table*}

\begin{table*}
	\caption{Fits of $\xi_x$ to Eq.~(\ref{Equ_xix2}) for an $x$ line in the critical phase.}~\label{Tab_xix2}
	\centering
	\scalebox{1.0}{
		\tabcolsep 0.1in
		\begin{tabular}{c|c|c|c|c|c}
			\hline
			\hline      
			{\rm Parameters}   & $L_{\rm min}$ & $\chi^2/{\rm DOF}$   & $a$       & $b$    & $\omega$      \\
			\hline
			${\rm P_1}$ & 32 & 24.57/5 & 1.03(2) & -0.908(6) & 0.65(4) \\
			& 48 & 12.24/4 & 0.92(2) & -0.96(3)  & 0.93(9) \\
			& 64 & 7.02/3  & 0.86(2) & -1.2(1)   & 1.3(2)  \\
			& 80 & 0.79/2  & 0.80(2) & -2.3(8)   & 2.0(3)  \\
			& 96 & 0.14/1  & 0.79(2) & -3.7(27)  & 2.4(6)  \\
			${\rm P_2}$   & 32 & 32.66/5 & 0.92(1) & -0.82(1)  & 0.93(5) \\
			& 48 & 10.43/4 & 0.85(1) & -1.08(9)  & 1.4(1)  \\
			& 64 & 2.30/3  & 0.82(1) & -1.7(4)   & 1.9(2)  \\
			& 80 & 2.30/2  & 0.82(2) & -1.7(8)   & 1.9(4)  \\
			& 96 & 1.19/1  & 0.80(2) & -3.7(35)  & 2.6(8)  \\
			${\rm P_3}$ & 48 & 6.88/4  & 1.9(2)  & -1.8(1)   & 0.43(7) \\
			& 64 & 0.91/3  & 1.5(1)  & -1.68(4)  & 0.7(2)  \\
			& 80 & 0.17/2  & 1.7(4)  & -1.7(2)   & 0.5(3)  \\
			& 96 & 0.14/1  & 1.6(5)  & -1.66(5)  & 0.6(5)  \\
			${\rm P_4}$   & 32 & 46.61/5 & 2.1(1)  & -1.8(1)   & 0.34(4) \\
			& 48 & 4.48/4  & 1.43(5) & -1.51(3)  & 0.84(9) \\
			& 64 & 4.11/3  & 1.39(7) & -1.6(1)   & 0.9(2)  \\
			& 80 & 3.57/2  & 1.5(2)  & -1.45(8)  & 0.7(3)  \\
			& 96 & 0.04/1  & 1.24(8) & -2.7(16)  & 1.6(6)  \\
			\hline
			\hline
		\end{tabular}
	}
\end{table*}

\begin{table*}
	\caption{Fits of $G_x$ to Eq.~(\ref{Equ_Gx1}) for an $x$ line in the critical phase.}~\label{Tab_Gx1}
	\centering
	\scalebox{1.0}{
		\tabcolsep 0.1in
		\begin{tabular}{c|c|c|c|c|c}
			\hline
			\hline      
			{\rm Parameters}   & $L_{\rm min}$ & $\chi^2/{\rm DOF}$   & $a$       & $l_0$    & $\hat{q}$      \\
			\hline
			${\rm P_1}$ & 32 & 30.56/5 & 0.613(2) & 6.8(3)   & 0.063(2)  \\
			& 48 & 3.39/4  & 0.598(2) & 11.5(10) & 0.049(2)  \\
			& 64 & 2.19/3  & 0.594(4) & 13.8(23) & 0.045(4)  \\
			& 80 & 0.18/2  & 0.61(2)  & 7.6(45)  & 0.06(1)   \\
			& 96 & 0.10/1  & 0.62(4)  & 5.7(75)  & 0.06(3)   \\
			${\rm P_2}$   & 32 & 7.49/5  & 0.651(2) & 7.1(4)   & 0.050(1)  \\
			& 48 & 2.10/4  & 0.643(3) & 9.5(11)  & 0.044(3)  \\
			& 64 & 2.00/3  & 0.645(6) & 8.8(24)  & 0.045(5)  \\
			& 80 & 1.45/2  & 0.638(9) & 12.4(56) & 0.039(8)  \\
			& 96 & 1.08/1  & 0.65(3)  & 7.1(92)  & 0.05(2)   \\
			${\rm P_3}$ & 32 & 17.22/5 & 0.754(1) & 7.2(3)   & 0.0437(9) \\
			& 48 & 8.31/4  & 0.748(2) & 9.5(8)   & 0.039(2)  \\
			& 64 & 5.76/3  & 0.742(3) & 12.1(19) & 0.035(3)  \\
			& 80 & 0.27/2  & 0.732(4) & 20.9(42) & 0.026(3)  \\
			& 96 & 0.14/1  & 0.730(6) & 23.6(84) & 0.025(5)  \\
			${\rm P_4}$   & 32 & 17.33/5 & 0.774(1) & 7.3(3)   & 0.0361(8) \\
			& 48 & 10.95/4 & 0.769(2) & 9.4(9)   & 0.032(2)  \\
			& 64 & 3.77/3  & 0.761(3) & 14.3(21) & 0.026(2)  \\
			& 80 & 1.01/2  & 0.755(3) & 21.3(46) & 0.021(3)  \\
			& 96 & 0.04/1  & 0.76(1)  & 13.2(92) & 0.027(8) \\
			\hline
			\hline
		\end{tabular}
	}
\end{table*}

\begin{table*}
	\caption{Fits of $G_x$ to Eq.~(\ref{Equ_Gx2}) for an $x$ line in the critical phase.}~\label{Tab_Gx2}
	\centering
	\scalebox{1.0}{
		\tabcolsep 0.1in
		\begin{tabular}{c|c|c|c|c}
			\hline
			\hline      
			{\rm Parameters}   & $L_{\rm min}$ & $\chi^2/{\rm DOF}$   & $a$      & $\tilde{q}$      \\
			\hline
			${\rm P_1}$ & 8   & 89646.95/9  & 0.69645(6) & 0.04380(3) \\
			& 16  & 22113.65/8  & 0.68110(8) & 0.03750(4) \\
			& 24  & 6205.57/7   & 0.6704(1)  & 0.03334(5) \\
			& 32  & 2340.91/6   & 0.6635(2)  & 0.03079(6) \\
			& 48  & 566.30/5    & 0.6500(4)  & 0.0261(1)  \\
			& 64  & 162.15/4    & 0.6422(5)  & 0.0234(2)  \\
			& 80  & 22.54/3     & 0.6343(8)  & 0.0208(3)  \\
\hline
			\hline
		\end{tabular}
	}
\end{table*}

\begin{table*}[h]
	\caption{Fits of $\langle M^2_{\rm P} \rangle$ to Eq.~(\ref{XP_K1}) near $K_2$ for $W=1.5$.}~\label{Tab_Xp}
	\centering
	\scalebox{0.8}{
		\tabcolsep 0.1in
		\begin{tabular}{c|c|c|c|c|c|c|c|c|c|c|c}
			\hline
			\hline 
			 $L_{\rm min}$ & $\chi^2/{\rm DOF}$   & $C_1$       & $a_0$        & $a_1$         & $C_2$       & $a_2$          & $a_3$         & $d_1$        & $d_2$        & $n_0$      & $n_1$        \\
			\hline
			8  & 43.85/8 & 2.177(5) & 0.89243(8) & -0.0594(5) & 1.60(3) & -0.00419(5) & -0.00029(7) & 0.0955(8) & -0.175(3)   & 0.224(8) & -0.008(2) \\
			16 & 13.06/5 & 2.22(1)  & 0.8918(1)  & -0.0590(8) & 1.63(6) & -0.00414(9) & -0.0004(1)  & 0.085(2)  & -0.10(1)    & 0.23(2)  & -0.006(3) \\
			32 & 0.50/2  & 2.32(4)  & 0.8906(5)  & -0.060(3)  & 1.5(2)  & -0.0043(3)  & -0.0002(3)  & 0.05(1)   & 0.4(2)      & 0.20(6)  & -0.01(1)  \\
			8  & 60.37/9 & 2.173(5) & 0.89250(8) & -0.0578(2) & 1.72(2) & -0.00408(4) & -0.00053(4) & 0.0959(8) & -0.175(3)   & 0.255(4) & -        \\
			16 & 17.08/6 & 2.22(1)  & 0.8918(1)  & -0.0574(2) & 1.75(2) & -0.00399(5) & -0.00054(4) & 0.085(2)  & -0.10(1)    & 0.264(5) & -        \\
			32 & 1.39/3  & 2.32(4)  & 0.8905(5)  & -0.0575(3) & 1.74(2) & -0.00401(6) & -0.00050(4) & 0.05(1)   & 0.4(2)      & 0.264(6) & -        \\
			64 & 0.00/0  & -2.3(21) & 0.97(6)    & -0.06(1)   & 2.1(5)  & -0.0037(7)  & -0.001(2)   & 6.0(67)   & -103.1(538) & 0.1(3)   & -       \\
			\hline
			\hline
		\end{tabular}
	}
\end{table*}

\begin{table*}[h]
	\caption{Fits of $G_z$ to Eq.~(\ref{K1}) ($\eta=1/4$, $\epsilon=1.11996-K$) near $K_2$ for $W=0.8$ and $1.5$.}~\label{Tab_3}
	\centering
	\scalebox{0.8}{
		\tabcolsep 0.1in
		\begin{tabular}{c|c|c|c|c|c|c|c|c|c|c|c|c}
			\hline
			\hline 
			$W$                    & $L_{\rm min}$ & $\chi^2/{\rm DOF}$   & $C_1$       & $a_0$        & $a_1$         & $C_2$       & $a_2$          & $a_3$         & $d_1$        & $d_2$        & $n_0$      & $n_1$        \\
			\hline
			0.8 & 8  & 20.02/14 & 4.72(4)  & 0.7753(4) & -0.0515(7) & 2.07(7)  & -0.0033(2) & -0.0006(1) & 0.067(2)  & -0.096(9) & 0.23(2) & -0.000(3) \\
			& 16 & 16.40/10 & 4.61(10) & 0.7764(9) & -0.0509(9) & 2.14(10) & -0.0033(3) & -0.0006(1) & 0.079(10) & -0.18(7)  & 0.25(3) & 0.007(8)  \\
			& 32 & 9.27/6   & 4.6(3)   & 0.776(3)  & -0.048(2)  & 2.5(2)   & -0.0031(3) & -0.0004(1) & 0.08(5)   & -0.3(7)   & 0.37(8) & 0.02(1)   \\
			& 8  & 20.02/15 & 4.72(4)  & 0.7753(4) & -0.0515(6) & 2.07(6)  & -0.0033(2) & -0.0006(1) & 0.067(2)  & -0.096(9) & 0.23(2) & -         \\
			& 16 & 17.04/11 & 4.62(10) & 0.7762(9) & -0.0513(8) & 2.10(8)  & -0.0032(2) & -0.0006(1) & 0.078(10) & -0.17(7)  & 0.24(2) & -         \\
			& 32 & 12.71/7  & 4.6(3)   & 0.777(3)  & -0.051(1)  & 2.2(1)   & -0.0031(3) & -0.0005(1) & 0.09(5)   & -0.3(7)   & 0.26(5) & -         \\
			1.5 & 8  & 20.33/14 & 4.65(4)  & 0.7759(4) & -0.0504(7) & 2.26(7)  & -0.0032(2) & -0.0005(1) & 0.116(2)  & -0.175(9) & 0.31(2) & 0.011(7)  \\
			& 16 & 14.53/10 & 4.52(9)  & 0.7773(9) & -0.0496(9) & 2.36(10) & -0.0032(2) & -0.0004(1) & 0.131(9)  & -0.28(6)  & 0.34(3) & 0.018(8)  \\
			& 32 & 11.64/6  & 4.5(3)   & 0.777(3)  & -0.048(2)  & 2.6(2)   & -0.0030(3) & -0.0003(1) & 0.13(5)   & -0.3(7)   & 0.43(8) & 0.03(1) \\
			\hline
			\hline
		\end{tabular}
	}
\end{table*}

\begin{table*}[h]
	\caption{Fits of $G_z$ to Eq.~(\ref{eq2}) at $K_2$ for $W=0.8$ and $1.5$.}~\label{Tab_4}
	\centering
	\scalebox{1.0}{
		\tabcolsep 0.1in
		\begin{tabular}{c|c|c|c|c|c|c|c}
			\hline
			\hline 					
			$W$ & $L_{\rm min}$ & $\chi^2/{\rm DOF}$ & $\eta$       & $C_1$       & $a_0$        & $d_1$       & $d_2$       \\
			\hline
			0.8 & 8  & 0.14/1 & 0.250(2) & 4.8(16) & 0.77(2)   & 0.06(3)  & -0.07(6) \\
			& 8  & 0.14/2 & 1/4     & 4.75(7) & 0.7750(7) & 0.063(4) & -0.08(2) \\
			& 16 & 0.11/1 & 1/4     & 4.7(2)  & 0.775(2)  & 0.07(2)  & -0.1(1)  \\
			1.5 & 8  & 0.05/1 & 0.249(2) & 5.0(16) & 0.77(2)   & 0.11(3)  & -0.16(6) \\
			& 8  & 0.13/2 & 1/4     & 4.61(6) & 0.7764(7) & 0.119(4) & -0.18(2) \\
			& 16 & 0.08/1 & 1/4     & 4.6(2)  & 0.777(2)  & 0.12(2)  & -0.2(1) \\
			\hline
			\hline
		\end{tabular}
	}
\end{table*}

\begin{table*}[h]
	\caption{Fits of $G_z$ to Eq.~(\ref{QLR}) for ${\rm P_1}$, ${\rm P_2}$, ${\rm P_3}$ and ${\rm P_4}$.}~\label{Tab_5}
	\centering
	\scalebox{1.0}{
		\tabcolsep 0.1in
		\begin{tabular}{c|c|c|c|c|c|c}
			\hline
			\hline 					
			Parameters  & $L_{\rm min}$ & $\chi^2/{\rm DOF}$ & $\eta$        & $a$        & $b$       & $\omega$       \\
			\hline
			${\rm P_1}$         & 8      & 3.27/2                    & 0.13396(5) & 0.9387(2) & 0.076(8) & 1.50(7) \\
			& 16     & 2.53/1                    & 0.1339(1)  & 0.9382(8) & 0.04(2)  & 1.2(3)  \\
			${\rm P_2}$         & 8      & 1.64/2                    & 0.09293(3) & 0.9567(1) & 0.06(1)  & 1.8(1)  \\
			& 16     & 0.78/1                    & 0.0929(1)  & 0.9563(6) & 0.01(2)  & 1.2(6)  \\
			${\rm P_3}$         & 8      & 1.49/2                    & 0.13401(5) & 0.9390(2) & 0.086(6) & 1.40(5) \\
			& 16     & 1.35/1                    & 0.1341(1)  & 0.9392(6) & 0.10(5)  & 1.5(2)  \\
			${\rm P_4}$         & 8      & 1.42/2                    & 0.09292(3) & 0.9567(1) & 0.054(8) & 1.62(9) \\
			& 16     & 0.00/1                    & 0.0928(1)  & 0.9562(7) & 0.02(1)  & 1.1(4) \\
			\hline
			\hline
		\end{tabular}
	}
\end{table*}

\begin{table*}[h]
	\caption{Fits of $G_{\rm P}$ to Eq.~(\ref{K1}) ($\eta=1/4$, $\epsilon=1.11996-K$) near $K_2$ for $W=0.8$ and $1.5$.}~\label{Tab_6}
	\centering
	\scalebox{0.8}{
		\tabcolsep 0.1in
		\begin{tabular}{c|c|c|c|c|c|c|c|c|c|c|c|c}
			\hline
			\hline 
			$W$                    & $L_{\rm min}$ & $\chi^2/{\rm DOF}$   & $C_1$       & $a_0$        & $a_1$         & $C_2$       & $a_2$          & $a_3$         & $d_1$        & $d_2$        & $n_0$      & $n_1$        \\
			\hline
			0.8 & 8  & 26.28/14 & 7.4(3)   & 0.495(2) & -0.035(2) & 2.6(3)  & -0.0021(6)  & -0.0004(3) & 0.037(6) & -0.21(2)  & 0.14(5) & 0.01(2)  \\
			& 16 & 22.66/10 & 6.5(7)   & 0.500(4) & -0.034(3) & 2.7(4)  & -0.0022(6)  & -0.0003(2) & 0.08(3)  & -0.5(2)   & 0.17(9) & 0.02(2)  \\
			& 32 & 12.91/6  & 16.3(83) & 0.46(2)  & -0.039(7) & 1.4(9)  & -0.003(2)   & -0.0007(9) & -0.3(1)  & 4.9(20)   & -0.1(2) & -0.01(5) \\
			& 8  & 26.40/15 & 7.4(3)   & 0.495(1) & -0.035(2) & 2.5(2)  & -0.0020(4)  & -0.0004(3) & 0.037(6) & -0.21(2)  & 0.13(5) & -        \\
			& 16 & 23.21/11 & 6.5(7)   & 0.499(4) & -0.035(2) & 2.5(3)  & -0.0019(6)  & -0.0004(3) & 0.07(3)  & -0.5(2)   & 0.14(7) & -        \\
			& 32 & 13.00/7  & 16.3(83) & 0.46(2)  & -0.038(4) & 1.6(5)  & -0.003(2)   & -0.0005(6) & -0.3(1)  & 4.9(20)   & -0.1(1) & -        \\
			& 64 & 0.67/3   & -3.6(1)  & 0.598(5) & -0.037(7) & 0.2(8)  & -0.012(10)  & -0.008(7)  & 7.9(2)   & 38.8(839) & -1.4(3) & -        \\
\hline
			\hline
		\end{tabular}
	}
\end{table*}

\begin{table*}[h]
	\caption{Fits of $G_{\rm P}$ to Eq.~(\ref{eq2}) at $K_2$ for $W=0.8$ and $1.5$.}~\label{Tab_7}
	\centering
	\scalebox{1.0}{
		\tabcolsep 0.1in
		\begin{tabular}{c|c|c|c|c|c|c|c}
			\hline
			\hline 					
			$W$ & $L_{\rm min}$ & $\chi^2/{\rm DOF}$ & $\eta$       & $C_1$       & $a_0$        & $d_1$       & $d_2$       \\
			\hline
			0.8 & 16 & 0.00/0 & 0.271(7)  & -2.1(4)  & 0.65(3)  & 1.12(9) & 4.9(106) \\
			& 16 & 0.20/1 & 0.2676(8) & -1.81(5) & 0.634(2) & 1.06(8) & -        \\
			& 32 & 0.77/1 & 1/4      & 5.4(8)   & 0.505(5) & 0.09(3) & -        \\
			1.5 & 8  & 1.91/1 & 0.255(6)  & 2.8(33)  & 0.61(5)  & 0.2(1)  & -0.4(2)  \\
			& 8  & 2.43/2 & 1/4      & 6.7(3)   & 0.562(2) & 0.12(1) & -0.27(4) \\
			& 16 & 2.29/1 & 1/4      & 7.1(11)  & 0.560(6) & 0.11(5) & -0.1(3) \\
			\hline
			\hline
		\end{tabular}
	}
\end{table*}

\begin{table*}[h]
	\caption{Fits of $G_{\rm P}$ to Eq.~(\ref{QLR}) for ${\rm P_1}$, ${\rm P_2}$, ${\rm P_3}$ and ${\rm P_4}$.}~\label{Tab_8}
	\centering
	\scalebox{1.0}{
		\tabcolsep 0.1in
		\begin{tabular}{c|c|c|c|c|c|c}
			\hline
			\hline 					
			Parameters  & $L_{\rm min}$ & $\chi^2/{\rm DOF}$ & $\eta$        & $a$        & $b$       & $\omega$       \\
			\hline
			${\rm P_1}$ & 8  & 2.41/2 & 0.1342(5)  & 0.681(2)   & -0.04(1)  & 1.1(2)  \\
			& 16 & 1.10/1 & 0.14(3)    & 0.7(3)     & -0.1(3)   & 0.2(7)  \\
			& 8  & 2.62/3 & 0.1344(2)  & 0.6819(4)  & -0.039(2) & 1       \\
			& 16 & 2.15/2 & 0.1342(3)  & 0.6815(8)  & -0.036(5) & 1       \\
			& 32 & 0.60/1 & 0.1351(7)  & 0.684(3)   & -0.07(3)  & 1       \\
			${\rm P_2}$ & 8  & 0.43/2 & 0.0933(4)  & 0.726(1)   & -0.08(1)  & 1.2(1)  \\
			& 16 & 0.32/1 & 0.094(2)   & 0.728(7)   & -0.05(5)  & 0.9(7)  \\
			& 8  & 2.29/3 & 0.0939(1)  & 0.7284(4)  & -0.064(2) & 1       \\
			& 16 & 0.32/2 & 0.0936(3)  & 0.7274(8)  & -0.057(5) & 1       \\
			& 32 & 0.32/1 & 0.0936(7)  & 0.727(3)   & -0.06(3)  & 1       \\
			${\rm P_3}$ & 8  & 0.18/2 & 0.1339(3)  & 0.7581(10) & 0.04(1)   & 1.2(3)  \\
			& 16 & 0.03/1 & 0.1341(4)  & 0.759(2)   & 0.2(8)    & 1.9(22) \\
			& 8  & 1.18/3 & 0.1336(1)  & 0.7569(3)  & 0.026(2)  & 1       \\
			& 16 & 0.35/2 & 0.1337(2)  & 0.7574(6)  & 0.023(4)  & 1       \\
			& 32 & 0.02/1 & 0.1340(5)  & 0.758(2)   & 0.01(2)   & 1       \\
			& 32 & 0.45/2 & 0.13432(9) & 0.7596(3)  & -        & -      \\
			& 64 & 0.05/1 & 0.1342(2)  & 0.7591(8)  & -        & -      \\
			${\rm P_4}$ & 8  & 0.49/3 & 0.09289(9) & 0.8017(3)  & 0.000(1)  & 1       \\
			& 16 & 0.45/2 & 0.0929(2)  & 0.8018(5)  & -0.001(3) & 1       \\
			& 32 & 0.42/1 & 0.0930(4)  & 0.802(2)   & -0.00(2)  & 1       \\
			& 8  & 0.49/4 & 0.09289(3) & 0.80172(8) & -        & -      \\
			& 16 & 0.48/3 & 0.09289(4) & 0.8017(1)  & -        & -      \\
			& 32 & 0.46/2 & 0.09290(8) & 0.8017(2)  & -        & -      \\
			& 64 & 0.45/1 & 0.0929(2)  & 0.8018(7)  & -        & - \\
			\hline
			\hline
		\end{tabular}
	}
\end{table*}

\bibliography{papers}